\documentclass[fleqn,usenatbib]{mnras}

\usepackage{newtxtext,newtxmath}

\usepackage[T1]{fontenc}

\DeclareRobustCommand{\VAN}[3]{#2}
\let\VANthebibliography\thebibliography
\def\thebibliography{\DeclareRobustCommand{\VAN}[3]{##3}\VANthebibliography}

\usepackage{graphicx}	
\usepackage{amsmath}	

\title[{\it JWST} Spectroscopy of Early Galaxies]{{\it JWST}/NIRSpec Spectroscopy of $z=7-9$ Star Forming Galaxies with CEERS: New Insight into Bright Ly$\alpha$ Emitters in Ionized Bubbles}

\author[M. Tang et al.]{
Mengtao Tang$^{1}$\thanks{E-mail: tangmtasua@arizona.edu},
Daniel P. Stark$^{1}$,
Zuyi Chen$^{1}$,
Charlotte Mason$^{2,3}$,
Michael Topping$^{1}$,
Ryan Endsley$^{4}$,\newauthor 
Peter Senchyna$^{5}$, 
Ad\`ele Plat$^{1}$,
Ting-Yi Lu$^{2,3}$,
Lily Whitler$^{1}$,
Brant Robertson$^{6}$, \&
St\'ephane Charlot$^{7}$
\\
\vspace{0in}\\
$^{1}$Steward Observatory, University of Arizona, 933 N Cherry Ave, Tucson, AZ 85721, USA \\
$^{2}$Cosmic Dawn Center (DAWN) \\
$^{3}$Niels Bohr Institute, University of Copenhagen, Jagtvej 128, 2200 K{\o{}}benhavn N, Denmark \\
$^{4}$Department of Astronomy, University of Texas, Austin, TX 78712, USA \\
$^{5}$The Observatories of the Carnegie Institution for Science, 813 Santa Barbara Street, Pasadena, CA 91101, USA \\
$^{6}$Department of Astronomy and Astrophysics, University of California, Santa Cruz, 1156 High Street, Santa Cruz, CA 95064, USA \\
$^{7}$Sorbonne Universit\'e, CNRS, UMR 7095, Institut d'Astrophysique de Paris, 98 bis bd Arago, 75014 Paris, France
}

\date{Accepted XXX. Received YYY; in original form ZZZ}

\pubyear{2023}

\begin{document}
\label{firstpage}
\pagerange{\pageref{firstpage}--\pageref{lastpage}}
\maketitle

\begin{abstract}
We describe new {\it JWST}/NIRSpec observations of galaxies at $z\gtrsim7$ taken from the CEERS survey. Previous observations of this area have revealed associations of Ly$\alpha$ emitters at redshifts ($z=7.5$, $7.7$, $8.7$) where the intergalactic medium (IGM) is thought to be mostly neutral, leading to suggestions that these systems are situated in large ionized bubbles. We identify 21 $z\gtrsim7$ galaxies with robust redshifts in the CEERS dataset, including 10 in the Ly$\alpha$ associations. Their spectra are indicative of very highly ionized and metal poor gas, with line ratios (O32 $=17.84$ and Ne3O2 $=0.89$, linear scale) and metallicity ($12+\log{(\rm{O/H})}=7.84$) that are rarely seen at lower redshifts. We find that the most extreme spectral properties are found in the six $z\gtrsim7$ Ly$\alpha$ emitters in the sample. Each has a hard ionizing spectrum indicating that their visibility is likely enhanced by efficient ionizing photon production. Ly$\alpha$ velocity offsets are found to be very large ($\gtrsim300$~km~s$^{-1}$), likely also contributing to their detectability. We find that Ly$\alpha$ in $z\gtrsim7$ galaxies is $6-12\times$ weaker than in lower redshift samples with matched rest-optical spectral properties. If the bubbles around the Ly$\alpha$ emitters are relatively small ($\lesssim0.5-1$~pMpc), we may expect such significant attenuation of Ly$\alpha$ in these ionized regions. We discuss several other effects that may contribute to weaker Ly$\alpha$ emission at $z\gtrsim7$. Deep spectroscopy of fainter galaxies in the vicinity of the Ly$\alpha$ emitters will better characterize the physical scale of the ionized bubbles in this field.
\end{abstract}

\begin{keywords}
dark ages, reionization, first stars - cosmology: observations - galaxies: evolution - galaxies: high-redshift
\end{keywords}




\section{Introduction} \label{sec:intro}

The reionization of intergalactic hydrogen is an important landmark in the early history of structure formation. The details of how and when this process came to pass depends sensitively on the nature of the first energetic objects emitting hydrogen ionizing radiation. Over the past two decades, concerted observational efforts have been devoted to constraining the history of reionization and the nature of early ionizing sources \citep{Dijkstra2014,Stark2016,Robertson2022}. The process is thought to be driven by the radiation from both massive stars \citep[e.g.,][]{Robertson2015,Stanway2016,Dayal2018,Finkelstein2019,Naidu2020} and active galactic nuclei (AGN) \citep[e.g.,][]{Haardt2015,Matsuoka2018,Kulkarni2019,Dayal2020}. Measurements of the Thomson scattering optical depth faced by cosmic microwave background (CMB) photons suggest a significant component of free electrons are present by $z\sim8$ \citep{Planck2016,Planck2020}. Quasar absorption spectra tell us that the intergalactic medium (IGM) is partially neutral at $z\simeq7$ \citep[e.g.,][]{Greig2017,Davies2018,Wang2020,Yang2020a} and mostly ionized by $z\simeq5.5-6$ \citep[e.g.,][]{Fan2006,McGreer2015,Yang2020b}. 

Observations of Ly$\alpha$ emission from star-forming galaxies provide a complementary picture. While strong Ly$\alpha$ is commonly seen in $z\simeq6$ galaxies \citep[e.g.,][]{Stark2011,Curtis-Lake2012,Cassata2015,DeBarros2017,Jiang2017}, it is very rare at $z\simeq7-8$ \citep[e.g.,][]{Fontana2010,Stark2010,Treu2013,Caruana2014,Pentericci2014,Tilvi2014,Hoag2019,Mason2019,Jung2020}. The equivalent width (EW) distribution of Ly$\alpha$ shows marked evolution toward weaker line emission between $z\simeq6$ and $z\simeq8$ \citep[e.g.,][]{Ono2012,Schenker2014,Tilvi2014,Jung2018,Jung2020}, as would be expected if the IGM is significantly neutral at $z\gtrsim7$ ($x_{\rm{HI}}\gtrsim0.6$; e.g., \citealt{Mesinger2015,Zheng2017,Mason2018a,Mason2019,Hoag2019,Whitler2020}).

As wider-area infrared imaging surveys emerged over the last decade, attention has turned to massive reionization-era galaxies that sit at the bright end of the ultraviolet (UV) luminosity function (M$_{\rm{UV}}<-21$). Many of these UV-bright galaxies were found to have Ly$\alpha$ emission at $z\simeq7.5-9$ \citep[e.g.,][]{Ono2012,Finkelstein2013,Oesch2015,Zitrin2015,Roberts-Borsani2016,Stark2017,Larson2022}. Initial studies suggested that Ly$\alpha$ was fairly ubiquitous in this population \citep{Stark2017}. Subsequent surveys over larger volumes have confirmed that the Ly$\alpha$ EW distribution in the UV-bright population shows no significant decline over $6<z<8$ \citep[e.g.][]{Endsley2021b,Jung2022,Roberts-Borsani2023a}, suggesting that Ly$\alpha$ photons emitted by massive UV-luminous galaxies are minimally impacted by the neutral IGM at $z\simeq7$.

The luminosity-dependence of the evolving Ly$\alpha$ visibility can be understood if the brightest galaxies tend to trace rare ionized regions in the IGM. If the bubbles are large enough, Ly$\alpha$ will redshift far enough into the damping wing before encountering neutral hydrogen, boosting the transmission of the line through the IGM \citep[e.g.,][]{Mason2018b,Qin2022}. Galaxies in overdensities produce an excess number of ionizing photons, so the largest bubbles are expected to be found in overdense regions \citep[e.g.,][]{Barkana2004,Furlanetto2004,Iliev2006,Castellano2016,Garaldi2022,Kannan2022,Leonova2022,Lu2023}. Whether the bright Ly$\alpha$ emitters at $z\gtrsim7$ are tracing such large ionized bubbles is not clear. Several studies have recently reported discovery of excess numbers of bright galaxies in the vicinity of $z\gtrsim7$ Ly$\alpha$ emitters \citep[e.g.,][]{Castellano2018,Tilvi2020,Leonova2022,Endsley2022a,Jung2022}, suggesting possible overdensities in these regions. Some of these bright neighboring galaxies have also been shown to exhibit strong Ly$\alpha$ emission \citep[e.g.,][]{Jung2020,Jung2022,Endsley2022a}, as might be expected if the ionized regions extend over very large volumes.

Much of this progress has been focused on the Cosmic Assembly Near-infrared Deep Extragalactic Legacy Survey (CANDELS; \citealt{Grogin2011,Koekemoer2011}) Extended Groth Strip (EGS; \citealt{Davis2007}) field, with groups of Ly$\alpha$ emitters confirmed at $z\simeq7.5$, $z\simeq7.7$, and $z\simeq8.7$. The first Ly$\alpha$ emitting galaxies were identified as bright dropouts selected based on the presence of intense [O~{\small III}]+H$\beta$ emission \citep{Roberts-Borsani2016}. Keck spectroscopic follow-up of this initial sample revealed Ly$\alpha$ emission at $z=7.48$ \citep{Roberts-Borsani2016,Stark2017}, $z=7.73$ \citep{Oesch2015}, and $z=8.68$ \citep{Zitrin2015}. Subsequent work has revealed additional Ly$\alpha$ emitters within $5-10$ physical Mpc (pMpc) of each of these galaxies \citep{Tilvi2020,Larson2022,Jung2022}, potentially indicating the presence of several extended ionized regions. Deep imaging with the {\it Hubble Space Telescope} ({\it HST}) suggests that this field may have a significant overdensity of galaxies at these redshifts \citep{Leonova2022}, as would be required to power such large ionized regions. 

Previous data do not strongly constrain the size of the ionized bubbles around the $z\gtrsim7$ Ly$\alpha$ emitters in the EGS. The physical scale of ionized regions at a given redshift will depend on the progress of reionization and the nature of the sources driving the process \citep[e.g.,][]{Furlanetto2004,McQuinn2007,Mason2020,Hutter2021,Leonova2022,Smith2022}. If the typical bubble sizes are large at $z\simeq7.7-8.8$, we may expect the entire $5-10$~pMpc region spanned by the Ly$\alpha$ emitters at each redshift to be ionized. But the existing observations are also consistent with there being smaller bubbles ($\lesssim1$~pMpc) centered on the individual bright Ly$\alpha$ emitting galaxies. In this latter case, the visibility of Ly$\alpha$ may be boosted by other factors. Intense radiation fields can increase the production efficiency of Ly$\alpha$ photons, enhancing the likelihood of detection. Outflows can result in Ly$\alpha$ profiles that are redshifted from systemic, boosting the transmission through the IGM \citep[e.g.,][]{Stark2017,Mason2018b,Endsley2022b}. Unfortunately little is known about the importance of either of these factors in regulating Ly$\alpha$ detection rates in the reionization era, limiting our ability to interpret $z\gtrsim7$ observations in the context of bubble sizes.

Spectroscopy of $z>6$ galaxies with {\it JWST} \citep{Gardner2023} has begun providing important insights into the physical properties of systems in the reionization era \citep[e.g.,][]{Schaerer2022b,Sun2022,Sun2023,Taylor2022,Wang2022,Backhaus2023,Bunker2023,Cameron2023,Carnall2023,Curtis-Lake2023,Fujimoto2023,Hsiao2023,Jung2023,Larson2023,Matthee2023,Nakajima2023,Roberts-Borsani2023b,Sanders2023a,Saxena2023b,Shapley2023,Tacchella2023,Trump2023}. In this paper, we investigate new {\it JWST}/NIRSpec \citep{Ferruit2022,Jakobsen2022} observations of galaxies at $z\gtrsim7$ in the EGS field taken as part of the Cosmic Evolution Early Release Science (CEERS\footnote{\url{https://ceers.github.io/}}) program (Finkelstein et al in prep., see also \citealt{Finkelstein2022,Finkelstein2023}). This {\it JWST} dataset opens many new avenues of characterizing early galaxies and their surrounding IGM. We investigate the spectroscopic properties of a sample of 21 galaxies at $z\gtrsim7$ with robust emission line detections in the CEERS observations. Using this database, we explore how the galaxies in the EGS with Ly$\alpha$ detections at $z\gtrsim7$ compare to the general population. We investigate the physics regulating Ly$\alpha$ production and escape, leveraging new constraints on ionizing photon production efficiency, Ly$\alpha$ velocity offsets, and the Ly$\alpha$ escape fraction. The observations include 10 galaxies in the three Ly$\alpha$ associations, improving characterization of the ionized regions surrounding these systems.

The organization of this paper is as follows. In Section~\ref{sec:data} we describe the {\it JWST}/NIRSpec observations (Section~\ref{sec:observation}), spectral energy distributions of galaxies in our spectroscopic sample (Section~\ref{sec:sed}), rest-frame optical (Section~\ref{sec:opt_lines}) and UV emission line measurements (Section~\ref{sec:uv_lines}). Based on CEERS spectroscopic results, we present the physical properties of galaxies in the three groups of Ly$\alpha$ emitting galaxies at $z>7$ in the CEERS field in Section~\ref{sec:lae}. We use the results to discuss the implications for Ly$\alpha$ visibility in the reionization era in Section~\ref{sec:discussion}. Finally, we summarize our conclusions in Section~\ref{sec:summary}. Throughout the paper we adopt a $\Lambda$-dominated, flat universe with $\Omega_{\Lambda}=0.7$, $\Omega_{\rm{M}}=0.3$, and H$_0=70$~km~s$^{-1}$~Mpc$^{-1}$. All magnitudes in this paper are quoted in the AB system \citep{Oke1983} and all EWs are quoted in the rest frame.


\section{Data and analysis} \label{sec:data}

\subsection{NIRSpec Observations and Reduction} \label{sec:observation}

We use the publicly available CEERS {\it JWST}/NIRSpec spectra which are centered on the EGS field. The CEERS NIRSpec observations of the EGS field include 6 pointings using the medium resolution (MR) grating with 3 grating/filter pairs (G140M/F100LP, G235M/F170LP, and G395M/F290LP) covering $1-5\mu$m and 4 additional pointings using the lower resolution prism providing simultaneous coverage of $1-5\mu$m. Details of the CEERS NIRSpec observations are summarized in the CEERS public Phase 2 PDF\footnote{\url{https://www.stsci.edu/jwst/phase2-public/1345.pdf}}. In brief, each grating and prism was observed with three exposures of 13 groups each, using the standard three-shutter slits for MR grating observations and three-point nodding. The total exposure time of each grating and prism is 2889 seconds. The spectral resolution of MR grating is $R\simeq1000$ (corresponding to a velocity resolution $\sigma\simeq130$~km~s$^{-1}$), and of prism is $R\simeq100$ ($\sigma\simeq1300$~km~s$^{-1}$). Overall there are 319 galaxies placed on the 6 MR grating pointings (52 $-$ 55 galaxies on each pointing) and 466 galaxies on the 3 prism pointings (151 $-$ 161 galaxies on each pointing). Targets were selected from Lyman break galaxies identified using the {\it HST} \citep{Bouwens2015,Bouwens2021,Finkelstein2015} and {\it JWST}/NIRCam \citep{Rieke2023} imaging \citep{Endsley2023a,Finkelstein2023,Whitler2023a}, spanning a wide photometric redshift range of $z\simeq0.5-12$. The CEERS collaboration prioritize targets that are at redshifts where key emission lines or continuum will be detected at $1-5\mu$m that allow the measurement of redshifts and line diagnostics.

The 2D NIRSpec spectra were reduced following the methods described in \citet{Shapley2023}. Individual uncalibrated detector exposures were first passed through the {\it JWST} \texttt{calwebb\_detector1} pipeline \footnote{\url{https://jwst-pipeline.readthedocs.io/}}. This step implements masking of all pixels that are saturated, subtraction of signal due to bias and dark current, and masking of `snowballs' and `showers' resulting from cosmic rays. The resulting images from this step were then corrected for striping. This is done by estimating and subtracting the $1/f$ noise in each image. The 2D spectrum of each slit on the micro-shutter assembly (MSA; \citep{Ferruit2022}) was then cut out. We applied the flat-field correction, photometry calibration, and the wavelength solution using the up-to-date calibration reference data system (CRDS) context (\texttt{jwst\_1027.pmap}). Each slitlet was then rectified, and interpolated onto a common wavelength grid for its grating and filter combination. The calibrated spectra were finally combined following the three-shutter dither pattern, excluding pixels that had been masked in a previous step of the reduction. The 2D error spectra were calculated as a combination of the variance from Poisson noise, readout noise, flat-fielding, and variance between exposures, all summed in quadrature.

\begin{figure*}
\begin{center}
\includegraphics[width=\linewidth]{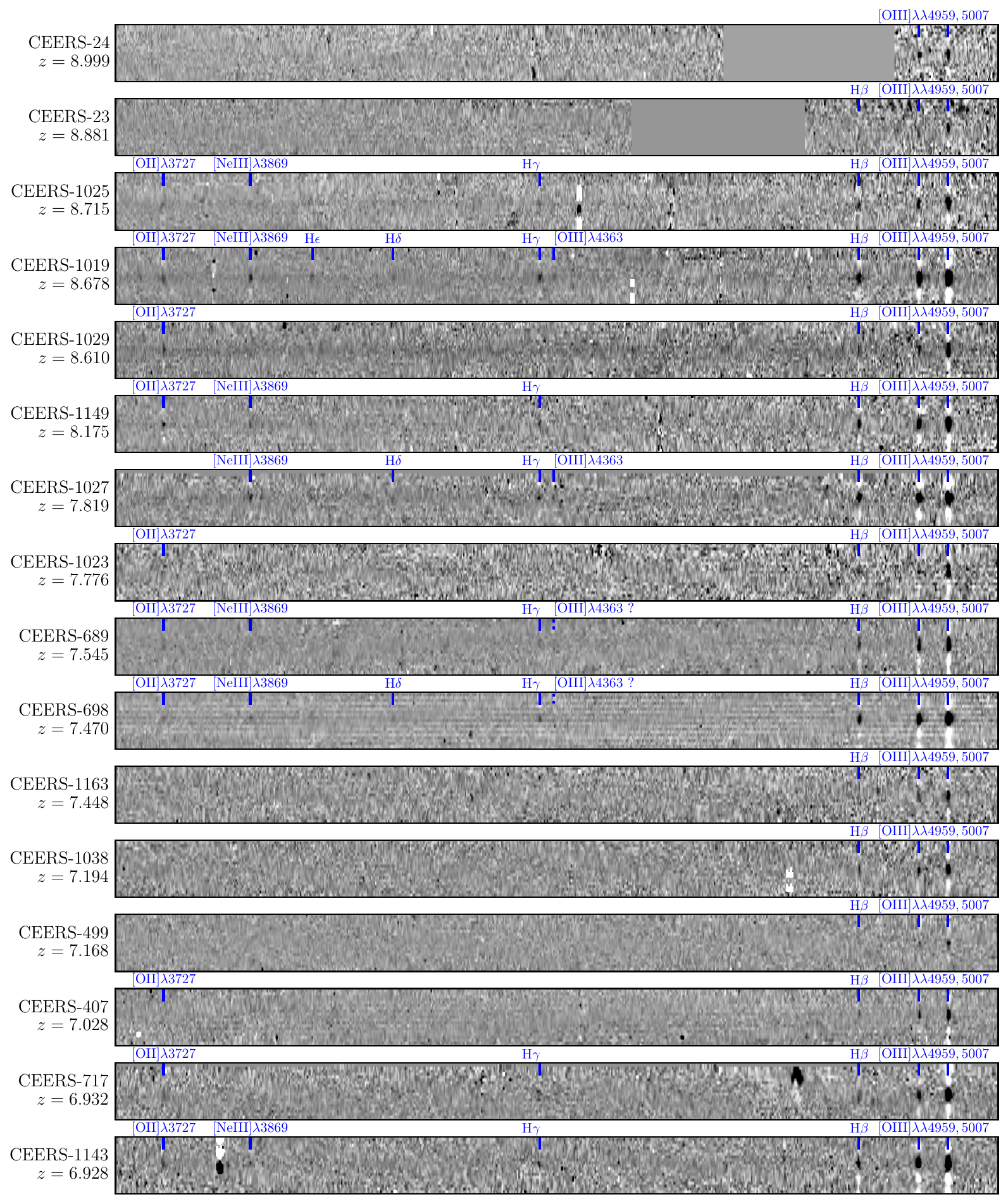}
\caption{Medium resolution ($R\sim1000$) 2D NIRSpec spectra of the sixteen galaxies at $z\gtrsim7$ at rest-frame optical wavelengths. Blue lines show the rest-frame optical emission lines detected on 2D spectra. The two tentative detections of [O~{\scriptsize III}]$\lambda4363$ in CEERS-689 and CEERS-698 are shown by blue dotted lines.}
\label{fig:spectra_opt_2d}
\end{center}
\end{figure*}

\begin{figure}
\begin{center}
\includegraphics[width=\linewidth]{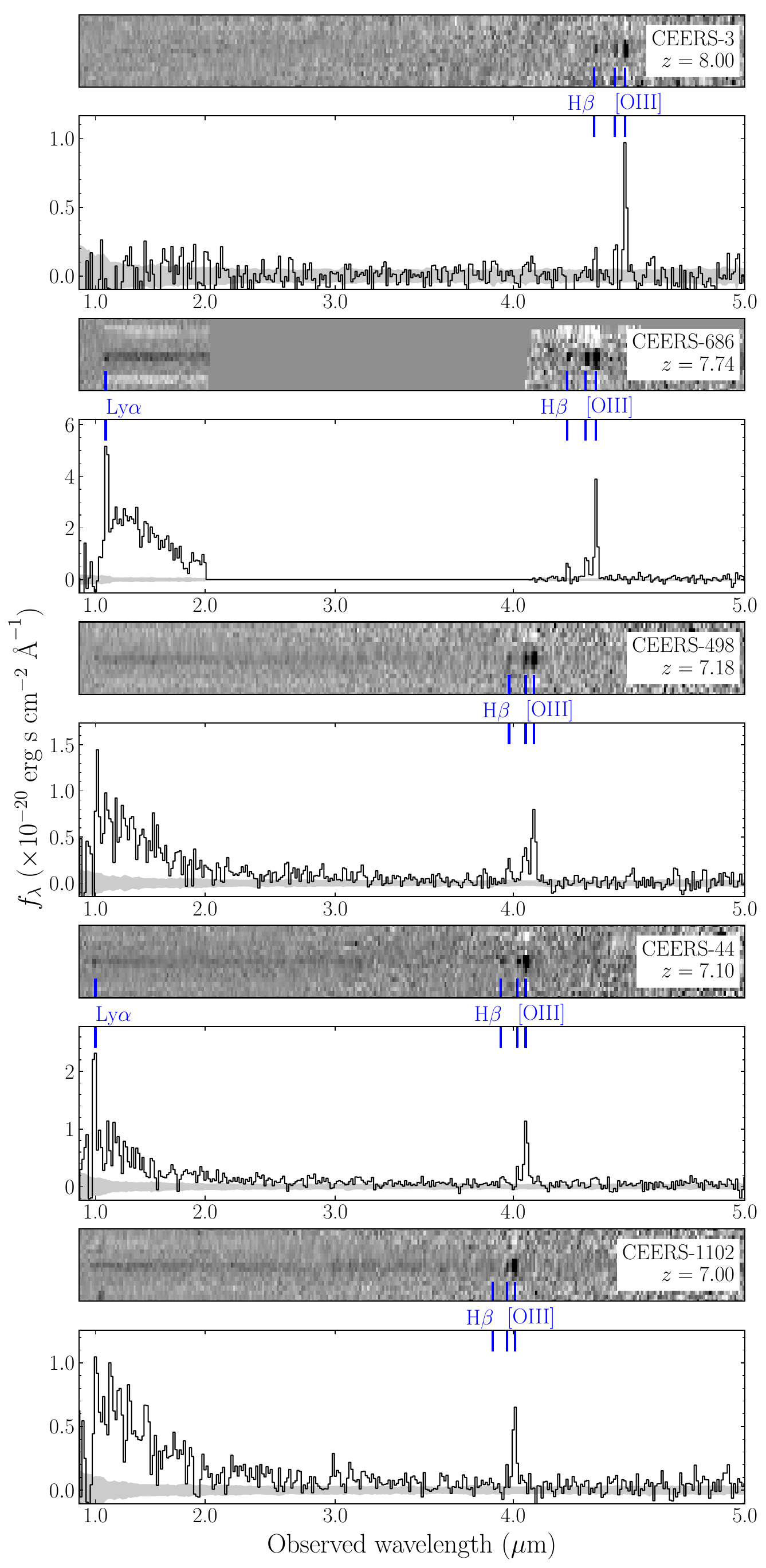}
\caption{Low resolution ($R\sim100$) 2D (top of each panel) and 1D (bottom of each panel) NIRSpec prism spectra of the five galaxies at $z>7$ with prism observations. Blue lines mark the position of emission lines detected in the spectrum of each object. The grey shaded regions show the uncertainty of flux.}
\label{fig:spectra_prism}
\end{center}
\end{figure}

Every reduced spectrum (low and medium resolution) was visually examined by a small group of the co-authors (MT and ZC). We searched for emission lines in all galaxies with spectra, assigning spectroscopic redshifts where lines are detected. In total we found 21 unique galaxies (16 with MR grating spectra and additional 5 with prism spectra) with robust emission line detections placing them confidently at $z\gtrsim7$. In Fig.~\ref{fig:spectra_opt_2d} and Fig.~\ref{fig:spectra_prism} we show the 2D NIRSpec spectra with emission line detections of the sixteen galaxies with MR grating observations and the other five with prism observations, respectively. Additional sources are found with detections of single or lower S/N lines which may be at $z\gtrsim7$, but we limit our analysis here to the 21 galaxies that we can robustly place in our chosen redshift range. Redshift confirmation is typically achieved via detection of the [O~{\small III}]$\lambda\lambda4959,5007$ doublet and one or more of the hydrogen Balmer lines (i.e., H$\beta$, H$\gamma$), but we additionally detect Ly$\alpha$ and fainter UV lines (e.g., C~{\small III}]) in several cases. These 21 objects range in redshift between $z=6.928$ and $z=8.999$, with a median redshift of $z=7.545$. The spectroscopic sample includes 10 galaxies in the 3 associations of Ly$\alpha$ emitters (LAEs) at $z=7.48$, $z=7.73$, and $z=8.68$, which we will discuss in more detail in Section~\ref{sec:lae}. An overview of our $z\gtrsim7$ spectroscopic catalog is given in Table~\ref{tab:source}.

\begin{table*}
\centering
\begin{tabular}{ccccccccc}
\hline
ID & $z_{\rm{spec}}$ & R.A. & Decl. & NIRCam/F150W & WFC3/F160W & M$_{\rm{UV}}$ & optical lines$^{\rm{a}}$ & Ref. \\
\hline
\hline
CEERS-24$^{\rm{b}}$ & $8.999$ & 14:19:35.336 & +52:50:37.87 & $27.93\pm0.12$ & ... & $-19.39$ & [O~{\scriptsize III}] \\
\hline
CEERS-23$^{\rm{b}}$ & $8.881$ & 14:19:36.301 & +52:50:49.19 & $28.44\pm0.23$ & ... & $-18.87$ & H$\beta$, [O~{\scriptsize III}] \\
\hline
CEERS-1025$^{\rm{b}}$ & $8.715$ & 14:19:52.211 & +52:55:58.63 & $26.15\pm0.02$ & ... & $-21.12$ & 
\begin{tabular}{@{}c@{}}[O~{\scriptsize II}], [Ne~{\scriptsize III}], \\ H$\gamma$, H$\beta$, [O~{\scriptsize III}]\end{tabular} & \\
\hline
CEERS-1019$^{\rm{b}}$ & $8.678$ & 14:20:08.494 & +52:53:26.38 & $24.85\pm0.02$ & ... & $-22.42$ & 
\begin{tabular}{@{}c@{}}[O~{\scriptsize II}], [Ne~{\scriptsize III}], H$\epsilon$, H$\delta$, \\ H$\gamma$, [O~{\scriptsize III}]$\lambda4363$, H$\beta$, [O~{\scriptsize III}]\end{tabular} & [1] \\
\hline
CEERS-1029$^{\rm{b}}$ & $8.610$ & 14:20:52.503 & +53:04:11.50 & ... & $25.63\pm0.07$ & $-21.63$ & [O~{\scriptsize II}], H$\beta$, [O~{\scriptsize III}] & [2] \\
\hline
CEERS-1149$^{\rm{b}}$ & $8.175$ & 14:20:21.531 & +52:57:58.26 & ... & $26.58\pm0.15$ & $-20.60$ & 
\begin{tabular}{@{}c@{}}[O~{\scriptsize II}], [Ne~{\scriptsize III}], \\ H$\gamma$, H$\beta$, [O~{\scriptsize III}]\end{tabular} & \\
\hline
CEERS-3$^{\rm{c}}$ & $8.00$ & 14:20:01.245 & +52:59:47.69 & $28.61\pm0.28$ & ... & $-18.54$ & H$\beta$, [O~{\scriptsize III}] & \\
\hline
CEERS-1027$^{\rm{b}}$ & $7.819$ & 14:19:31.919 & +52:50:25.50 & $26.50\pm0.03$ & ... & $-20.60$ & 
\begin{tabular}{@{}c@{}}[Ne~{\scriptsize III}], H$\delta$, H$\gamma$, \\ {[}O~{\scriptsize III}]$\lambda4363$, H$\beta$, [O~{\scriptsize III}]\end{tabular} & \\
\hline
CEERS-1023$^{\rm{b}}$ & $7.776$ & 14:20:45.219 & +53:02:01.13 & ... & $26.23\pm0.16$ & $-20.87$ & [O~{\scriptsize II}], H$\beta$, [O~{\scriptsize III}] & \\
\hline
CEERS-686$^{\rm{c}}$ & $7.74$ & 14:20:36.207 & +52:59:22.42 & ... & $26.44\pm0.10$ & $-20.66$ & H$\beta$, [O~{\scriptsize III}] & [3] \\
\hline
CEERS-689$^{\rm{b}}$ & $7.545$ & 14:19:59.773 & +52:56:31.12 & ... & $24.91\pm0.18$ & $-22.14$ & 
\begin{tabular}{@{}c@{}}[O~{\scriptsize II}], [Ne~{\scriptsize III}], H$\gamma$, \\ {[}O~{\scriptsize III}]$\lambda4363$, H$\beta$, [O~{\scriptsize III}]\end{tabular} & \\
\hline
CEERS-698$^{\rm{b}}$ & $7.470$ & 14:20:12.076 & +53:00:26.79 & ... & $25.18\pm0.04$ & $-21.86$ & 
\begin{tabular}{@{}c@{}}[O~{\scriptsize II}], [Ne~{\scriptsize III}], H$\delta$, H$\gamma$, \\ {[}O~{\scriptsize III}]$\lambda4363$, H$\beta$, [O~{\scriptsize III}]\end{tabular} & [4],[5] \\
\hline
CEERS-1163$^{\rm{b}}$ & $7.448$ & 14:19:57.712 & +52:58:19.16 & ... & $26.80\pm0.17$ & $-20.24$ & [O~{\scriptsize II}], H$\beta$, [O~{\scriptsize III}] & \\
\hline
CEERS-1038$^{\rm{b}}$ & $7.194$ & 14:20:09.527 & +52:54:05.75 & $27.73\pm0.13$ & ... & $-19.25$ & H$\beta$, [O~{\scriptsize III}] & \\
\hline
CEERS-498$^{\rm{c}}$ & $7.18$ & 14:19:15.131 & +52:50:03.30 & $26.76\pm0.03$ & ... & $-20.21$ & H$\beta$, [O~{\scriptsize III}] & \\
\hline
CEERS-499$^{\rm{b}}$ & $7.168$ & 14:19:15.121 & +52:50:03.01 & $30.00\pm0.54$ & ... & $-16.97$ & H$\beta$, [O~{\scriptsize III}] & \\
\hline
CEERS-44$^{\rm{c}}$ & $7.10$ & 14:20:00.268 & +53:00:40.57 & $27.59\pm0.06$ & ... & $-19.37$ & H$\beta$, [O~{\scriptsize III}] & \\
\hline
CEERS-407$^{\rm{b}}$ & $7.028$ & 14:19:21.436 & +52:52:57.23 & $27.99\pm0.09$ & ... & $-18.95$ & [O~{\scriptsize II}], H$\beta$, [O~{\scriptsize III}] & \\
\hline
CEERS-1102$^{\rm{c}}$ & $7.00$ & 14:20:21.851 & +52:57:15.43 & ... & $26.90\pm0.22$ & $-20.04$ & H$\beta$, [O~{\scriptsize III}], H$\alpha$ & \\
\hline
CEERS-717$^{\rm{b}}$ & $6.932$ & 14:20:19.537 & +52:58:19.85 & ... & $25.28\pm0.06$ & $-21.64$ & [O~{\scriptsize II}], H$\gamma$, H$\beta$, [O~{\scriptsize III}] & \\
\hline
CEERS-1143$^{\rm{b}}$ & $6.928$ & 14:20:18.482 & +52:58:10.22 & ... & $26.88\pm0.18$ & $-20.04$ & 
\begin{tabular}{@{}c@{}}[O~{\scriptsize II}], [Ne~{\scriptsize III}], \\ H$\gamma$, H$\beta$, [O~{\scriptsize III}]\end{tabular} & \\
\hline
\multicolumn{9}{l}{Note:} \\
\multicolumn{9}{l}{a. Rest-frame optical emission lines detections. [O~{\scriptsize II}], [Ne~{\scriptsize III}], and [O~{\scriptsize III}] refer to [O~{\scriptsize II}]$\lambda3727$, [Ne~{\scriptsize III}]$\lambda3869$, and [O~{\scriptsize III}]$\lambda\lambda4959,5007$, respectively.} \\
\multicolumn{9}{l}{b. Emission lines detected in NIRSpec MR grating spectra.} \\
\multicolumn{9}{l}{c. Emission lines detected in NIRSpec prism spectra.} \\
\multicolumn{9}{l}{References: [1]. \protect\citet{Zitrin2015}; [2]. \citet{Larson2022}; [3]. \citet{Jung2022}; [4]. \citet{Roberts-Borsani2016}; [5]. \citet{Stark2017}.} \\
\end{tabular}
\caption{List of $z\gtrsim7$ galaxies identified in the CEERS NIRSpec MR grating or prism spectra, including four previously-identified Ly$\alpha$ emitting galaxies at $z>7$ in the literature in the reference column. Spectroscopic redshifts are derived using the central wavelengths of the strongest rest-frame optical emission lines. If the object is observed in CEERS imaging survey, the {\it JWST}/NIRCam F150W magnitude is provided. Otherwise the {\it HST}/WFC3 F160W magnitude is provided. The absolute UV magnitude is converted from the available broadband photometry near rest-frame $1500$~\AA. We list the rest-frame optical emission lines detected in the NIRSpec data for each source.}
\label{tab:source}
\end{table*}

The flux calibration was performed to the 2D spectrum of each source using the PHOTOM reference file\footnote{\url{https://jwst-pipeline.readthedocs.io/en/stable/jwst/photom/reference_files.html\#photom-reffile}} in the {\it JWST} data reduction pipeline. We computed slit loss corrections following the path-loss correction\footnote{\url{https://jwst-pipeline.readthedocs.io/en/latest/jwst/pathloss/index.html}} step in the pipeline. Since the galaxies in our sample are not significantly extended, we apply the path-loss correction assuming a point source instead of an uniform extended source. The path-loss correction factor depends on wavelength and is $\sim1.0-1.4\times$ for the 21 galaxies in our sample. Nevertheless, the emission line ratios discussed in this paper (described below) do not change significantly ($<10$~per~cent) before and after the path-loss correction. We explored path-loss corrections assuming an uniform source as well and find these have a negligible effect on our analysis based on line ratios. We also explored a different slit loss correction than the path-loss correction procedures applied by {\it JWST} pipeline. To do this, we first extracted a postage stamp of each galaxy from the {\it HST} F160W image \citep{Skelton2014}. We then smooth the postage stamp and fit the smoothed image with a 2D Gaussian profile. The {\it HST} point spread function (PSF) is deconvolved from the profile to obtain the intrinsic morphology of the sources. The deconvolved profile is then convolved with the wavelength-dependent {\it JWST} PSF computed using WebbPSF\footnote{\url{https://webbpsf.readthedocs.io/}} \citep{Perrin2014}. Using the wavelength-dependent profile, we place a box on each source profile with the same length and width and position angle of the NIRSpec MSA micro-shutter according to the position of the source in the shutter. Finally we compute the ratio of the light inside the mimic shutter box to the total light of the source profile and derive the slit-loss correction factor as a function of wavelength. We find broadly similar results to the pipeline with emission line ratios that are still stable before and after applying this wavelength-dependent slit-loss correction.

\begin{figure*}
\begin{center}
\includegraphics[width=\linewidth]{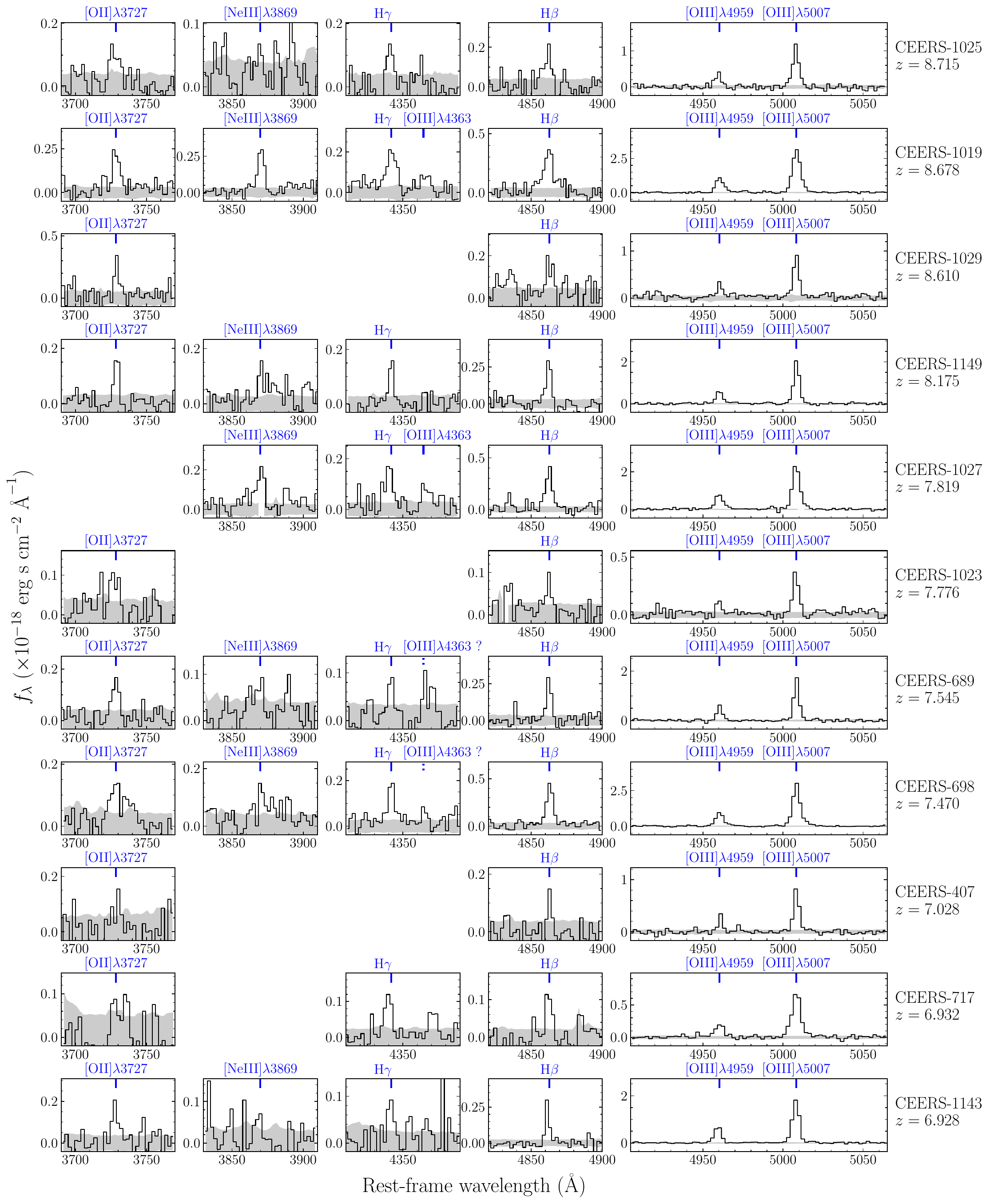}
\caption{Medium resolution ($R\sim1000$) NIRSpec 1D spectra of rest-frame optical emission lines of $z\gtrsim7$ galaxies with more than three detected lines ([O~{\small III}], H$\beta$, and [O~{\small II}] or [Ne~{\small III}] or H$\gamma$; eleven objects). The two tentative detections of [O~{\scriptsize III}]$\lambda4363$ in CEERS-689 and CEERS-698 are shown by blue dotted lines. The spectra have been shifted to the rest frame. The grey shaded regions show the flux uncertainty. Spectra of the remaining  five galaxies with three and fewer rest-frame optical line detections are shown in Fig.~\ref{fig:spectra_opt_1d_less} in Appendix~\ref{sec:appendix}.}
\label{fig:spectra_opt_1d}
\end{center}
\end{figure*}

For each source in our spectroscopic catalog, we extract the 1D spectrum from the calibrated 2D spectrum using a boxcar extraction. The extraction aperture is designed to match the emission line profile along the spatial direction and the typical width is $\sim6$ pixels. Systemic redshifts are derived for each galaxy using the central wavelengths of the strongest rest-frame optical emission lines. Fig.~\ref{fig:spectra_opt_1d} shows 1D profiles of strong rest-frame optical emission lines detected in MR grating spectra. We test the consistency of the wavelength calibration in each of the three bands of the MR gratings by comparing the systemic redshift derived from non-resonant emission lines seen in each setup. The redshifts measured between the three settings are nearly identical ($\Delta z<0.0005$), indicating that the wavelength calibration should be reliable between different bands. We then derive emission line fluxes for each object. The line fluxes are computed through fitting a Gaussian profile to emission lines with S/N $>5$. For lines with lower S/N ($<5$) we calculate the fluxes using direct integration. Four of the galaxies with MR grating spectra are also observed in the low-resolution prism mode. The line fluxes of strong rest-frame optical emission lines are consistent in both spectra within $2\sigma$ uncertainty. For undetected emission lines, we derive a $3\sigma$ upper limit of the flux by summing the error spectrum in quadrature over $\sim700$~km~s$^{-1}$. This integration range is chosen to be consistent with the upper bound of line widths found for rest-frame UV and optical emission lines of extreme emission line galaxies \citep[e.g.,][]{Maseda2014,Mainali2020,Tang2022}. 

The $5\sigma$ limiting line flux sensitivity of the three NIRSpec gratings are $\sim3.0\times10^{-18}$ erg s$^{-1}$ cm$^{-2}$ for G140M/F100LP, $\sim1.9\times10^{-18}$ erg s$^{-1}$ cm$^{-2}$ for G235M/F170LP, and $\sim1.2\times10^{-18}$ erg s$^{-1}$ cm$^{-2}$ for G395M/F290LP. Considering the [O~{\small III}]$\lambda5007$ line, this flux sensitivity corresponds to $5\sigma$ limiting EW $=25$~\AA, $62$~\AA, $160$~\AA, and $390$~\AA\ for a galaxy with rest-frame optical continuum of $25$, $26$, $27$, and $28$ AB mag at $z=8$. Given that we typically require detection of both components of the [O~{\small III}] doublet and H$\beta$ for redshift confirmation, our actual [O~{\small III}] EW limit is several times larger than these values. We also test the consistency of the flux calibration among the three bands of the MR gratings. To do this, we choose sources at redshifts where the same strong lines are detected in two bands simultaneously (e.g., H$\beta$ detected in both G235M and G395M at $z\simeq5.1$, and H$\beta$ detected in both G140M and G235M at $z\simeq2.7$). Then we measure the flux of the same emission line in each band, and we find that the line fluxes measured in different bands are consistent within the $1\sigma$ uncertainty. In this work, we will primarily focus on emission line ratios that are not sensitive to absolute flux calibration, but probe the ionizing spectra, the gas excitation conditions, and the gas-phase metallicity. We will consider the following ratios and use the following definitions: O3 ($\equiv$ [O~{\small III}]$\lambda5007$ / H$\beta$), O32 ($\equiv$ [O~{\small III}]$\lambda\lambda4959,5007$ / [O~{\small II}]$\lambda\lambda3726,3729$), Ne3O2 ($\equiv$ [Ne~{\small III}]$\lambda3869$ / [O~{\small II}]$\lambda\lambda3726,3729$), R23 ($\equiv$ ([O~{\small III}]$\lambda\lambda4959,5007+$ [O~{\small II}]$\lambda\lambda3726,3729$) / H$\beta$), and [O~{\small III}]$\lambda4363$ / [O~{\small III}]$\lambda5007$. 


\subsection{Spectral Energy Distributions of $z\gtrsim7$ Galaxies} \label{sec:sed}

\begin{figure*}
\begin{center}
\includegraphics[width=\linewidth]{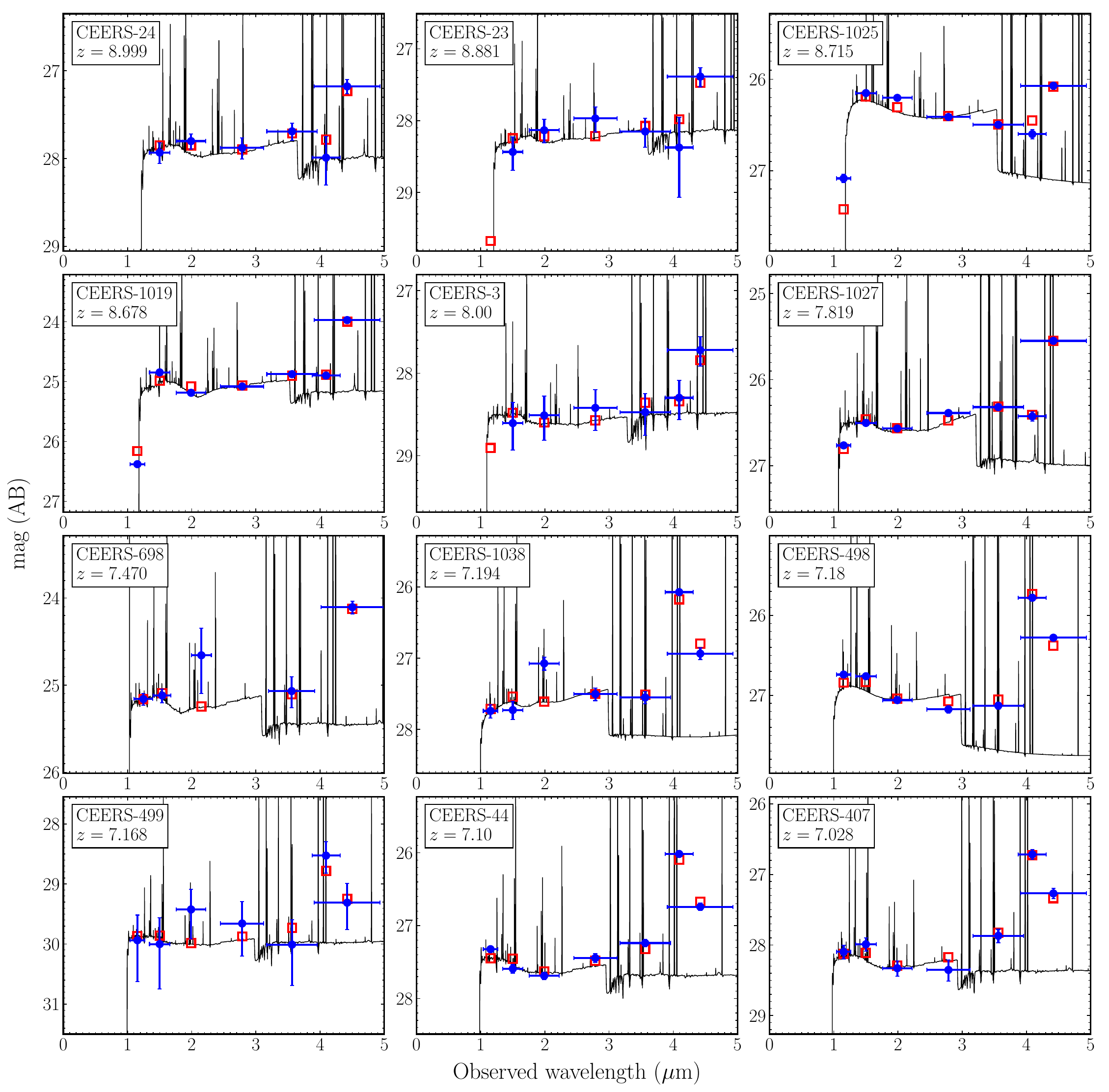}
\caption{Spectral energy distributions (blue circles) with the best-fit \textsc{beagle} stellar population synthesis models (black lines) of the 12 galaxies at $z>7$ with broadband photometry measurements spanning from rest-frame UV to optical in the CEERS NIRSpec sample. The best-fit spectra and synthetic photometry (red squares) are derived from the posterior median. For CEERS-24, CEERS-23, and CEERS-3, their F115W fluxes are non-detections. To avoid introducing uncertainties from Lyman series emission and absorption, we do not fit F115W flux for galaxies at $z>7.5$. The SEDs show large flux excess at NIRCam F410M or F444W, or IRAC $[4.5]$, indicating intense [O~{\small III}]+H$\beta$ emission lines.}
\label{fig:sed}
\end{center}
\end{figure*}

We use available broadband spectral energy distributions (SEDs) to investigate the stellar population properties of the spectroscopic sample. The SEDs are also critical for characterization of the [O~{\small III}]+H$\beta$ EWs based on flux excesses in the $3-5\mu$m photometry. The imaging-based SED analysis in this paper will focus primarily on objects with NIRCam measurements, given the vastly improved photometric constraints that are achieved relative to what is possible with {\it HST} and {\it Spitzer}. A subset of the galaxies in our sample (11 of 21) overlap with the region of the EGS that has been observed by NIRCam as part of the CEERS ERS program. These include sources covered in the original NIRCam pointings taken in summer 2022 (CEERS-23, CEERS-24, CEERS-3, CEERS-1027, CEERS-498, CEERS-499, CEERS-44, CEERS-407) and three additional sources observed in NIRCam imaging undertaken in December 2022 (CEERS-1019, CEERS-1025, and CEERS-1038). The imaging associated with the latter sources has been reduced in the same manner as described in \citet{Endsley2023a} and the reader is directed to that paper for details. For these sources, we compute SEDs using CEERS photometry catalogs presented in \citet{Endsley2023a}. The CEERS NIRCam imaging was taken in six broad-band filters (F115W, F150W, F200W, F277W, F356W, and F444W) and one medium-band filter (F410M). The flux of each band was measured in elliptical aperture with a \citet{Kron1980} factor of $k=1.2$, and then was corrected to the total flux by multiplying the ratio of flux measured in $k=2.5$ to $k=1.2$ aperture in F200W. Flux from neighboring objects that is in the aperture was also subtracted. The derived NIRCam H-band magnitudes (F150W) range between $30.00$ and $24.85$ in these 11 galaxies, implying absolute UV magnitude (M$_{\rm{UV}}$) between $-16.97$ and $-22.42$. The SEDs are shown in Fig.~\ref{fig:sed}. For these 11 galaxies with NIRCam photometry, color excesses are readily apparent in the long-wavelength photometry, revealing the characteristic signature of strong rest-frame optical line emission. We will come back to quantify the implied [O~{\small III}]+H$\beta$ EWs later in the subsection.


For the 10 sources that lack NIRCam photometry, we collate {\it HST} and {\it Spitzer} photometry. The {\it HST} Wide Field Camera 3 (WFC3) IR photometry is computed for each source using the mosaics produced as part of the Complete Hubble Archive for Galaxy Evolution (CHArGE) project \citep{Kokorev2022}. The CHArGE mosaics include the {\it HST}/WFC3 data in the EGS field, and are matched to the Gaia astrometric frame with a pixel scale of 80~mas~pixel$^{-1}$. Details of the CHArGE data used in CEERS have been described in \citet{Chen2023} and \citet{Endsley2023a}. The derived H$_{\rm{160}}$-band magnitudes range between $26.90$ and $24.91$, implying M$_{\rm{UV}}$ between $-20.04$ and $-22.14$. At these magnitudes, the {\it Spitzer} measurements are mostly not useful given the limited sensitivity of the IRAC mosaics across the EGS (typical $5\sigma$ limit of $m\simeq25.2$ at $3.6\mu$m). Only one galaxy in this subset of 10 (CEERS-698) has {\it Spitzer}/IRAC flux density measurements with S/N in excess of 5 in both $[3.6]$ and $[4.5]$. We add this $z=7.470$ source (one of the original \citealt{Roberts-Borsani2016} galaxies) to our sub-sample for SED-based analysis. For the remaining sources, the absence of significant rest-frame optical constraints keeps stellar population modeling from yielding meaningful information. While we do investigate the emission line properties of this subset of 9 galaxies, we do not include their imaging-based SED fits in our subsequent analysis.

We fit the SEDs of the 12 sources (eleven with {\it JWST}/NIRCam photometry and the one with high S/N {\it HST}+{\it Spitzer} photometry) with rest-frame UV to optical photometry using the Bayesian galaxy SED modelling and interpreting tool BayEsian Analysis of GaLaxy sEds (\textsc{beagle}, version 0.20.4; \citealt{Chevallard2016}). The \textsc{beagle} set up and SED fitting procedures are mainly described in \citet{Endsley2023a} and are briefly summarized in the following. \textsc{beagle} utilizes a combination of the latest version of \citet{Bruzual2003} stellar population synthesis models and the \citet{Gutkin2016} photoionization models of star-forming galaxies with the {\small CLOUDY} code \citep{Ferland2013}. We assume a constant star formation history (CSFH), allow the galaxy age to vary between $1$~Myr and the age of the Universe at the given spectroscopic redshift with a log-uniform prior. We adopt a \citet{Chabrier2003} initial mass function (IMF) with a stellar mass range of $0.1-300\ M_{\odot}$ and allow the metallicity to vary in the range $-2.2\le\log{(Z/Z_{\odot})}\le-0.3$ ($Z_{\odot}=0.01524$; \citealt{Caffau2011}). The interstellar metallicity is set to equal to the stellar metallicity with dust-to-metal mass ratio fixed to $\xi_{\rm{d}}=0.3$ for ease of comparison with previous SED studies (although we note we will take a different approach when modeling the NIRSpec data). We put an upper limit on metallicity of $0.5\ Z_{\odot}$ to avoid uncommon solutions of near-solar metallicity in reionization-era galaxies. The ionization parameter (the ratio of the number density of ionizing photons to the number density of hydrogen atoms in the H~{\small II} regions; \citealt{Penston1990}) $U$ is adjusted in the range $-4.0\le \log{U}\le-1.0$. We adopt log-uniform priors for $Z$ and $U$. We assume the Small Magellan Cloud (SMC) extinction curve \citep{Pei1992} to account for the dust attenuation and the $V$-band optical depth $\tau_{V}$ is allowed to vary between $0.001$ and $5$ with a log-uniform prior. Finally, we adopt the prescription of \citet{Inoue2014} to include the absorption of IGM. When fitting the SEDs, we remove fluxes in filters that lie blueward of Ly$\alpha$ to avoid introducing the uncertain flux contribution from Lyman series emission and absorption. The best-fit model with observed SED of each source in presented in Fig.~\ref{fig:sed}.

The SED fitting results allow us to examine the slit-loss correction adopted on NIRSpec spectra. Here we compare the [O~{\small III}]~$\lambda5007$ fluxes (which is the brightest emission line with the highest S/N at rest-frame optical in our sample) inferred from the best-fit SED models and the slit-loss corrected [O~{\small III}]~$\lambda5007$ fluxes measured by NIRSpec. We find that both fluxes are consistent within $1\sigma$ confidence interval, indicating that the slit-loss correction adopted in this work is reliable.

The physical properties of the models that reproduce the observed SEDs are shown in Table~\ref{tab:sed_fitting}. The SEDs suggest minimal dust reddening, with $V$-band optical depths ranging between $\tau_V$ of $0.003$ and $0.12$, with a median of $0.03$. The inferred CSFH ages are very young, ranging between $1.5$ and $27$~Myr, with a median ($5$~Myr) that is $2-14$ times lower than the typical age at these redshifts \citep[e.g.,][]{Tacchella2022,Endsley2023a,Endsley2023b,Whitler2023a,Whitler2023b}. The young ages reflect SEDs dominated by a recent burst or upturn in star formation history, with the light in the rest-frame UV and optical powered by a very young stellar population (though AGN can also contribute; see \citealt{Larson2023}). At these young ages, the rest-frame optical continuum is weak and emission lines are strong, leading to significant color excesses from very large EW rest-frame optical emission lines. The rest-frame [O~{\small III}]+H$\beta$ EWs implied by the SEDs range between $879$ and $3276$~\AA, with a median of $1994$~\AA. These [O~{\small III}]+H$\beta$ EWs reveal that our CEERS spectroscopic sample lies at the more extreme part of the $z\simeq7-9$ population, with values that correspond to the upper $43$~per~cent of the [O~{\small III}]+H$\beta$ EW distribution \citep{Endsley2023a}. The Ly$\alpha$ emitters (EW$_{\rm{Ly}\alpha}>9$~\AA\footnote{In this work we choose $9$~\AA\ as the EW threshold of Ly$\alpha$ emitters because it equals to the $5\sigma$ Ly$\alpha$ EW limit that can be detected in a UV-bright galaxy (M$_{\rm{UV}}<-21$, or equivalently NIRCam/F150W $<26$) at $z=8$ with CEERS NIRSpec observations.}) in our sample show even larger [O~{\small III}]+H$\beta$ EWs $=1786-3276$~\AA, corresponding to the upper $11$~per~cent of the [O~{\small III}]+H$\beta$ EW distribution. This suggests that our CEERS spectroscopic sample is not representative of the $z\simeq7-9$ population, a result that is not surprising given that selection on [O~{\small III}] is necessary for redshift confirmation. This bias must be taken into account when interpreting the spectroscopic properties of the sample. 

\begin{table*}
\centering
\begin{tabular}{ccccccccc}
\hline
ID & $z_{\rm{spec}}$ & $\log{(M_{\star}/M_{\odot})}^{\rm{a}}$ & $\log{(t_{\rm{age}}/\rm{yr})}$ & $\tau_{V}$ & EW$_{\rm{[OIII]+H}\beta}$ (\AA) & $\log{(\xi_{\rm{ion}}/\rm{erg}^{-1}\rm{Hz})}$ & $\log{(M_{\star}/M_{\odot})}^{\rm{b}}$ & $\log{(M_{\star}/M_{\odot})}^{\rm{c}}$ \\
\hline
CEERS-24 & $8.999$ & $7.48^{+0.69}_{-0.13}$ & $6.70^{+0.86}_{-0.38}$ & $0.050^{+0.064}_{-0.042}$ & $2085^{+2781}_{-1122}$ & $25.80^{+0.13}_{-0.22}$ & $8.58^{+0.40}_{-0.47}$ & $7.88^{+0.66}_{-0.31}$ \\
CEERS-23 & $8.881$ & $7.62^{+0.65}_{-0.37}$ & $7.07^{+0.77}_{-0.64}$ & $0.094^{+0.110}_{-0.086}$ & $1175^{+1267}_{-528}$ & $25.71^{+0.14}_{-0.19}$ & $8.78^{+0.17}_{-0.27}$ & $8.24^{+0.68}_{-0.56}$ \\
CEERS-1025 & $8.715$ & $7.86^{+0.03}_{-0.02}$ & $6.20^{+0.12}_{-0.13}$ & $0.003^{+0.005}_{-0.001}$ & $1787^{+313}_{-468}$ & $25.80^{+0.01}_{-0.01}$ & $8.44^{+0.10}_{-0.10}$ & $8.31^{+0.08}_{-0.09}$ \\
CEERS-1019 & $8.678$ & $8.67^{+0.11}_{-0.09}$ & $6.69^{+0.13}_{-0.25}$ & $0.005^{+0.010}_{-0.003}$ & $2599^{+642}_{-459}$ & $26.00^{+0.14}_{-0.10}$ & $9.69^{+0.06}_{-0.07}$ & $8.60^{+0.02}_{-0.01}$ \\
CEERS-3 & $8.00$ & $7.67^{+0.59}_{-0.58}$ & $7.43^{+0.78}_{-0.69}$ & $0.034^{+0.097}_{-0.030}$ & $879^{+531}_{-378}$ & $25.61^{+0.16}_{-0.16}$ & $8.49^{+0.19}_{-0.19}$ & $8.23^{+0.26}_{-0.63}$ \\
CEERS-1027 & $7.819$ & $7.82^{+0.12}_{-0.06}$ & $6.53^{+0.29}_{-0.32}$ & $0.056^{+0.039}_{-0.040}$ & $3276^{+975}_{-1082}$ & $25.79^{+0.04}_{-0.04}$ & $8.28^{+0.32}_{-0.10}$ & $8.15^{+0.15}_{-0.10}$ \\
CEERS-698 & $7.470$ & $8.69^{+0.13}_{-0.10}$ & $6.58^{+0.23}_{-0.32}$ & $0.040^{+0.083}_{-0.035}$ & $2564^{+1879}_{-1001}$ & $25.82^{+0.14}_{-0.10}$ & $9.48^{+0.28}_{-0.26}$ & $8.73^{+0.71}_{-0.24}$ \\
CEERS-1038 & $7.194$ & $7.42^{+0.07}_{-0.08}$ & $6.22^{+0.17}_{-0.14}$ & $0.119^{+0.036}_{-0.047}$ & $2650^{+420}_{-432}$ & $25.80^{+0.02}_{-0.02}$ & $7.95^{+0.21}_{-0.18}$ & $7.64^{+0.13}_{-0.11}$ \\
CEERS-498 & $7.18$ & $7.52^{+0.03}_{-0.03}$ & $6.17^{+0.13}_{-0.11}$ & $0.003^{+0.005}_{-0.001}$ & $2840^{+206}_{-199}$ & $25.81^{+0.02}_{-0.01}$ & $7.93^{+0.14}_{-0.13}$ & $7.74^{+0.07}_{-0.08}$ \\
CEERS-499 & $7.168$ & $6.79^{+0.75}_{-0.44}$ & $7.22^{+0.87}_{-0.75}$ & $0.019^{+0.072}_{-0.016}$ & $1045^{+969}_{-486}$ & $25.64^{+0.16}_{-0.17}$ & $7.66^{+0.27}_{-0.21}$ & $7.01^{+0.54}_{-0.39}$ \\
CEERS-44 & $7.10$ & $7.56^{+0.24}_{-0.23}$ & $7.01^{+0.22}_{-0.27}$ & $0.005^{+0.014}_{-0.004}$ & $1786^{+254}_{-271}$ & $25.72^{+0.05}_{-0.05}$ & $8.63^{+0.10}_{-0.11}$ & $8.54^{+0.15}_{-0.17}$ \\
CEERS-407 & $7.028$ & $7.17^{+0.33}_{-0.15}$ & $6.72^{+0.45}_{-0.46}$ & $0.005^{+0.016}_{-0.003}$ & $1902^{+1844}_{-453}$ & $25.80^{+0.10}_{-0.09}$ & $8.11^{+0.18}_{-0.22}$ & $7.75^{+0.40}_{-0.32}$ \\
\hline
\multicolumn{9}{l}{Note:} \\
\multicolumn{9}{l}{a. Stellar mass derived from CSFH \textsc{beagle} SED fitting.} \\
\multicolumn{9}{l}{b. Stellar mass derived from non-parametric SFH \textsc{prospector} SED fitting with continuity prior.} \\
\multicolumn{9}{l}{c. Stellar mass derived from non-parametric SFH \textsc{prospector} SED fitting with bursty continuity prior.} \\
\end{tabular}
\caption{Posterior median and $1\sigma$ confidence interval of physical properties derived from \textsc{beagle} SED fitting for CEERS NIRSpec galaxies at $z>7$. Stellar mass, stellar age (assuming constant SFH), dust attenuation at rest-frame $V$ band $\tau_{V}$, [O~{\scriptsize III}]+H$\beta$ EW, and ionizing photon production efficiency $\xi_{\rm{ion}}$ are provided. The last two columns show stellar masses derived from non-parametric SFH \textsc{prospector} SED fitting with continuity prior and bursty continuity prior. We only consider galaxies with high S/N photometry over the rest-frame UV (near-infrared) and rest-frame optical (mid-infrared).}
\label{tab:sed_fitting}
\end{table*}

The derived stellar masses range between $6.2\times10^6-4.9\times10^8\ M_\odot$ in our fiducial CSFH \textsc{beagle} models. These masses correspond to the very young stellar population that dominates the SED. As has been discussed in the recent literature \citep[e.g.,][]{Roberts-Borsani2020,Laporte2021,Tacchella2022,Whitler2023a,Whitler2023b}, older populations can easily be hidden underneath the light of the young stellar population, increasing the stellar mass by over an order of magnitude. To determine upper bounds on the stellar mass in these galaxies, we fit the SEDs using models which incorporate non-parametric star formation histories (SFHs). We follow an approach that is very similar to that described in \citet{Whitler2023a}. In brief, we fit the SEDs with \textsc{prospector} \citep{Leja2019,Johnson2021}, which is based on the Flexible Stellar Population Synthesis code \citep{Conroy2009,Conroy2010} and the nebular emission models of \citet{Byler2017}. We adopt a \citet{Chabrier2003} IMF with a mass range of $0.1-300\ M_{\odot}$, a SMC dust attenuation law \citep{Pei1992} with a log-uniform prior of $V$-band optical depth $\tau_V$, and the \citet{Inoue2014} IGM attenuation model, which are similar to the CSFH SED fitting with \textsc{beagle}. Here, we consider models with non-parametric SFH in \textsc{prospector}, which are piecewise constant functions in time. We adopt eight age bins spanning from the time of observation to the lookback time corresponding to a formation redshift $z_{\rm{form}}$. We allow $z_{\rm{form}}$ to vary between $15$ and $30$. The two most recent time bins of the SFH are fixed to $0-3$~Myr and $3-10$~Myr, and the remaining six bins are spaced evenly in logarithmic lookback time. For non-parametric SFH prior, we use the built-in `continuity' prior in \textsc{prospector} and also a `bursty' version of the continuity prior \citep{Tacchella2022,Whitler2023b}. The continuity prior allows a smoothly evolving star formation rate (SFR) over time, while the bursty continuity prior allows more sharp changes in SFR between time bins. The results reveal stellar masses assuming continuity prior (bursty continuity prior) that range between $4.6\times10^7-4.9\times10^9\ M_\odot$ ($1.0\times10^7-5.4\times10^8\ M_\odot$), roughly $\sim7$ ($\sim2$) times larger than the CSFH models. We list these values in Table~\ref{tab:sed_fitting}. While our results do not depend sensitively on the stellar masses, we will consider the range between the \textsc{beagle} CSFH and \textsc{prospector} non-parametric SFH models (with continuity prior) as the allowed range for each source.

\begin{figure}
\begin{center}
\includegraphics[width=\linewidth]{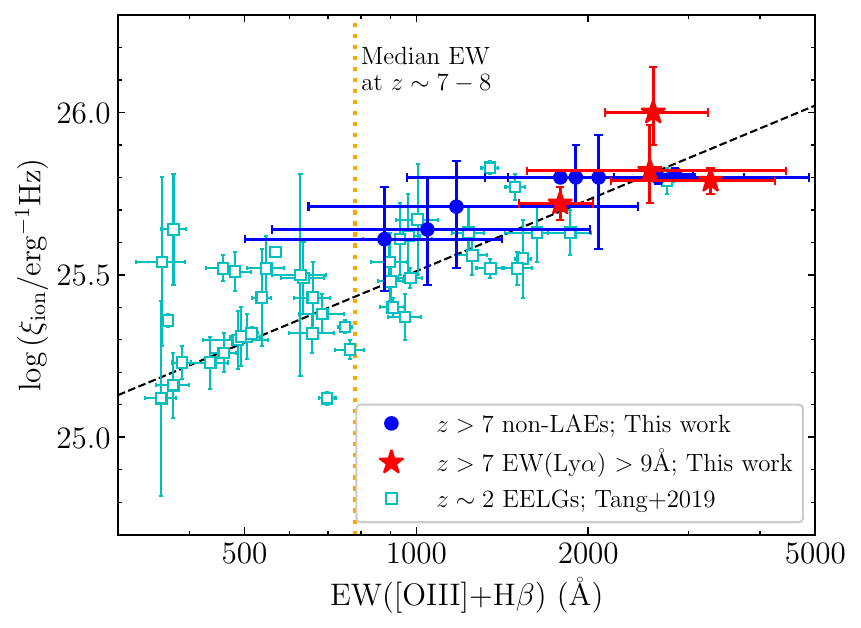}
\caption{Correlation between ionizing photon production efficiency $\xi_{\rm{ion}}$ and [O~{\scriptsize III}]+H$\beta$ EW for CEERS NIRSpec galaxies at $z\gtrsim7$. The EW and $\xi_{\rm{ion}}$ are derived from SED fitting using \textsc{beagle} \citep{Chevallard2016}, and the posterior median and $1\sigma$ confidence interval are plotted. Ly$\alpha$ emitters (LAEs; EW$_{\rm{Ly}\alpha}>9$~\AA) and non-LAEs at $z>7$ are shown by red stars and blue circles, respectively. The median [O~{\scriptsize III}]+H$\beta$ EW of typical $z\sim7-8$ star-forming galaxies ($\approx780$~\AA; \citealt{Endsley2023a}) is shown by the vertical orange dotted line. We also overplot EELGs (i.e., systems with similarly large [O~{\scriptsize III}]+H$\beta$ EWs as $z>7$ galaxies) at $z\sim2$ \citep{Tang2019} in cyan squares as comparison. The $\xi_{\rm{ion}}-$ EW$_{\rm{[OIII]+H}\beta}$ relation at $z\sim2$ from \citet{Tang2019} is presented as black dashed line.}
\label{fig:xi_ion}
\end{center}
\end{figure}

The observed SEDs also constrain the hydrogen ionizing photon production efficiency ($\xi_{\rm{ion}}$), equal to the ratio of the hydrogen ionizing photon production rate ($\dot{N}_{\rm{ion}}$) and the far-UV continuum luminosity at rest-frame $1500$~\AA\ ($L_{\rm{UV}}$). Because ionizing photons are reprocessed into recombination lines, $\xi_{\rm{ion}}$ also sets the production efficiency of Ly$\alpha$ photons. Galaxies with larger $\xi_{\rm{ion}}$ will have larger Ly$\alpha$ luminosities for fixed SFR, potentially boosting their visibility. In this paper, we define $\xi_{\rm{ion}}$ as the hydrogen ionizing photon production rate per unit intrinsic $L_{\rm{UV}}$, the observed UV luminosity at rest-frame $1500$~\AA\ (including nebular and stellar continuum) corrected for dust attenuation (see \citealt{Chevallard2018} and \citealt{Tang2019} for definitions of various $\xi_{\rm{ion}}$). This is the most commonly used definition of $\xi_{\rm{ion}}$ in literature. The derived values of the ionizing photon production efficiency range between $\log{(\xi_{\rm{ion}}/\rm{erg}^{-1}\ \rm{Hz})}$ $=25.61$ and $26.00$. At lower redshifts, $\xi_{\rm{ion}}$ is known to correlate with the derived [O~{\small III}]+H$\beta$ EW \citep{Chevallard2018,Tang2019,Onodera2020}. In Fig.~\ref{fig:xi_ion}, we plot the inferred ionizing photon production efficiencies versus [O~{\small III}]+H$\beta$ EW, overlaying the $z\gtrsim7$ results on the $z\simeq2$ extreme emission line galaxy (EELG) sample from \citet{Tang2019}. While the $z\gtrsim7$ galaxies are consistent with the lower redshift relation, they are situated in a region of the diagram that is rarely seen in lower redshifts samples of EELGs. These results suggest that the $z\gtrsim7$ Ly$\alpha$ emitters are among the most efficient ionizing agents known in normal star-forming galaxies. This result is not surprising given the very young stellar populations which dominate the SEDs of these systems. At sufficiently young ages under a CSFH, the B star population has yet to build up its contribution to the UV continuum luminosity, leading to very large values of $\xi_{\rm{ion}}$. We will discuss implications of the ionizing production for Ly$\alpha$ visibility in Section~\ref{sec:discussion}.

In the end of this section, we explore the impact of non-zero ionizing photon escape fraction ($f_{\rm{esc}}$) to the physical parameters derived from \textsc{beagle} SED fitting. For galaxies to be the primary ionizing agents that are responsible for cosmic reionization, the ionizing photon escape fraction is required to be $f_{\rm{esc}}\simeq0.1-0.2$ \citep[e.g.,][]{Robertson2015,Finkelstein2019,Naidu2020}. We fit the SEDs of the 12 galaxies with {\it JWST}/NIRCam or high S/N {\it HST}+{\it Spitzer} photometry with CSFH \textsc{beagle} models (assuming ionization-bounded nebula) while varying $f_{\rm{esc}}$ as a free parameter. The parameter spaces of other physical quantities (e.g., stellar mass, age, dust attenuation, ionization parameter, metallicity) remain the same and we allow $f_{\rm{esc}}$ to vary in the range $0-0.9$ assuming a uniform prior. {The SED fitting results show $f_{\rm{esc}}=0.08-0.6$ with a median value of $f_{\rm{esc}}=0.2$, but the uncertainty of each fitted $f_{\rm{esc}}$ is large (with a typical $1\sigma$ confidence interval of $\pm0.2$).} Other physical parameters remain consistent with the values derived from \textsc{beagle} models with $f_{\rm{esc}}=0$ except for stellar age. The stellar ages inferred from non-zero $f_{\rm{esc}}$ models are on average $0.2$~dex younger than those inferred from models with $f_{\rm{esc}}=0$. This is because when some of the ionizing photons produced in galaxies escape directly into the IGM (and thus are not reprocessed into nebular emission), it requires younger stellar populations to produce more ionizing photons in order to reproduce the same nebular emission line strength.

\subsection{Rest-Frame Optical Emission Lines in $z\gtrsim7$ Galaxies} \label{sec:opt_lines}

In this section, we characterize the rest-frame optical line ratios and discuss what they tell us about the ionized gas properties of our spectroscopic sample. The reader is also directed to a variety of other papers on the rest-frame optical lines in the reionization era \citep[e.g.,][]{ArrabalHaro2023,Bunker2023,Cameron2023,Fujimoto2023,Hsiao2023,Jung2023,Larson2023,Mascia2023,Nakajima2023,Sanders2023a,Sanders2023b,Saxena2023a,Saxena2023b}. Our primary interest is in the nature of the galaxies in the CEERS spectroscopic sample, with a particular focus on those with Ly$\alpha$ emission (which we will come back to discuss in Section~\ref{sec:lae} and Section~\ref{sec:discussion}). A detailed examination of the ionized gas properties of the general population at $z\gtrsim7$ is beyond the scope of the current study and will eventually require deeper data that is less biased to strong [O~{\small III}] emitters. In what follows, we first consider the gas properties implied by photoionization model fits. We then quantify the rest-frame optical line ratios of our sample, comparing to galaxies at lower redshifts.

\begin{figure}
\begin{center}
\includegraphics[width=\linewidth]{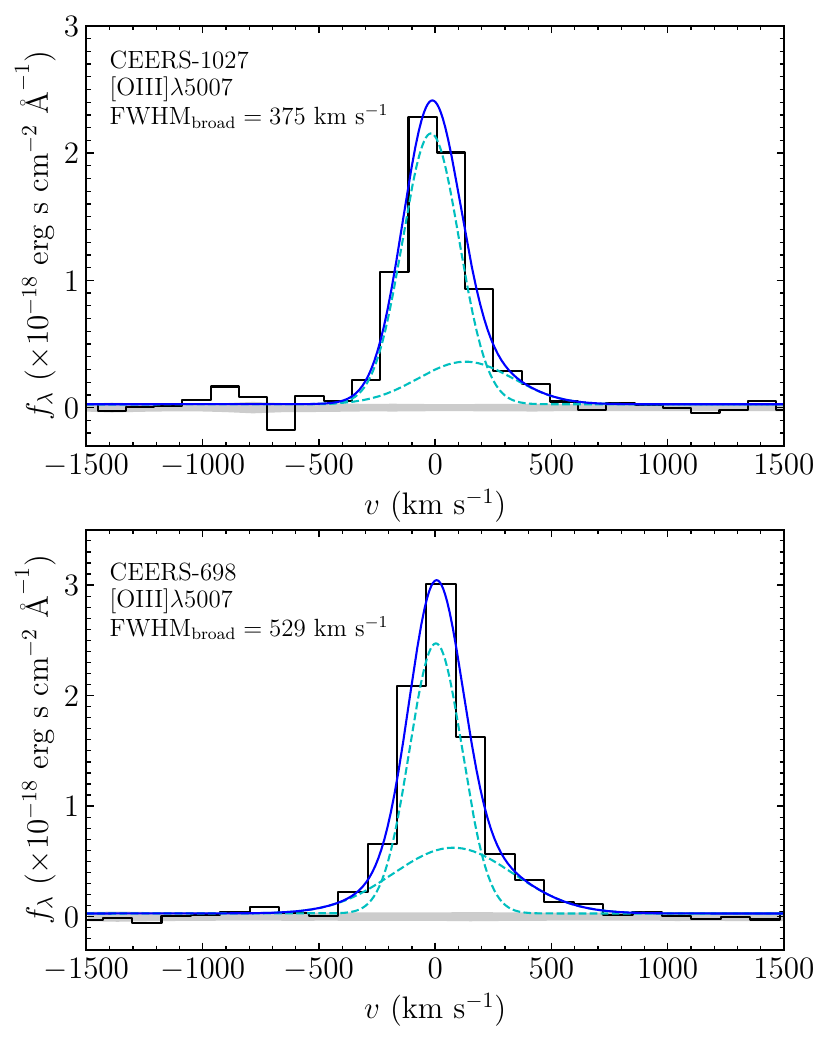}
\caption{[O~{\scriptsize III}]$\lambda5007$ velocity profiles of CEERS-1027 (top panel) and CEERS-698 (bottom panel). Broad emission line components are detected for both objects. The observed line flux (uncertainty) is shown by black lines (grey shaded regions). Each line profile is fitted by a narrow and a broad Gaussian profile, which are shown by cyan dashed lines. The blue lines present the sum of the narrow and the broad Gaussian profiles.}
\label{fig:spectra_oiii5007}
\end{center}
\end{figure}

A summary of the CEERS rest-frame optical emission line measurements is given in Table~\ref{tab:opt_line}. We detect the [O~{\small III}]$\lambda\lambda4959,5007$ doublet in all 21 sources and H$\beta$ in 20 sources in our sample. In additional 11 (7) galaxies, we detect the [O~{\small II}] ([Ne~{\small III}]) lines, allowing a range of ionization and excitation-sensitive line ratio diagnostics to be derived. We detect H$\gamma$ (H$\delta$) in 9 (3) galaxies, allowing constraints on Balmer line ratios. In a further two sources (CEERS-1019 and CEERS-1027), we detect the auroral [O~{\small III}]$\lambda4363$ (S/N $=3-4$), with tentative [O~{\small III}]$\lambda4363$ features (S/N $=2$) seen in additional two galaxies (CEERS-689 and CEERS-698). When combined with the other strong lines, the auroral lines allow constraints on the gas-phase oxygen abundance via the direct method \citep[e.g.,][]{Izotov2006,Sanders2020}. In two Ly$\alpha$ emitting galaxies (CEERS-1027 and CEERS-698), we also detect broad emission components. The intrinsic full widths at half maximum (FWHM) after subtracting the spectral resolution measured by the instrument team ($R\sim1000$, corresponding to spectral FWHM $=300$~km~s$^{-1}$; \citealt{Jakobsen2022}) for the broad [O~{\small III}]$\lambda5007$ lines of these two galaxies (Fig.~\ref{fig:spectra_oiii5007}) are $375$~km~s$^{-1}$ and $529$~km~s$^{-1}$, which comprise $21$ and $37$~per~cent of the total [O~{\small III}]$\lambda5007$ fluxes. Such broad [O~{\small III}] emission may indicate highly ionized gas outflows driven by stellar feedback associated with very young stellar populations (see Table~\ref{tab:sed_fitting}), which is similar as seen in a young ($\sim3$~Myr), Lyman continuum emitting galaxy at lower redshift \citep{Mainali2022}. Fast outflows may also scatter Ly$\alpha$ photons redwards as they backscatter off outflowing gas \citep[e.g.,][]{Verhamme2006,Steidel2010}, as we will discuss in Section~\ref{sec:uv_lines}.

\begin{table*}
\centering
\begin{tabular}{ccccccccc}
\hline
ID & $z_{\rm{spec}}$ & O3 & O32 & R23 & Ne3O2 & H$\gamma$/H$\beta$ & [O~{\scriptsize III}]$\lambda4363$/[O~{\scriptsize III}]$\lambda5007$ \\
\hline
CEERS-24 & $8.999$ & ... & $>3.98$ & ... & ... & ... & ... \\
CEERS-23 & $8.881$ & $3.80\pm1.95$ & $>3.87$ & $<6.66$ & ... & ... & ... \\
CEERS-1025 & $8.715$ & $4.78\pm1.68$ & $8.12\pm3.15$ & $7.09\pm2.50$ & $0.25\pm0.26$ & $0.668\pm0.355$ & ... \\
CEERS-1019 & $8.678$ & $6.46\pm1.01$ & $18.06\pm5.99$ & $8.98\pm1.42$ & $1.08\pm0.48$ & $0.467\pm0.123$ & $0.025\pm0.007$ \\
CEERS-1029 & $8.610$ & $5.25\pm2.42$ & $4.81\pm1.25$ & $8.02\pm3.71$ & $<0.46$ & ... & ... \\
CEERS-1149 & $8.175$ & $6.99\pm1.35$ & $18.77\pm6.06$ & $9.81\pm1.92$ & $0.77\pm0.38$ & $0.497\pm0.179$ & ... \\
CEERS-3 & $8.00$ & $5.49\pm2.20$ & $>3.68$ & $<8.87$ & ... & ... & ... \\
CEERS-1027 & $7.819$ & $6.04\pm0.91$ & $>30.22$ & $<8.33$ & $>1.80$ & $0.394\pm0.135$ & $0.025\pm0.012$ \\
CEERS-1023 & $7.776$ & $4.59\pm2.33$ & $2.80\pm1.25$ & $7.97\pm4.12$ & $<0.82$ & ... & ... \\
CEERS-686 & $7.74$ & $5.25\pm0.41$ & ... & ... & ... & ... & ... \\
CEERS-689 & $7.545$ & $6.00\pm0.99$ & $10.13\pm2.67$ & $8.84\pm1.48$ & $0.71\pm0.46$ & $0.333\pm0.174$ & $0.059\pm0.032$ \\
CEERS-698 & $7.470$ & $6.41\pm0.60$ & $14.72\pm4.40$ & $9.03\pm0.88$ & $0.87\pm0.49$ & $0.358\pm0.085$ & $0.014\pm0.008$ \\
CEERS-1163 & $7.448$ & $3.67\pm2.91$ & $>4.20$ & $<6.73$ & ... & ... & ... \\
CEERS-1038 & $7.194$ & $3.79\pm1.18$ & $>4.43$ & $<6.09$ & ... & ... & ... \\
CEERS-498 & $7.18$ & $3.57\pm0.30$ & $>23.45$ & $<5.53$ & ... & ... & ... \\
CEERS-499 & $7.168$ & $3.66\pm3.38$ & $>3.43$ & $<6.49$ & ... & ... & ... \\
CEERS-44 & $7.10$ & $4.52\pm0.50$ & $>18.93$ & $<6.01$ & ... & ... & ... \\
CEERS-407 & $7.028$ & $6.55\pm1.96$ & $5.96\pm3.91$ & $10.00\pm3.13$ & $<0.59$ & ... & ... \\
CEERS-1102 & $7.00$ & $4.09\pm0.58$ & $2.12\pm0.26$ & $8.01\pm1.16$ & ... & ... & ... \\
CEERS-717 & $6.932$ & $6.07\pm1.97$ & $5.58\pm2.76$ & $9.19\pm3.06$ & $<0.30$ & $0.935\pm0.421$ & ... \\
CEERS-1143 & $6.928$ & $9.68\pm1.42$ & $14.46\pm5.70$ & $14.25\pm2.15$ & $0.36\pm0.36$ & $0.390\pm0.286$ & ... \\
Composite & $7.7$ & $6.81\pm0.26$ & $17.84\pm1.71$ & $9.42\pm0.36$ & $0.89\pm0.13$ & $0.463\pm0.038$ & $0.022\pm0.005$ \\
\hline
\end{tabular}
\caption{Rest-frame optical emission line ratios of $z\gtrsim7$ galaxies identified in CEERS NIRSpec spectra. Line ratios constraining ionizing spectra (O32 and Ne3O2), oxygen abundance and excitation (O3 and R23), dust attenuation (H$\gamma$/H$\beta$), and electron temperature ([O~{\scriptsize III}]$\lambda4363$/[O~{\scriptsize III}]$\lambda5007$) are provided. For galaxies without significant [O~{\scriptsize II}] or [Ne~{\scriptsize III}] measurements, $3\sigma$ upper limits are given for O32, R23, or Ne3O2 ratios. In the last row, we provide the line ratios measured from the composite spectrum of the 16 MR grating sources at $z>7$.}
\label{tab:opt_line}
\end{table*}

We first must consider the dust attenuation corrections for the emission line ratios. In Section~\ref{sec:sed}, we demonstrated that the SEDs of our sample are typically best fit with little to no dust reddening of the stellar continuum, with median UV slope $\beta=-2.4$ and median $\tau_V=0.03$ in the \textsc{beagle} models. Only 1 of 21 sources (CEERS-1023) shows evidence of significant reddening in its rest-frame UV photometry with UV slope $\beta=-0.9$. The blue UV slopes we find in the majority of the sample are consistent with expectations for $z\simeq7-9$ galaxies \citep[e.g.,][]{McLure2011,Dunlop2012,Finkelstein2012,Rogers2013,Bouwens2014,Jiang2020,Bhatawdekar2021,Topping2022}. Measurement of the hydrogen Balmer line ratios (i.e., H$\gamma$/H$\beta$) enables derivation of the attenuation facing the emission lines. Assuming the case B recombination and $T_{\rm{e}}=10^4$~K gas, we expect the intrinsic H$\gamma$/H$\beta$ ratio to be $0.468$ \citep{Osterbrock2006}. The presence of dust will preferentially attenuate H$\gamma$ relative to H$\beta$, decreasing the flux ratio. At $z\simeq2$, galaxies with similar [O~{\small III}] EWs to those in our spectroscopic sample have negligible attenuation, with H$\alpha$/H$\beta$ ratios nearly identical to the case B value \citep{Tang2019}. The H$\gamma$/H$\beta$ line ratios in individual galaxies in our sample are highly uncertain (median error $=0.17$) and thus not surprisingly span a significant range ($0.33-0.93$). To more robustly assess the typical level of attenuation in our sample, we create a composite spectrum by stacking the NIRSpec MR grating spectra. We first shift individual spectra to the rest-frame using the systemic redshifts measured from [O~{\small III}]$\lambda5007$. Each spectrum is then interpolated to a common rest-frame wavelength scale of $1$~\AA\ and normalized by individual [O~{\small III}]$\lambda5007$ luminosities. Finally the spectra are stacked using inverse variance weighted luminosities in each wavelength bin (i.e., weighted by $1/\sigma^2$ where $\sigma$ is the standard deviation of individual spectrum). 

\begin{figure*}
\begin{center}
\includegraphics[width=\linewidth]{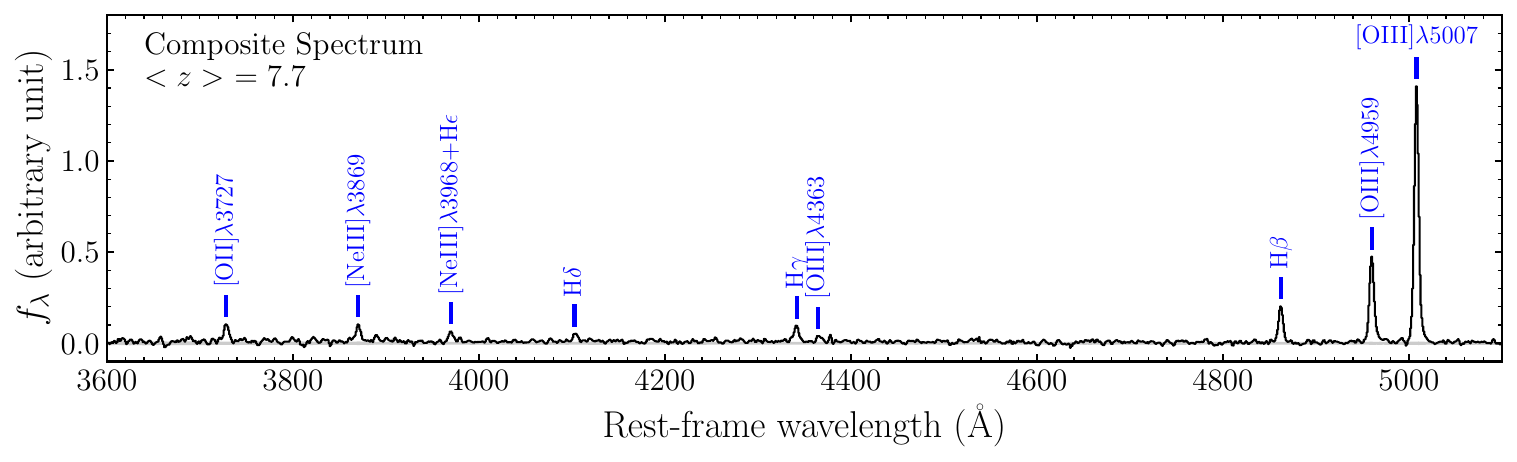}
\caption{Composite spectrum of the 16 galaxies at $z\gtrsim7$ with MR grating spectra at rest-frame optical wavelengths. Each individual source spectrum is normalized by [O~{\small III}]$\lambda5007$ luminosity and stacked using inverse variance weighted luminosities in each wavelength bin. The average redshift of the 16 galaxies is $<z>\ =7.7$. The grey shaded regions show the uncertainty of flux. The detected strong rest-frame optical emission lines are marked by blue lines.}
\label{fig:spectra_stack}
\end{center}
\end{figure*}

We show the resulting $z\simeq7-9$ composite spectrum in Fig.~\ref{fig:spectra_stack}. We measure H$\gamma$/H$\beta=0.463\pm0.038$ in the composite, consistent with the case B recombination value and hence implying negligible attenuation ($E(B-V)=0.02$; assuming the \citealt{Cardelli1989} curve). We also consider a stack of the 3 $z\gtrsim7$ Ly$\alpha$ emitting galaxies (CEERS-1019, CEERS-1027, and CEERS-698). In this composite, we measure H$\gamma$/H$\beta=0.458\pm0.046$, also consistent with negligible attenuation ($E(B-V)=0.04$). Based on these results, we will not apply dust corrections to our fiducial line ratio measurements. However we will comment on the influence that such corrections may have on our results, considering in particular CEERS-1023, which shows modest reddening of its stellar continuum as mentioned above.

As a first exploration of the gas properties of our sample, we use \textsc{beagle} \citep{Chevallard2016} to fit available rest-frame optical emission lines including [O~{\small II}]$\lambda3727$, H$\gamma$, [O~{\small III}]$\lambda4363$, H$\beta$, [O~{\small III}]$\lambda4959$, and [O~{\small III}]$\lambda5007$. We use a similar \textsc{beagle} CSFH model setup to that described in Section~\ref{sec:sed} and in Plat et al. (in prep.), allowing the ionization parameter $U$ and the interstellar metallicity $Z_{\rm{ISM}}$ to vary in the range $-4<\log{U}<-1$ and $-2.2<\log{(Z/Z_{\odot})}<0.24$ with log-uniform priors, and the dust-to-metal ratio $\xi_{\rm{d}}$ to vary in $0.1<\xi_{\rm{d}}<0.5$ with uniform prior. The stellar population age is allowed to vary between $1$~Myr and $10$~Myr with a log-uniform prior because the hydrogen-ionizing spectrum (hereafter ``ionizing spectrum'') reaches steady state after $10$~Myr of CSFH. In modeling the emission line ratios, we must take into account observations that have demonstrated that high-redshift galaxies are typically alpha-enhanced \citep[e.g.,][]{Steidel2016,Strom2017,Strom2018,Strom2022,Shapley2019,Sanders2020,Topping2020,Cullen2021,Runco2021}, as expected given the delay in iron production (relative to oxygen and other alpha elements). Given that it is the iron abundance in massive star atmospheres that regulates the emergent ionizing spectrum whereas the oxygen acts as an important coolant in the $T\simeq10^4$~K ionized gas, this implies that the stellar population metallicity is likely to be lower than the oxygen abundance in the H~{\small II} regions\footnote{The comparison between metallicities and elemental abundances is based on a solar abundance pattern, i.e., the stellar metallicity relative to solar metallicity, $Z_{\star}/Z_{\odot}$, is lower than the oxygen abundance relative to the solar oxygen abundance, [O/H]/[O/H]$_{\odot}$.}. Motivated by these results, we consider \textsc{beagle} photoionization models where the stellar metallicity (tracing the iron abundance [Fe/H]) is set to $1/5$ of the interstellar metallicity (tracing the oxygen abundance [O/H]). This corresponds to the theoretical limit for type II supernovae \citep{Nomoto2006}. We have also considered \textsc{beagle} models where the stellar metallicity is matched to the interstellar metallicity, and we confirm that our main results are not strongly dependent on these assumptions. Finally, motivated by the Balmer line ratios in the composite spectra described above, we assume the emission lines are not attenuated, setting $\tau_V=0$ in our fits. However we will also explore how sensitive our results are to this assumption. 

\begin{figure*}
\begin{center}
\includegraphics[width=\linewidth]{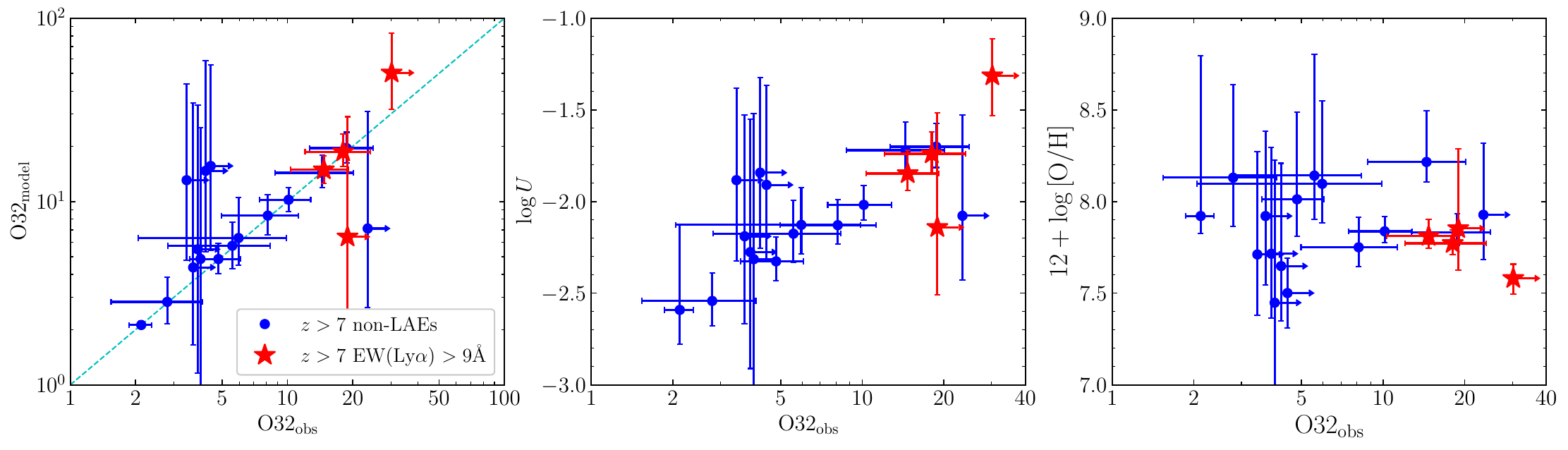}
\caption{Comparison between O32 measured from NIRSpec spectra and the results derived from the best-fit photoionization models using \textsc{beagle} (assuming $\alpha$-enhanced models with stellar metallicity $Z_{\star}=0.2Z_{\rm{ISM}}$, where $Z_{\rm{ISM}}$ is interstellar metallicity). Each panel shows observed O32 versus O32 derived from photoionization models (left), ionization parameter $\log{U}$ (middle), and gas-phase oxygen abundance (right). LAEs (EW$_{\rm{Ly}\alpha}>9$~\AA) and non-LAEs (EW$_{\rm{Ly}\alpha}<9$~\AA) at $z\gtrsim7$ are shown by red stars and blue circles, respectively.}
\label{fig:model}
\end{center}
\end{figure*}

We summarize the \textsc{beagle} model results in Table~\ref{tab:photoionization}. The observed rest-frame optical flux ratios are generally well-reproduced by the models. As an example, we show the consistency between the observed O32 indices and the preferred model O32 values in the left panel of Fig.~\ref{fig:model}. We plot the model-preferred ionization parameters and gas-phase oxygen abundances in the middle and right panels of Fig.~\ref{fig:model}, respectively. The spectroscopic sample is characterized by high ionization parameters (median $\log{U}=-2.11$) and metal poor gas (median $12+\log{(\rm{O/H})}=7.84$). If we consider the 8 galaxies with the largest O32 ratios in the sample (O32 $>10$), we find models prefer similarly metal poor gas (median $12+\log{(\rm{O/H})}=7.84$) and even higher ionization parameters ($\log{U}=-1.80$). Similar results are seen when we consider the 4 $z\gtrsim7$ galaxies with strong Ly$\alpha$ emission (EW $>9$~\AA). Here we find $\log{U}=-2.14$ to $-1.31$ and $12+\log{(\rm{O/H})}=7.58$ to $7.85$. By fitting the rest-frame optical emission lines measured from the composite MR grating spectrum, we also find a high ionization parameter ($\log{U}=-1.76$) and metal-poor gas ($12+\log{(\rm{O/H})}=7.74$). It is clear from this analysis that while the entire $z\gtrsim7$ sample is quite extreme in its nature, the Ly$\alpha$ emitting galaxies stand out as having among the most metal poor gas and largest ionization parameters (see red symbols in Fig.~\ref{fig:model}). These model results are not strongly dependent on our assumption of zero attenuation. If we instead allow $\tau_V$ to take on a slightly larger value that is still consistent with the blue UV slopes in the SEDs (i.e., $\tau_V=0.1$), we find that the gas-phase metallicities and ionization parameters change by less than $0.1$ dex. The gas-phase metallicities of our sources are consistent with those found in other studies of the $z\gtrsim7$ population with similar stellar masses \citep[e.g.,][]{Cameron2023,Curti2023,Nakajima2023,Sanders2023b}, and they are well below what is typically seen in UV-selected galaxies at $z\simeq2$ ($12+\log{(\rm{O/H})}=8.1-8.5$; \citealt{Steidel2016,Sanders2021}). Taken at face value, these results suggest that very metal poor gas is common in $z\gtrsim7$ galaxies dominated by recent upturns in star formation (and/or combined with AGN contributions; e.g., \citealt{Larson2023}) which power strong [O~{\small III}]+H$\beta$ emission.

\begin{table*}
\centering
\begin{tabular}{ccccccc}
\hline
ID & $z_{\rm{spec}}$ & O32$_{\rm{obs}}$ & O32$_{\rm{model}}$ & $\log{U}$ & $12+\log{(\rm{O/H})}_{\rm{model}}$ & $12+\log{(\rm{O/H})}_{\rm{direct}}$ \\
\hline
CEERS-24 & $8.999$ & $>3.98$ & $4.86^{+20.48}_{-4.20}$ & $-2.31^{+0.79}_{-1.04}$ & $7.45^{+0.78}_{-0.59}$ & ... \\
CEERS-23 & $8.881$ & $>3.87$ & $5.49^{+28.00}_{-4.33}$ & $-2.28^{+0.72}_{-0.64}$ & $7.72^{+0.58}_{-0.35}$ & ... \\
CEERS-1025 & $8.715$ & $8.12\pm3.15$ & $8.38^{+2.53}_{-1.79}$ & $-2.13^{+0.14}_{-0.10}$ & $7.75^{+0.16}_{-0.11}$ & ... \\
CEERS-1019 & $8.678$ & $18.06\pm5.99$ & $18.68^{+4.64}_{-3.08}$ & $-1.74^{+0.12}_{-0.10}$ & $7.77^{+0.07}_{-0.06}$ & $7.73^{+0.17}_{-0.14}$ \\
CEERS-1029 & $8.610$ & $4.81\pm1.25$ & $4.85^{+1.06}_{-0.83}$ & $-2.33^{+0.13}_{-0.11}$ & $8.01^{+0.48}_{-0.20}$ & ... \\
CEERS-1149 & $8.175$ & $18.77\pm6.06$ & $19.54^{+4.33}_{-3.33}$ & $-1.70^{+0.13}_{-0.11}$ & $7.83^{+0.10}_{-0.07}$ & ... \\
CEERS-3 & $8.00$ & $>3.68$ & $4.37^{+30.17}_{-2.72}$ & $-2.19^{+0.66}_{-0.48}$ & $7.92^{+0.46}_{-0.38}$ & ... \\
CEERS-1027 & $7.819$ & $>30.22$ & $50.34^{+32.34}_{-18.58}$ & $-1.31^{+0.20}_{-0.22}$ & $7.58^{+0.08}_{-0.09}$ & $7.74^{+0.28}_{-0.19}$ \\
CEERS-1023 & $7.776$ & $2.80\pm1.25$ & $2.83^{+1.03}_{-0.68}$ & $-2.54^{+0.15}_{-0.14}$ & $8.13^{+0.51}_{-0.27}$ & ... \\
CEERS-689 & $7.545$ & $10.13\pm2.67$ & $10.22^{+1.69}_{-1.43}$ & $-2.02^{+0.10}_{-0.08}$ & $7.84^{+0.08}_{-0.06}$ & $7.66^{+0.34}_{-0.18}$ \\
CEERS-698 & $7.470$ & $14.72\pm4.40$ & $14.91^{+2.76}_{-2.32}$ & $-1.85^{+0.13}_{-0.09}$ & $7.81^{+0.09}_{-0.07}$ & $8.00^{+0.30}_{-0.19}$ \\
CEERS-1163 & $7.448$ & $>4.20$ & $14.70^{+43.98}_{-9.40}$ & $-1.84^{+0.52}_{-0.41}$ & $7.65^{+0.56}_{-0.30}$ & ... \\
CEERS-1038 & $7.194$ & $>4.43$ & $15.62^{+40.18}_{-10.19}$ & $-1.91^{+0.54}_{-0.41}$ & $7.50^{+0.19}_{-0.19}$ & ... \\
CEERS-498 & $7.18$ & $>23.45$ & $7.12^{+23.87}_{-4.48}$ & $-2.08^{+0.55}_{-0.35}$ & $7.93^{+0.39}_{-0.24}$ & ... \\
CEERS-499 & $7.168$ & $>3.43$ & $13.07^{+30.70}_{-8.28}$ & $-1.88^{+0.50}_{-0.44}$ & $7.71^{+0.56}_{-0.33}$ & ... \\
CEERS-44 & $7.10$ & $>18.93$ & $6.42^{+22.60}_{-4.10}$ & $-2.14^{+0.63}_{-0.37}$ & $7.85^{+0.43}_{-0.23}$ & ... \\
CEERS-407 & $7.028$ & $5.96\pm3.91$ & $6.32^{+4.21}_{-1.81}$ & $-2.13^{+0.20}_{-0.16}$ & $8.10^{+0.45}_{-0.21}$ & ... \\
CEERS-1102 & $7.00$ & $2.12\pm0.26$ & $2.12^{+0.10}_{-0.10}$ & $-2.59^{+0.46}_{-0.19}$ & $7.92^{+0.87}_{-0.10}$ & ... \\
CEERS-717 & $6.932$ & $5.58\pm2.76$ & $5.73^{+1.98}_{-1.42}$ & $-2.18^{+0.18}_{-0.16}$ & $8.14^{+0.66}_{-0.24}$ & ... \\
CEERS-1143 & $6.928$ & $14.46\pm5.70$ & $14.34^{+3.63}_{-2.46}$ & $-1.72^{+0.15}_{-0.12}$ & $8.22^{+0.28}_{-0.11}$ & ... \\
Composite & $7.7$ & $17.84\pm1.71$ & $17.76^{+1.18}_{-1.10}$ & $-1.76^{+0.06}_{-0.09}$ & $7.74^{+0.05}_{-0.04}$ & $7.83^{+0.02}_{-0.03}$ \\
\hline
\end{tabular}
\caption{Best-fit parameters (O32, ionization parameter $\log{U}$, and gas-phase oxygen abundance $12+\log{(\rm{O/H})}$) derived from \textsc{beagle} modeling of rest-frame optical emission line ratios for both individual galaxies at $z\gtrsim7$ and the composite MR grating spectrum. We adopt $\alpha$-enhanced models with stellar metallicity lower than interstellar metallicity, $Z_{\star}=0.2Z_{\rm{ISM}}$ (Plat et al. in prep.). In the last column, we also list the oxygen abundance derived using direct-method for the four galaxies with [O~{\scriptsize III}]$\lambda4363$ detections (see Table~\ref{tab:source} and \ref{tab:opt_line}; the [O~{\scriptsize III}]$\lambda4363$ detections in CEERS-689 and CEERS-698 are tentative with S/N $\simeq2$). The oxygen abundance derived from photoionization models and direct-method is consistent.}
\label{tab:photoionization}
\end{table*}

The gas-phase oxygen abundance can also be constrained with the direct method in the two galaxies with robust detections of the [O~{\small III}]$\lambda4363$ auroral emission line (CEERS-1019, CEERS-1027). We follow the procedures in \citet{Izotov2006} to derive the direct $T_{\rm{e}}$ oxygen abundance. We use the observed [O~{\small III}]$\lambda4363$/[O~{\small III}]$5007$ ratios to compute the O$^{++}$ zone electron temperature ($T_{\rm{e}}$(O~{\small III})) with {\small PYTHON} package {\small PyNeb}\footnote{\url{http://research.iac.es/proyecto/PyNeb//}}. We assume an electron density of $n_{\rm{e}}=250$~cm$^{-3}$ \citep{Sanders2016}, and we have confirmed that the derived electron temperature and oxygen abundance change negligibly over $n_{\rm{e}}=100-1000$~cm$^{-3}$. Since we do not have auroral [O~{\small II}]$\lambda\lambda7320,7330$ detections, we derive the O$^{+}$ zone electron temperature $T_{\rm{e}}$(O~{\small II}) using the relation from \citet{Campbell1986} and \citet{Garnett1992}: $T_{\rm{e}}$(O~{\small II}) $=0.7\times T_{\rm{e}}$(O~{\small III}) $+3000$~K. Then the O$^{++}$ abundance and O$^{+}$ abundance are derived using $T_{\rm{e}}$ and $n_{\rm{e}}$ with [O~{\small III}]$\lambda\lambda4959,5007$/H$\beta$ and [O~{\small II}]$\lambda\lambda3726,3729$/H$\beta$ ratios. The final oxygen abundance is O/H $=$ O$^{++}$/H $+$ O$^{+}$/H. The results indicate very metal poor H~{\small II} regions, with values consistent with the photoionization modeling described above. For CEERS-1019, the direct method suggests $12+\log{(\rm{O/H})}=7.73$ ($0.10\ Z_\odot$, where the solar metallicity corresponds to a gas-phase oxygen abundance $12+\log{(\rm{O/H})}=8.71$; \citealt{Gutkin2016}) whereas the \textsc{beagle} models suggest $12+\log{(\rm{O/H})}=7.77$. For CEERS-1027, the direct method indicates $12+\log{(\rm{O/H})}=7.74$ ($0.11\ Z_\odot$), which is again similar to the preferred \textsc{beagle} model value ($12+\log{(\rm{O/H})}=7.58$) within $1\sigma$ confidence interval. The oxygen abundances of these two galaxies are also consistent with the values derived from direct-$T_{\rm{e}}$ method reported in \citet{Sanders2023b}. 

\begin{figure}
\begin{center}
\includegraphics[width=\linewidth]{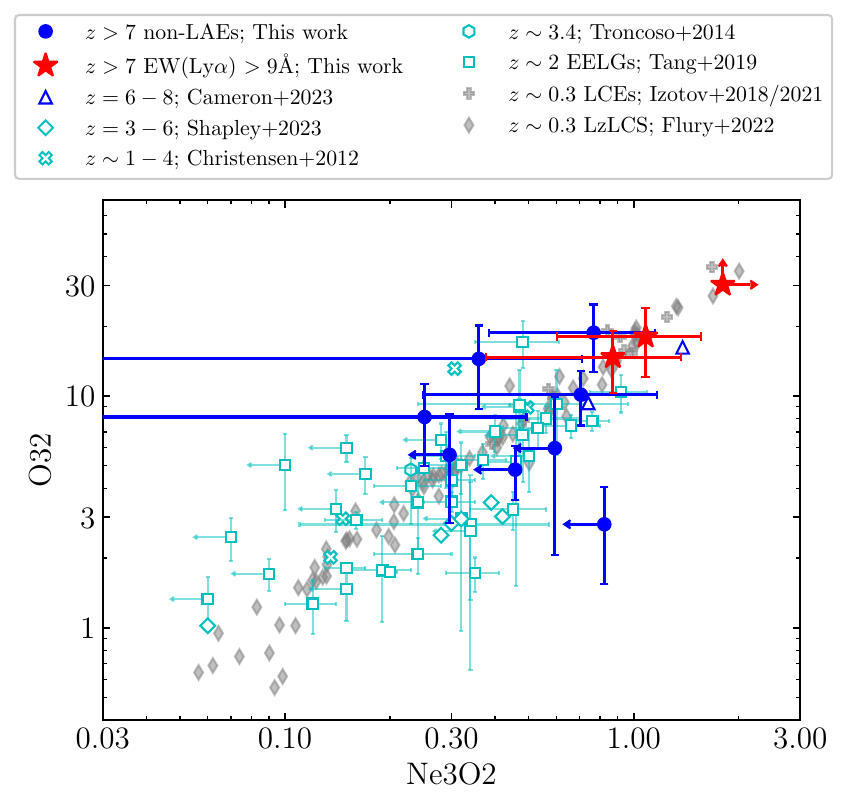}
\caption{Ionization sensitive line ratios O32 versus Ne3O2 for CEERS NIRSpec galaxies at $z\gtrsim7$. LAEs (EW$_{\rm{Ly}\alpha}>9$~\AA) and non-LAEs at $z\gtrsim7$ are shown by red stars and blue circles, respectively. We also plot galaxies at $z=6-8$ from \citet{Cameron2023} in open blue triangles. It is clear that Ly$\alpha$ emitting galaxies have larger O32 and Ne3O2 indicating harder ionizing spectra. As a comparison, we show the line ratios of star-forming galaxies at $z\sim1-4$ (open cyan crosses; \citealt{Christensen2012}), at $z\sim3.4$ (open cyan hexagons; \citealt{Troncoso2014}), and at $z\sim3-5$ (open cyan diamonds; \citealt{Shapley2023}), EELGs at $z\sim2$ (open cyan squares; \citealt{Tang2019}), as well as Lyman continuum emitters (LCEs) at $z\simeq0.2-0.4$ from \citet{Izotov2018,Izotov2021} and from the Low-Redshift Lyman Continuum Survey (LzLCS; \citealt{Flury2022}) in grey plus and diamond, respectively.}
\label{fig:o32_ne3o2}
\end{center}
\end{figure}

We also gain insight into the nature of our sample through comparison of line ratios to other {\it JWST} measurements at $z\gtrsim7$ and those at lower redshifts. The O32 index is sensitive to the ionization state of the nebular gas, providing a probe of the ionization parameter. The Ne3O2 index is nearly equivalent as O32 since both neon and oxygen are alpha elements and Ne$^{++}$ and O$^{++}$ have similar ionization potentials \citep[e.g.,][]{Perez-Montero2007,Levesque2014,Witstok2021}. Therefore, Ne3O2 is also sensitive to the ionization conditions of the H~{\small II} regions \citep[e.g.,][]{Levesque2014,Strom2018}, with the benefit of being slightly less dependent on the attenuation correction owing to the proximity of the two emission lines in wavelength. In Fig.~\ref{fig:o32_ne3o2}, we plot the O32 and Ne3O2 ratios of the $z\gtrsim7$ sample in CEERS. The O32 values range between $2.12$ and $>30.22$. We find an average value of O32 $=17.84$ from the composite spectrum (Fig.~\ref{fig:spectra_stack}), consistent with the average O32 found in $z\gtrsim7$ galaxies with {\it JWST} in literature (O32 $\simeq10-20$; e.g., \citealt{Cameron2023,Mascia2023,Sanders2023a,Saxena2023b}. This is about $4\times$ larger than values seen at $z\simeq3-5$ (O32 $\simeq3-5$; e.g., \citealt{Christensen2012,Troncoso2014,Shapley2023}) and about an order of magnitude larger than typical values seen at $z\simeq2$ (O32 $\simeq1-2$; e.g., \citealt{Steidel2016,Sanders2016,Sanders2021}, suggesting that the $z\gtrsim7$ spectroscopic sample is comprised of galaxies with extreme ionization conditions, consistent with the high ionization parameters preferred by \textsc{beagle}. The Ne3O2 ratios suggest a similar picture, with an average value (Ne3O2 $=0.89$) measured from composite spectrum (Fig.~\ref{fig:spectra_stack}) that is consistent with $z\gtrsim7$ measurements in literature (Ne3O2 $\simeq1$; e.g., \citealt{Trump2023,Cameron2023}) and well above the average at $z\simeq2$ (Ne3O2 $=0.10-0.15$; \citealt{Steidel2016,Sanders2021}) and at $z\simeq3-5$ (Ne3O2 $=0.2-0.4$; \citealt{Christensen2012,Troncoso2014,Shapley2017,Witstok2021,Shapley2023}). These conclusions are unchanged if we apply a modest attenuation correction ($\tau_V=0.1$) to the emission lines. The Ne3O2 ratios do not change at all, O32 is moderately reduced (average O32 $=14.31$) but still well above the typical values at $z\simeq2$.

\begin{figure}
\begin{center}
\includegraphics[width=\linewidth]{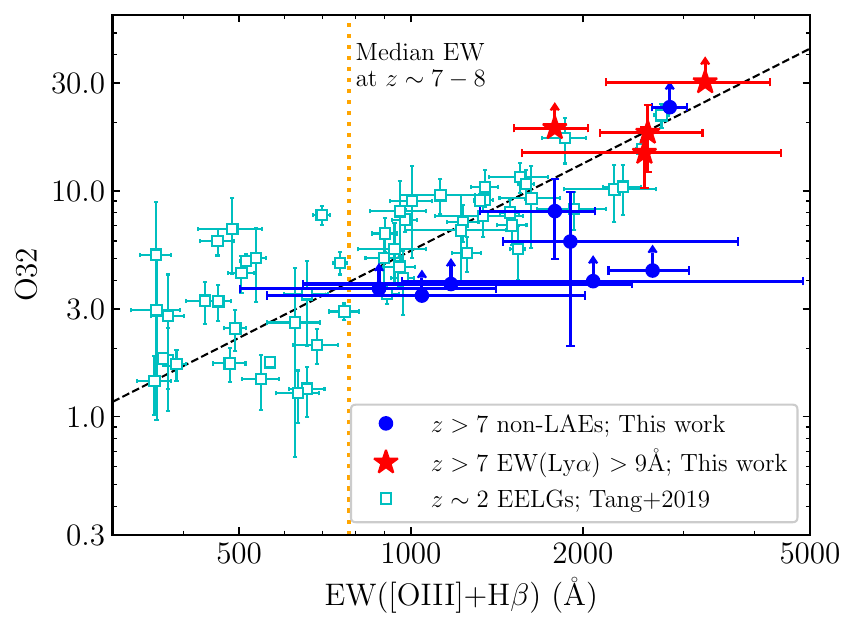}
\caption{Correlation between O32 and [O~{\scriptsize III}]+H$\beta$ EW for CEERS NIRSpec galaxies at $z>7$. EW$_{\rm{[OIII]+H}\beta}$ is derived from SED fitting. LAEs (EW$_{\rm{Ly}\alpha}>9$~\AA) and non-LAEs at $z>7$ are shown by red stars and blue circles, respectively. The median [O~{\scriptsize III}]+H$\beta$ EW of typical $z\sim7-8$ star-forming galaxies ($\approx780$~\AA; \citealt{Endsley2023a}) is shown by the vertical orange dotted line. O32 and EW$_{\rm{[OIII]+H}\beta}$ of EELGs at $z\sim2$ \citep{Tang2019} are overplotted as cyan squares. The O32 $-$ EW$_{\rm{[OIII]+H}\beta}$ relation at $z\sim2$ from \citet{Tang2019} is presented as black dashed line.}
\label{fig:o32_o3hbew}
\end{center}
\end{figure}

Both O32 and Ne3O2 are known to increase with the rest-frame optical emission line equivalent widths \citep[e.g.,][]{Tang2019,Sanders2020}. In Fig.~\ref{fig:o32_o3hbew}, we plot the O32 ratios of our sample versus the [O~{\small III}]+H$\beta$ EW for the 12 galaxies where we have robust SED constraints and O32 measurements. We overlay the $z\gtrsim7$ results on the $z\simeq2$ relation derived from spectroscopic follow-up of EELGs \citep{Tang2019}. At $z\simeq2$, it is seen that O32 increases from $2$ at [O~{\small III}]+H$\beta=300$~\AA\ to O32 $=20$ at [O~{\small III}]+H$\beta=3000$~\AA, reflecting the tendency for ionization parameter to be larger in galaxies with very young stellar populations (see discussion in \citealt{Tang2019} and Plat et al. in prep.). The $z\gtrsim7$ measurements are consistent with the $z\simeq2$ relation, albeit mostly sampling the large O32 ratios associated with the highest [O~{\small III}]+H$\beta$ EWs ($>1000$~\AA). In the context of O32 $-$ [O~{\small III}]+H$\beta$ EW relation, the presence of large O32 ratios in our sample largely reflects the young stellar population ages (and associated large [O~{\small III}]+H$\beta$ EWs) of the sources in the spectroscopic sample.

Based on Fig.~\ref{fig:o32_o3hbew}, the redshift evolution in the distribution of O32 (and other ionization-sensitive line ratios) will depend significantly on the evolution in the [O~{\small III}]+H$\beta$ EW distribution. Recent work has shown that the [O~{\small III}]+H$\beta$ EW of the overall population increases between $z\simeq2$ and $z\simeq7$, going from typical values of $250$~\AA\ in relatively massive galaxies at $z\simeq2$ \citep[e.g.,][]{Reddy2018,Boyett2022} to $780$~\AA\ at $z\simeq7$ \citep[e.g.,][]{Endsley2021a,Endsley2023a}. This trend will result in larger O32 ratios being more common at $z\gtrsim7$, undoubtedly contributing somewhat to the larger O32 ratios seen in our CEERS spectroscopic sample. Recent ALMA observations have also revealed that the ionization-sensitive far-infrared [O~{\small III}]$88\mu$m/[C~{\small II}]$158\mu$m ratios are ubiquitously high in $z>6$ galaxies, with values larger than in lower-redshift galaxies \citep[e.g.,][]{Harikane2020,Carniani2020}. It is likely that [O~{\small III}]$88\mu$m/[C~{\small II}]$158\mu$m also correlates with [O~{\small III}]+H$\beta$ EWs \citep[e.g.,][]{Witstok2022}. But the bias in our spectroscopic sample toward large [O~{\small III}]+H$\beta$ EWs (as discussed in Section~\ref{sec:sed}) further shifts our O32 distribution to extreme values. If $z\gtrsim7$ galaxies follow a similar relation between O32 and [O~{\small III}]+H$\beta$ EW as is seen at $z\simeq2$ (as appears to be the case in Fig.~\ref{fig:o32_o3hbew}), then we would expect representative $z\gtrsim7$ galaxy samples to show O32 extending $2-3\times$ lower than the values seen in the CEERS sample. Deeper rest-frame optical spectra will be required to obtain representative line ratio distributions for the full population of $z\gtrsim7$ galaxies. 

\begin{figure}
\begin{center}
\includegraphics[width=\linewidth]{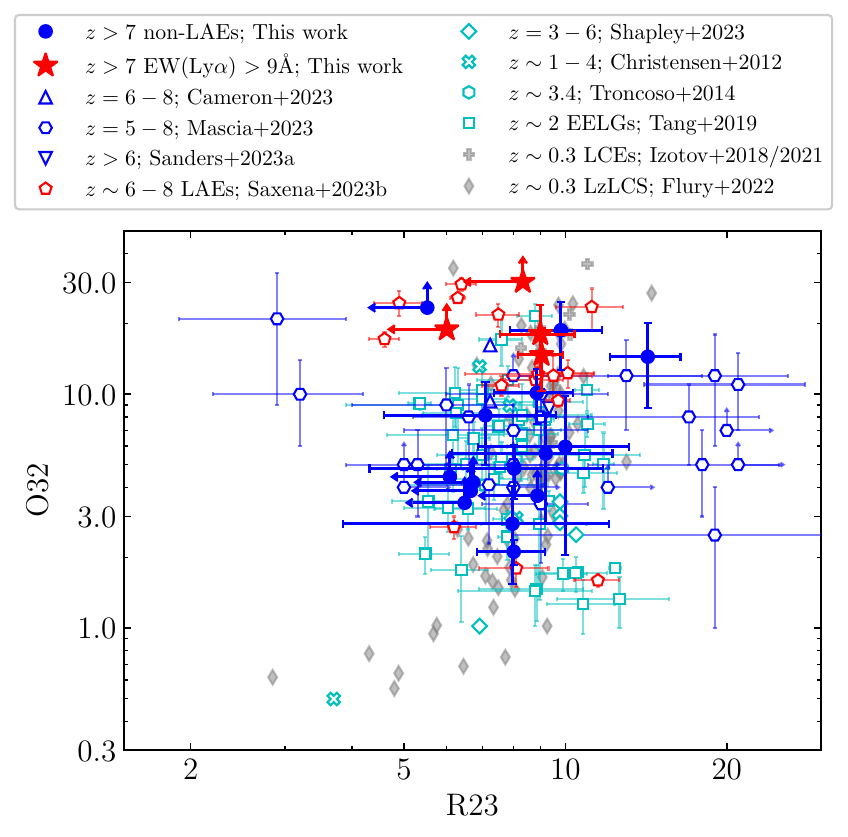}
\caption{O32 versus R23 ratio for CEERS NIRSpec galaxies at $z\gtrsim7$. LAEs (EW$_{\rm{Ly}\alpha}>9$~\AA) and non-LAEs at $z\gtrsim7$ are shown by red stars and blue circles, respectively. We also plot recent {\it JWST}/NIRSpec results at $z\sim5-9$ \citep{Cameron2023,Mascia2023,Sanders2023a} in open blue symbols, as well as Ly$\alpha$ emitters from \citet{Saxena2023b} in open red pentagons. As a comparison, we show the line ratios of star-forming galaxies at $z\sim1-5$ \citep{Christensen2012,Troncoso2014,Shapley2023} and EELGs at $z\sim2$ \citep{Tang2019} in open cyan symbols, as well as LCEs at $z\simeq0.2-0.4$ from \citet{Izotov2018,Izotov2021} and from LzLCS \citep{Flury2022} in grey symbols. Ly$\alpha$ emitting galaxies at $z\gtrsim7$ show larger O32 comparing to non-Ly$\alpha$ emitters at similar redshift as well as the lower redshift analogs.}
\label{fig:o32_r23}
\end{center}
\end{figure}

We also quantify the [O~{\small III}]$\lambda5007$/H$\beta$ flux ratio (O3) and the R23 index (([O~{\small II}]+[O~{\small III}])/H$\beta$) in Table~\ref{tab:opt_line}. We measure an average O3 value of $6.8$ from the composite spectrum of our CEERS $z\gtrsim7$ NIRSpec sample (Fig.~\ref{fig:spectra_stack}). This is consistent with the average O3 ratios measured in $z\gtrsim7$ galaxies in literature (O3 $\simeq5-7$; e.g., \citealt{Cameron2023,Nakajima2023,Sanders2023a,Trump2023}). It is only marginally greater than the O3 seen in typical $z\simeq2$ galaxies ($4.3$ in KBSS, $3.4-4.5$ in MOSDEF; \citealt{Steidel2016,Sanders2021}), where the range quoted in the MOSDEF sample corresponds to line ratios for stacks with stellar masses between $\simeq10^{9.5}\ M_\odot$ and $10^{10}\ M_\odot$. The extreme ionization conditions of the $z\gtrsim7$ galaxies contribute to the larger O3 values, but this is countered by the lower oxygen abundance of the $z\gtrsim7$ galaxies in our sample. The net effect is that O3 ratios are only slightly larger than those at $z\simeq2$. In Fig.~\ref{fig:o32_r23}, we plot the R23 indices of the $z\gtrsim7$. We find R23 values that range between $<6$ and $14$, with an average of $9.4$ measured from the composite spectrum (Fig.~\ref{fig:spectra_stack}). These are comparable to the R23 values seen in $z\simeq2$ star-forming galaxies (R23 $=8.5$ in KBSS and R23 $=8.9-9.1$ for $M_{\star}=10^{9.5}-10^{10}\ M_\odot$ galaxies in MOSDEF; \citealt{Steidel2016,Sanders2021}) and other $z\gtrsim7$ galaxies (R23 $\simeq7-10$; e.g., \citealt{Cameron2023,Mascia2023,Sanders2023a,Saxena2023b}). The origin of the similar R32 indices (in spite of very different gas-phase properties) is analogous to the explanation for O3. The lower metallicity of the $z\gtrsim7$ galaxies increases the excitation (boosting collisionally-excited lines relative to H$\beta$), but the effect this has on R23 is countered by the reduced oxygen abundance. Thus in the R23 versus O32 plane (Fig.~\ref{fig:o32_r23}), we primarily see the $z\gtrsim7$ galaxies shift toward larger O32 at fixed R23.

In summary, the CEERS NIRSpec observations have revealed a population of galaxies at $z\gtrsim7$ that is very different than that at $z\simeq2$, with significantly more metal poor gas and larger ionization parameters. The differences are most clear in the ionization-sensitive line ratios. The O32 values in our spectroscopic sample are often over an order of magnitude larger than those typical at $z\simeq2$. We find similar line ratios in samples of galaxies at lower redshifts with comparable [O~{\small III}]+H$\beta$ EWs as those in our sample. These include the most extreme tail of the EELG population at $z\sim2$ \citep[e.g.,][]{Tang2019,Du2020} and a subset of galaxies in the LzLCS survey \citep{Flury2022}. We will come back to directly compare the Ly$\alpha$ properties of these samples to the $z\gtrsim7$ galaxies in Section~\ref{sec:discussion}. 

\subsection{Rest-Frame UV Emission Lines in $z\gtrsim7$ Galaxies} \label{sec:uv_lines}

The emission lines in the rest-frame UV of $z\gtrsim7$ galaxies also provide insights into the Ly$\alpha$ transmission in the reionization era and the ionizing sources. NIRSpec data not only probe the Ly$\alpha$ EWs, but they also constrain the velocity profile and transmission of Ly$\alpha$, both valuable diagnostics for assessment of the impact of the IGM on the line. The fainter UV metal emission lines (i.e., C~{\small III}], C~{\small IV}) and He~{\small II} provide information on the stars, gas, and AGN activity in the reionization-era population, allowing investigation of whether $z\gtrsim7$ sources with Ly$\alpha$ are different from the general population. We describe the analysis associated with each of these measurements in this subsection.

\begin{figure}
\begin{center}
\includegraphics[width=\linewidth]{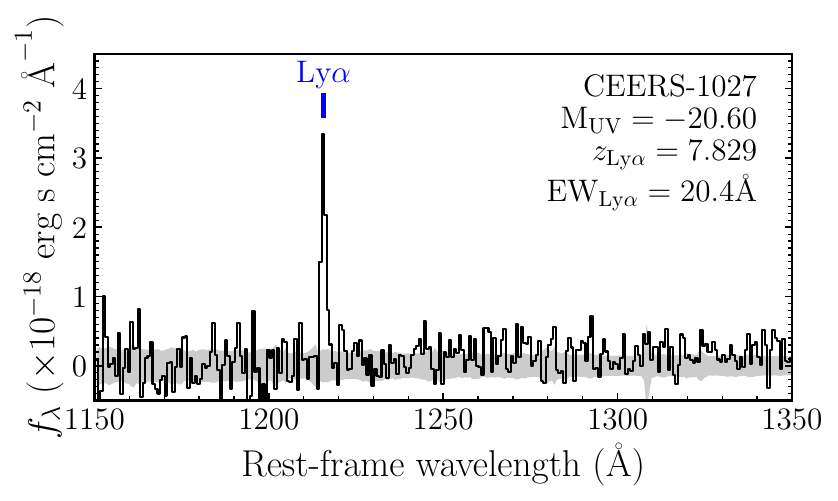}
\caption{Ly$\alpha$ emission line of the newly detected Ly$\alpha$ emitter CEERS-1027. The Ly$\alpha$ spectrum has been shifted to the rest-frame. The grey shaded regions show the uncertainty of flux. Medium-resolution Ly$\alpha$ spectra of the four objects with Ly$\alpha$ detections (CEERS-1019, CEERS-1149, CEERS-1027, CEERS-698) are shown in Fig.~\ref{fig:spectra_lya_2d} in Appendix~\ref{sec:appendix}.}
\label{fig:spectra_lya_1d}
\end{center}
\end{figure}

We detect Ly$\alpha$ emission in six $z>7$ galaxies in the CEERS spectroscopic sample (see Table~\ref{tab:lya_line}). Two of the six (CEERS-1027, Fig.~\ref{fig:spectra_lya_1d}, CEERS-44, Fig.~\ref{fig:spectra_prism}) are newly-confirmed as Ly$\alpha$ emitters in these observations. Both are strong Ly$\alpha$ emitters for $z\gtrsim 7$ \citep[e.g.,][]{Endsley2021b}, with rest-frame EWs of 77.6~\AA\ (CEERS-44) and 20.4~\AA\ (CEERS-1027). The other four are previously confirmed Ly$\alpha$ emitters, including CEERS-1019 \citep{Zitrin2015}, CEERS-1029 (\citealt{Larson2022}; tentative Ly$\alpha$ detection found in NIRSpec spectrum), CEERS-686 \citep{Jung2022}, and CEERS-698 \citep{Roberts-Borsani2016,Stark2017}. We notice that the Ly$\alpha$ emission lines of CEERS-1019, CEERS-1029, and CEERS-698 have lower S/N ($\simeq3-4$; Fig.~\ref{fig:spectra_lya_2d}). To test whether the emission features of these three objects are due to noise fluctuations, we follow the test in Chen et al. (in prep.) that randomly place $1000$ circular apertures with radius $=3$~pixels (consistent with the aperture we calculate Ly$\alpha$ flux) near the Ly$\alpha$ emission feature of each object and measure the aperture fluxes. For CEERS-1019 and CEERS-698, we find that no random aperture shows flux comparable to the emission feature. This indicates that the emission features of these two objects are more likely to be Ly$\alpha$ than random fluctuations. For CEERS-1029, there are a few random apertures with fluxes approaching the significance of the emission feature. Because the emission feature of CEERS-1029 lies at the expected position of Ly$\alpha$ (after taking into account the difference between air and vacuum wavelengths) based on the Ly$\alpha$ redshift reported in \citep{Larson2022}, we recognize it as a tentative Ly$\alpha$ emission. The low S/N Ly$\alpha$ detections of these three objects demonstrates that we are pushing the limits of NIRSpec spectra in CEERS observations. Future deep spectroscopy will help to improve these measurements.

Each of CEERS-1019, CEERS-1029, CEERS-686, and CEERS-698 show UV continuum emission in the CEERS spectra, allowing us to compute Ly$\alpha$ EWs. Here we compute Ly$\alpha$ EW using the average continuum flux between rest-frame $1225$~\AA\ and $1255$~\AA, which minimizes the contribution from nearby absorption features \citep[e.g.,][]{Kornei2010,Stark2010}. The derived values range between $4.2$~\AA\ (CEERS-1029) and $41.9$~\AA\ (CEERS-686). In three of these four galaxies, the Ly$\alpha$ EWs are consistent (within $1\sigma$ uncertainties) with the ground-based measurements from MOSFIRE. In the case of CEERS-1019, the offset is just outside of the $1\sigma$ confidence interval. The ground-based Ly$\alpha$ EW measurement is $28^{+14}_{-11}$~\AA, with the large uncertainty due to the presence of an OH sky line near the Ly$\alpha$ feature. The NIRSpec observations suggest the EW is much lower ($9.8\pm2.4$~\AA). The NIRSpec micro-shutter is centered on the two brightest clumps of the galaxy, as revealed in NIRCam imaging. It is possible that the MOSFIRE observations may have picked up additional Ly$\alpha$ emission associated with a fainter clump outside of the NIRSpec micro-shutter. Given the impact of the sky line on the ground-based Ly$\alpha$ measurement, we will adopt the NIRSpec value as fiducial in our analysis, but we will comment on how our results would change if the MOSFIRE value is correct.

\begin{table*}
\centering
\begin{tabular}{ccccccccc}
\hline
ID & $z_{\rm{sys}}$ & $z_{\rm{Ly}\alpha}$ & EW(Ly$\alpha$)$_{\rm{literature}}$ & EW(Ly$\alpha$)$_{\rm{NIRSpec}}$ & $\Delta v_{\rm{Ly}\alpha}$ & $f_{\rm{Ly}\alpha}/f_{\rm{H}\beta}$ & $f_{\rm{esc,Ly}\alpha}$ & Ref. \\
 & & & (\AA) & (\AA) & (km~s$^{-1}$) & & & \\
\hline
CEERS-23 & $8.881$ & ... & ... & ... & ... & $<10.50$ & $<0.422$ & \\
CEERS-1019 & $8.678$ & $8.693$ & $28.0^{+14.0}_{-11.0}$ & $9.8\pm2.4$ & $458\pm161$ & $0.87\pm0.25$ & $0.035\pm0.010$ & [1] \\
CEERS-1029 & $8.610$ & $8.672$ & $4.7^{+1.7}_{-1.7}$ & $4.2\pm1.3$ & $1938\pm162$ & $1.52\pm0.84$ & $0.061\pm0.034$ & [2] \\
CEERS-1149 & $8.175$ & ... & ... & $<15.0$ & ... & $<0.91$ & $<0.036$ & \\
CEERS-3 & $8.00$ & ... & ... & ... & ... & $<4.48$ & $<0.180$ & \\
CEERS-1027 & $7.819$ & $7.829$ & ... & $20.4\pm3.1$ & $323\pm18$ & $2.12\pm0.45$ & $0.085\pm0.018$ & \\
CEERS-1023 & $7.776$ & ... & ... & ... & ... & $<3.35$ & $<0.135$ & \\
CEERS-686 & $7.74$ & $7.86$ & $69.1^{+29.8}_{-19.9}$ & $41.9\pm1.6$ & ... & $8.09\pm0.69$ & $0.325\pm0.028$ & [3] \\
CEERS-689 & $7.545$ & ... & ... & $<26.1$ & ... & $<1.36$ & $<0.055$ & \\
CEERS-698 & $7.470$ & $7.485$ & $9.3^{+1.4}_{-1.4}$ & $9.5\pm3.1$ & $545\pm184$ & $1.12\pm0.38$ & $0.045\pm0.015$ & [4],[5] \\
CEERS-1163 & $7.448$ & ... & ... & ... & ... & $<3.61$ & $<0.145$ & \\
CEERS-1038 & $7.194$ & ... & ... & ... & ... & $<3.57$ & $<0.144$ & \\
CEERS-498 & $7.18$ & ... & ... & $<22.5$ & ... & $<1.84$ & $<0.074$ & \\
CEERS-499 & $7.168$ & ... & ... & ... & ... & $<3.80$ & $<0.153$ & \\
CEERS-44 & $7.10$ & $7.24$ & ... & $77.6\pm5.5$ & ... & $8.44\pm1.09$ & $0.339\pm0.044$ & \\
CEERS-1102 & $7.00$ & ... & ... & $<23.8$ & ... & $<5.15$ & $<0.207$ & \\
\hline
\multicolumn{9}{l}{References: [1]. \citet{Zitrin2015}; [2]. \citet{Larson2022}; [3]. \citet{Jung2022}; [4]. \citet{Roberts-Borsani2016}; [5]. \citet{Stark2017}.} \\
\end{tabular}
\caption{Ly$\alpha$ properties of CEERS spectroscopically confirmed $z>7$ galaxies. Overall the Ly$\alpha$ EWs measured in literature are consistent with the EWs measured from NIRSpec spectra within $1\sigma$ except for CEERS-1019. We adopt the redshift measured by fitting rest-frame optical emission lines (e.g., [O~{\scriptsize III}]$\lambda5007$) as the systemic redshift ($z_{\rm{sys}}$). The Ly$\alpha$ redshifts ($z_{\rm{Ly}\alpha}$) are measured from the peak of Ly$\alpha$ emission presented in NIRSpec spectra. The Ly$\alpha$ velocity offset $\Delta v_{\rm{Ly}\alpha}$ is derived by comparing $z_{\rm{sys}}$ and $z_{\rm{Ly}\alpha}$. We do not provide Ly$\alpha$ velocity offsets for CEERS-686 and CEERS-44 because their systemic redshifts are measured from emission lines detected in low-resolution ($R\sim100$) prism spectra and hence the uncertainty is large ($1\sigma>3000$~km~s$^{-1}$). The Ly$\alpha$ escape fraction ($f_{\rm{esc,Ly}\alpha}$) is computed from Ly$\alpha$ to H$\beta$ ratio, assuming an intrinsic H$\alpha$/H$\beta=2.86$ and Ly$\alpha$/H$\alpha=8.7$.}
\label{tab:lya_line}
\end{table*}

The transmission of Ly$\alpha$ through a partially neutral IGM depends on the velocity profile of the line as it emerges from the galaxy. Ly$\alpha$ is typically observed with a positive velocity offset relative to the systemic redshift of its host galaxy, owing largely to resonant scattering effects (whereby the redshifted component mostly has its origin in line photons that are backscattered from outflowing gas on the far side of the galaxy). The larger the velocity offset ($\Delta v_{\rm{Ly\alpha}}$), the further Ly$\alpha$ photons will be shifted into the damping wing profile before they encounter neutral hydrogen. If velocity offsets in galaxies at $z\gtrsim7$ are large enough, the transmission of Ly$\alpha$ will be less impacted by the neutral IGM \citep[e.g.,][]{Dijkstra2011,Stark2017,Mason2018b,Hashimoto2019,Endsley2022b}. 

Observational efforts to quantify the velocity offsets of reionization era galaxies have long been stunted by the challenge of measuring systemic redshifts with non-resonant emission lines. A small number of $z\gtrsim6$ $\Delta v_{\rm{Ly\alpha}}$ measurements have been made via detection of the faint C~{\small III}] emission line \citep{Stark2015,Stark2017,Hutchison2019}. Recent studies with ALMA have made further progress via detections of [C~{\small II}]$158\mu$m and [O~{\small III}]$88\mu$m fine structure lines \citep[e.g.,][]{Willott2015,Pentericci2016,Carniani2017,Hashimoto2019,Endsley2022b}. {\it JWST} spectroscopy has the potential to make velocity offset work much easier in the reionization era given the ease of efficiently detecting the strong rest-frame optical emission lines in large samples. 

\begin{figure}
\begin{center}
\includegraphics[width=\linewidth]{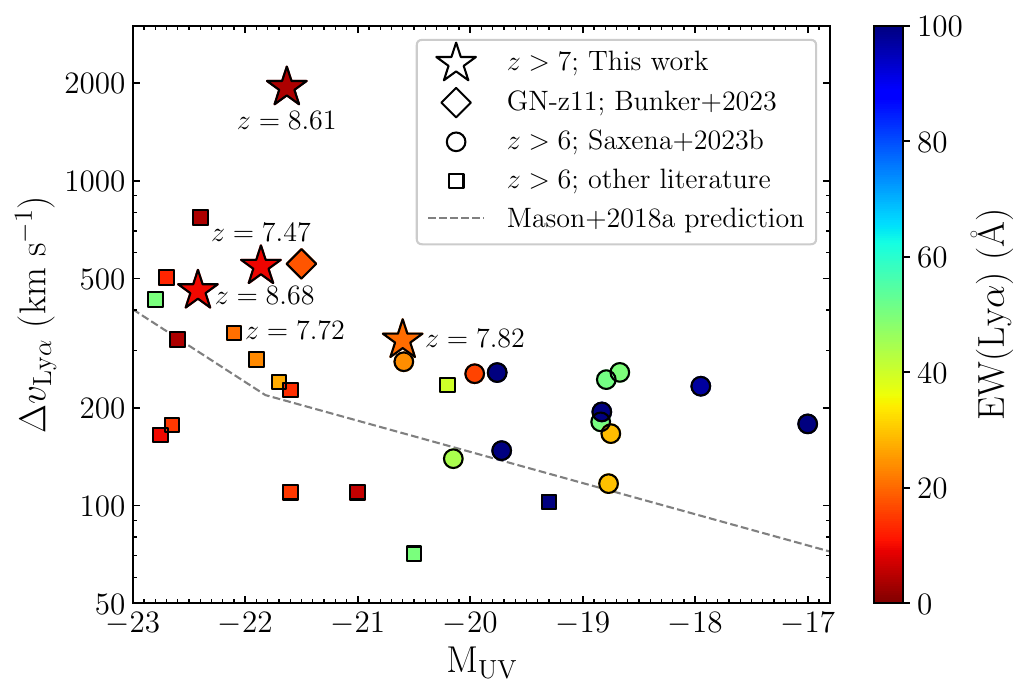}
\caption{Ly$\alpha$ velocity offset ($\Delta v_{\rm{Ly}\alpha}$) versus M$_{\rm{UV}}$ for galaxies at $z>6$, color coded by Ly$\alpha$ EW. Ly$\alpha$ emitting galaxies with CEERS NIRSpec observations are shown by stars. We overplot data of $z>6$ galaxies from literature (\citealt{Cuby2003,Pentericci2011,Pentericci2016,Pentericci2018,Vanzella2011,Willott2013,Willott2015,Maiolino2015,Oesch2015,Stark2015,Stark2017,Furusawa2016,Knudsen2016,Carniani2017,Laporte2017,Mainali2017,Hashimoto2019,Endsley2022b}; see \citealt{Endsley2022b} and the Table 4 therein) in squares, as well as {\it JWST} observations of GN-z11 ($z=10.603$; \citealt{Bunker2023}) in the diamond and Ly$\alpha$ emitters at $z>6.5$ from \citet{Saxena2023b} in circles. Prediction of the correlation between Ly$\alpha$ velocity offset and M$_{\rm{UV}}$ at $z=7$ from \citet{Mason2018a} is shown by grey dashed line. We mark the redshifts of $z>7.4$ Ly$\alpha$ emitting galaxies in the EGS field, including CEERS-1019 ($z=8.68$; \citealt{Zitrin2015}), CEERS-1029 ($z=8.61$; \citealt{Larson2022}), CEERS-1027 ($z=7.82$; this work), CEERS-698 ($z=7.47$; \citealt{Roberts-Borsani2016}), and EGS-zs8-1 ($z=7.72$; \citealt{Oesch2015,Stark2017}).}
\label{fig:lya_offset}
\end{center}
\end{figure}

CEERS spectroscopy provides medium resolution ($R\sim1000$) rest-frame optical emission line detections for 4 sources with Ly$\alpha$ emission, allowing us to estimate the Ly$\alpha$ velocity offsets. We calculate systemic redshifts ($z_{\rm{sys}}$) by simultaneously fitting the centroids of H$\beta$, [O~{\small III}]$\lambda4959$, and [O~{\small III}]$\lambda5007$ emission lines, and we calculate Ly$\alpha$ redshifts using the wavelength of the peak of Ly$\alpha$ emission in the NIRSpec spectra. The derived Ly$\alpha$ velocity offsets are on average large for these 4 objects, with $323$~km~s$^{-1}$ for CEERS-1027, $458$~km~s$^{-1}$ for CEERS-1019, $545$~km~s$^{-1}$ for CEERS-698, and $1938$~km~s$^{-1}$ for CEERS-1029. The Ly$\alpha$ velocity offsets of CEERS-1027, CEERS-1019, and CEERS-698 are consistent with those measured for $z>6$ galaxies with similar M$_{\rm{UV}}$ (Fig.~\ref{fig:lya_offset}), while the Ly$\alpha$ velocity offset of CEERS-1029 is much larger and we will discuss this object further in Section~\ref{sec:lae}. If the broad [O~{\small III}]$\lambda5007$ emission detected in CEERS-1027 and CEERS-698 is driven by outflows, the Ly$\alpha$ velocity offsets of these two galaxies are found to be similar to $2$ times of the outflow velocity measured from [O~{\small III}]$\lambda5007$ line width ($2v_{\rm{out}}=318$~km~s$^{-1}$ for CEERS-1027 and $=449$~km~s$^{-1}$ for CEERS-698, where $v_{\rm{out}}=\rm{FWHM}_{\rm{broad}}/2.355$). This is consistent with the predictions of the `shell-model' of Ly$\alpha$ radiation transfer for Ly$\alpha$ emission backscattering off an expanding shell \citep{Verhamme2006}. We note that due to the medium resolution of NIRSpec grating, the uncertainty of Ly$\alpha$ line center measurement and hence the velocity offset is large ($\Delta z_{\rm{Ly}\alpha}\simeq0.005$) for 3 sources except CEERS-1027, which has the brightest Ly$\alpha$ emission with the highest S/N ($=7$) among these 4 sources. 

To explore how the large Ly$\alpha$ velocity offsets of $z>7$ Ly$\alpha$ emitting galaxies impact the IGM transmission, we estimate the damping wing optical depth of Ly$\alpha$ as a function of velocity offset \citep{Miralda-Escude1998} using the methods described in \citet{Endsley2022b}. We consider these galaxies are in ionized regions, assuming the IGM inside the ionized region is completely ionized and outside is completely neutral. For CEERS-1027 ($\Delta v_{\rm{Ly}\alpha}=323$~km~s$^{-1}$), CEERS-1019 ($\Delta v_{\rm{Ly}\alpha}=458$~km~s$^{-1}$), and CEERS-698 ($\Delta v_{\rm{Ly}\alpha}=545$~km~s$^{-1}$), if they are in relatively large ionized regions ($R=1$~pMpc) the Ly$\alpha$ transmission through the neutral IGM can reach up to $50$~per~cent at their velocity offsets. Even in very small ionized regions ($R=0.1$~pMpc) their Ly$\alpha$ transmission is still significant ($\sim15-20$~per~cent) at velocity offset $\Delta v_{\rm{Ly}\alpha}\simeq300-500$~km~s$^{-1}$, and their Ly$\alpha$ visibility can still be boosted together with their enhanced ionizing photon (and hence Ly$\alpha$) production efficiencies (Fig.~\ref{fig:xi_ion}). For CEERS-1029 ($\Delta v_{\rm{Ly}\alpha}=1938$~km~s$^{-1}$)\footnote{We notice that the Ly$\alpha$ emission of CEERS-1029 shows two-component like feature in 2D spectrum (Fig.~\ref{fig:spectra_lya_2d}). These two components are $+1938$~km~s$^{-1}$ and $+2747$~km~s$^{-1}$ with respect to the systemic redshift, respectively. We use the peak of entire Ly$\alpha$ feature to compute the velocity offset relative to the $z_{\rm{sys}}$, which is $1938$~km~s$^{-1}$ (i.e., equivalent to the velocity offset of the first component).}, the Ly$\alpha$ transmission at its velocity offset is very large ($70$~per~cent) even in a small ionized region with $R=0.1$~pMpc. This will contribute significantly to its Ly$\alpha$ visibility. However, we note that although the IGM transmission can be significant at the measured Ly$\alpha$ velocity offsets of CEERS $z>7$ Ly$\alpha$ emitting galaxies, in Section~\ref{sec:discussion} we will discuss how the observed Ly$\alpha$ EWs may be lower than expected. This could imply that most of the Ly$\alpha$ flux was emitted at bluer wavelengths and attenuated by the IGM damping wing, thus only emission from the reddest part of the intrinsic Ly$\alpha$ profile is visible.

We also use the CEERS spectra to calculate Ly$\alpha$ escape fractions ($f_{\rm{esc,Ly}\alpha}$). We use the new H$\beta$ detections to estimate the Ly$\alpha$ flux expected in our spectra in absence of resonance scattering and dust absorption. For case B recombination and $T_{\rm{e}}=10^4$ K gas, the intrinsic Ly$\alpha$/H$\beta$ ratio is expected to be $24.9$ \citep{Osterbrock2006}. The Ly$\alpha$ emitters in our sample have observed ratios that range between $0.87$ and $8.44$. The inferred $f_{\rm{esc,Ly}\alpha}$ of the 6 galaxies with Ly$\alpha$ detections are $0.035-0.339$ with a median of $0.073$. The low implied Ly$\alpha$ escape fractions are consistent with expectations for typical UV-bright galaxies at $z\simeq2$ \citep[e.g.,][]{Hayes2011,Ciardullo2014,Matthee2016,Sobral2017} and also the Ly$\alpha$ emitting galaxies at $z\simeq6$ \citep{Ning2023}, but we will show in Section~\ref{sec:discussion} that they are significantly lower than is typically found in lower redshift galaxies with similar rest-frame optical line spectra.

After Ly$\alpha$, the next strongest emission lines in the rest-frame UV are fainter collisionally-excited features from highly-ionized metal species (i.e., C~{\small III}], C~{\small IV}, O~{\small III}]). These rest-frame UV metal lines are high energy transitions, requiring hard ionizing flux and large electron temperature (and hence metal poor gas) for significant excitation. They appear most prominently in UV spectra powered by very young stellar populations (under CSFH) that have yet to build up significant far-UV continuum luminosities from B stars \citep[e.g.,][]{Erb2010,Stark2014,Rigby2015,Senchyna2017,Berg2018,Mainali2020,Tang2021a}. Previous work has revealed a handful of intense C~{\small III}] and O~{\small III}] emitters in $z\gtrsim6$ galaxies \citep[e.g.,][]{Stark2015,Stark2017,Laporte2017,Mainali2017,Hutchison2019,Jiang2021,Topping2021}, with rest-frame EWs ($\simeq10-20$~\AA) that are more than an order of magnitude larger than what is common at $z\simeq1-3$ \citep[e.g.,][]{Shapley2003,Steidel2016,Du2017,Maseda2017}. 

One of the most intense C~{\small III}] emitters at $z\gtrsim7$ is EGS-zs8-1 (EW $=22$~\AA; \citealt{Stark2017}), the brightest galaxy in the $z=7.7$ Ly$\alpha$ association in the EGS field. While this galaxy was not targeted with CEERS spectroscopy, the dataset includes observations of many similarly intense optical line emitters, allowing us to determine if such strong C~{\small III}] is common at $z\gtrsim7$. We detect C~{\small III}] emission (S/N $=2-4$) in 3 galaxies in the NIRSpec observations (CEERS-1019, CEERS-1149, CEERS-1027; UV line detections are shown in Fig.~\ref{fig:spectra_uv_2d} in Appendix~\ref{sec:appendix}). Each of these systems has UV continuum detections in their spectra, allowing us to compute C~{\small III}] EWs. The values all point to very strong line emission ($10.9-16.0$~\AA; Table~\ref{tab:uv_line}) seen in young metal poor galaxies. In sources lacking C~{\small III}], we can place $3\sigma$ upper limits on the EW, with typical values of $6$~\AA\ for bright ($H=25$) galaxies and $24$~\AA\ for fainter ($H=26.5$) galaxies in our sample. 

\begin{table*}
\centering
\begin{tabular}{cccccc}
\hline
ID & $z_{\rm{spec}}$ & EW(C~{\scriptsize IV}$\lambda1549$) & EW(He~{\scriptsize II}) & EW(O~{\scriptsize III}]$\lambda1661$) & EW(C~{\scriptsize III}]$\lambda1908$) \\
 & & (\AA) & (\AA) & (\AA) & (\AA) \\
\hline
CEERS-1019 & $8.678$ & ... & ... & $8.0\pm4.9$ & $10.9\pm6.0$ \\
CEERS-1149 & $8.175$ & ... & ... & ... & $16.0\pm4.4$ \\
CEERS-1027 & $7.819$ & $5.7\pm2.1$ & $3.9\pm1.6$ & ... & $12.9\pm4.5$ \\
\hline
\end{tabular}
\caption{Equivalent widths of rest-frame UV emission lines detected in CEERS NIRSpec $z>7$ galaxies.}
\label{tab:uv_line}
\end{table*}

\begin{figure}
\begin{center}
\includegraphics[width=\linewidth]{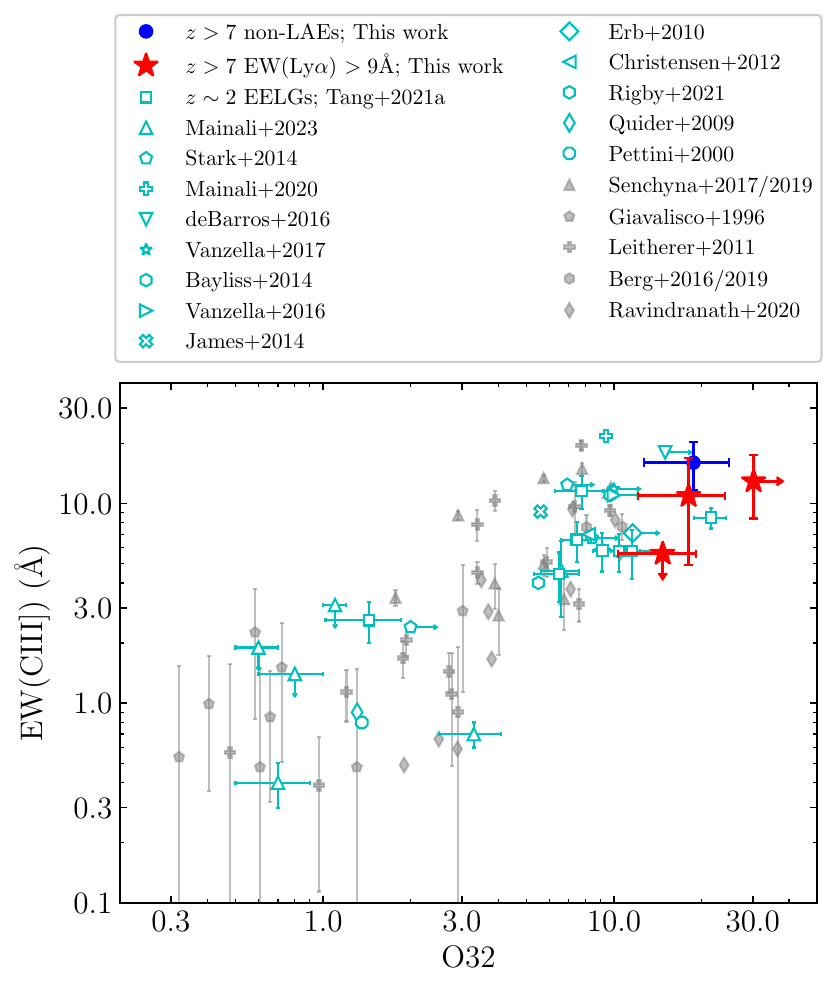}
\caption{C~{\scriptsize III}] EW as a function of O32 for CEERS NIRSpec galaxies at $z>7$. LAEs (EW$_{\rm{Ly}\alpha}>9$~\AA) and non-LAEs at $z>7$ are shown by red stars and blue circles, respectively. As a comparison, we overplot literature data of galaxies at $z\sim1-3$ (open cyan symbols; \citealt{Pettini2000,Quider2009,Erb2010,Christensen2012,Bayliss2014,James2014,Stark2014,deBarros2016,Vanzella2016,Vanzella2017,Mainali2020,Rigby2021,Tang2021a,Mainali2023}) and $z\simeq0$ galaxies (solid grey symbols; \citealt{Giavalisco1996,Leitherer2011,Berg2016,Berg2019a,Senchyna2017,Senchyna2019,Ravindranath2020}).}
\label{fig:c3ew_o32}
\end{center}
\end{figure}

The three galaxies in CEERS with intense C~{\small III}] emission also have the largest O32 ratios in the sample (Fig.~\ref{fig:c3ew_o32}), suggesting a connection between the UV line EWs and the ionization conditions of the interstellar medium (ISM). The connection between C~{\small III}] EW and O32 exists in data from $z\simeq0-3$ \citep{Mainali2023}. The same factors which lead to large C~{\small III}] EW (young stellar populations, metal poor gas) also lead to large O32. The hard ionizing spectra produced by young stars increases the O32 ratio. The electron temperature is larger in metal-poor gas, which results in higher O$^{++}$ to O$^{+}$ ratios as O$^{++}$ is the most efficient coolant \citep[e.g.,][]{Kewley2019}. Notably, the three LAEs with Ly$\alpha$ EW $>9$~\AA\ are situated in the upper right of the C~{\small III}] EW versus O32 diagram, again underlying their extreme nature relative to the galaxy population seen at $z\simeq7-9$. We verify that the \textsc{beagle} model fits described in Section~\ref{sec:opt_lines} can reproduce the C~{\small III}] EWs. When we add C~{\small III}] to the optical lines in our \textsc{beagle} model fits, we find that the inferred metallicities remain similarly low, whereas the ionization parameters increase modestly (median change of $0.3$ dex). 

\begin{figure*}
\begin{center}
\includegraphics[width=\linewidth]{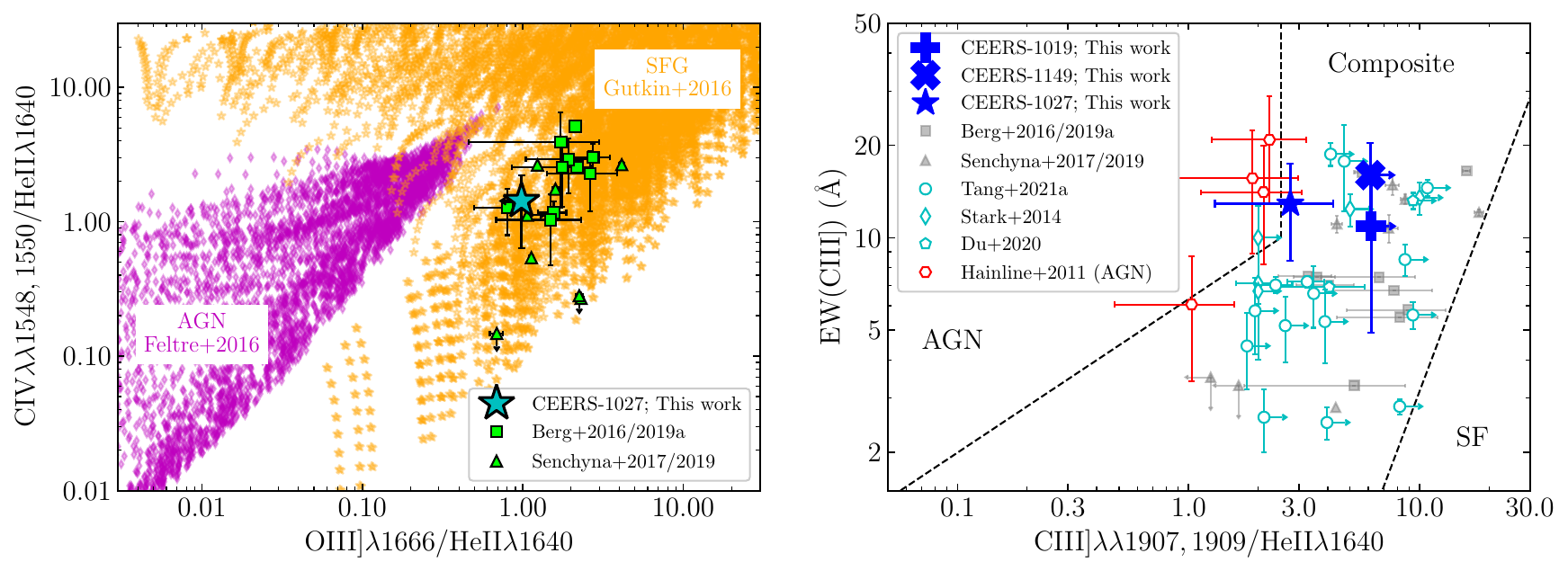}
\caption{UV diagnostics and the position of $z\gtrsim7$ CEERS galaxies. The left panel shows the O~{\scriptsize III}]$\lambda1666$/He~{\scriptsize II}$\lambda1640$ versus C~{\scriptsize IV}$\lambda\lambda1548,1550$/He~{\scriptsize II}$\lambda1640$ diagnostic diagram and the position of CEERS-1027 (the cyan star). We overplot the line ratios predicted from photoionization models for AGN (magenta squares; \citealt{Feltre2016,Mignoli2019}) and star-forming galaxy (SFG; orange stars; \citealt{Gutkin2016}), and the line ratios of extreme metal-poor galaxies at low redshift (green symbols; \citealt{Berg2016,Berg2019a,Senchyna2017,Senchyna2019}). The UV line ratios of CEERS-1027 are consistent with local metal-poor star-forming galaxies and fall in the region where massive stars are likely to be the dominant ionizing source. The right panel shows the C~{\scriptsize III}]$\lambda\lambda1907,1909$/He~{\scriptsize II}$\lambda1640$ versus C~{\scriptsize III}] EW diagram. The black dashed lines separate the regions dominated by AGN (left), composite of AGN and star formation (middle), and star formation (right) predicted by nebular emission models calculated in \citet{Hirschmann2019}. Models are calculated using prescriptions for star-forming galaxies \citep{Gutkin2016}, AGN \citep{Feltre2016}, and quiescent galaxies \citep{Hirschmann2017}. The three galaxies at $z\gtrsim7$ with C~{\scriptsize III}] detections (CEERS-1019, CEERS-1149, CEERS-1027) fall in the composite region. We overplot the diagnostics of star-forming galaxies at $z\sim2$ (open cyan symbols; \citealt{Stark2014,Du2020,Tang2021a}), local star-forming galaxies (grey symbols; \citealt{Berg2016,Berg2019a,Senchyna2017,Senchyna2019}), and $z\sim2-3$ AGN (open red hexagon) from \citet{Hainline2011}.}
\label{fig:uv_diagnostics}
\end{center}
\end{figure*}

One galaxy in our sample (CEERS-1027 at $z=7.819$) additionally shows faint detections of C~{\small IV} (EW $=5.7$~\AA) and He~{\small II} (EW $=3.9$~\AA) emission (Table~\ref{tab:uv_line}; see Fig.~\ref{fig:spectra_uv_2d} in Appendix~\ref{sec:appendix} for spectra). The C~{\small IV} and He~{\small II} line widths are narrow ($\sigma=65$~km~s$^{-1}$) and consistent with the widths of the other emission lines, so we interpret these as nebular in nature (i.e., not broad stellar features). Powering both lines requires a hard ionizing spectrum, with significant ionizing flux in excess of $48$~eV (C~{\small IV}) and $54$~eV (He~{\small II}). At lower redshifts it has been shown that both are often associated with metal poor galaxies \citep[e.g.,][]{Stark2014,Berg2016,Berg2018,Berg2019a,Senchyna2017,Senchyna2019,Senchyna2022,Tang2021a,Saxena2022,Schaerer2022a}, although the precise origin of the line emission may require contributions from additional ionizing sources (e.g., X-ray binaries, or AGN; e.g., \citealt{Hainline2011,Volonteri2017,Berg2019b,Plat2019,Saxena2020a,Saxena2020b,Berg2021,Olivier2022}). The optical lines indicate that CEERS-1027 is extremely metal poor ($12+\log{(\rm{O/H})}=7.58$), consistent with the values that are usually associated with C~{\small IV} and He~{\small II} emitters. To investigate the origin of the line emission in CEERS-1027, we compare the UV line ratios to expectations from photoionization models in the left panel of Fig.~\ref{fig:uv_diagnostics}. The C~{\small IV}/He~{\small II} ratio ($=1.4$) versus O~{\small III}]$\lambda1666$/He~{\small II} ratio ($<1.0$) is consistent with metal poor galaxies seen locally. The line ratios fall in the region of the diagram where massive stars are likely to be the dominant ionizing source, although the data do not rule out mixtures of AGN and massive stars, or contributions of energetic photons from X-ray binaries or intermediate mass black holes.

For the other two C~{\small III}] emitting galaxies in our sample (CEERS-1019 at $z=8.678$ and CEERS-1149 at $z=8.175$), we can investigate their ionizing sources in the C~{\small III}] EW - C~{\small III}]/He~{\small II} diagram \citep[e.g.,][]{Nakajima2018,Hirschmann2019} together with CEERS-1027. In the right panel of Fig.~\ref{fig:uv_diagnostics}, we show the C~{\small III}] EW versus C~{\small III}]/He~{\small II} of the three CEERS galaxies with intense C~{\small III}] emission. Based on the photoionization models studied in \citet{Hirschmann2019}, these three galaxies fall in the region where the line emission is likely dominated by a combination of massive stars and AGN. CEERS-1019 and CEERS-1149 present similar C~{\small III}] EWs and C~{\small III}]/He~{\small II} ratios as star-forming galaxies at $z\simeq0-3$ \citep[e.g.,][]{Stark2014,Berg2016,Berg2019a,Senchyna2017,Senchyna2019,Du2020,Tang2021a}, but again the data do not rule out a significant contribution of ionizing photons from AGN (e.g., for CEERS-1019; \citealt{Larson2023}).


\section{Reionization-Era Lyman-alpha Emitters} \label{sec:lae}

\begin{figure*}
\begin{center}
\includegraphics[width=0.33\linewidth]{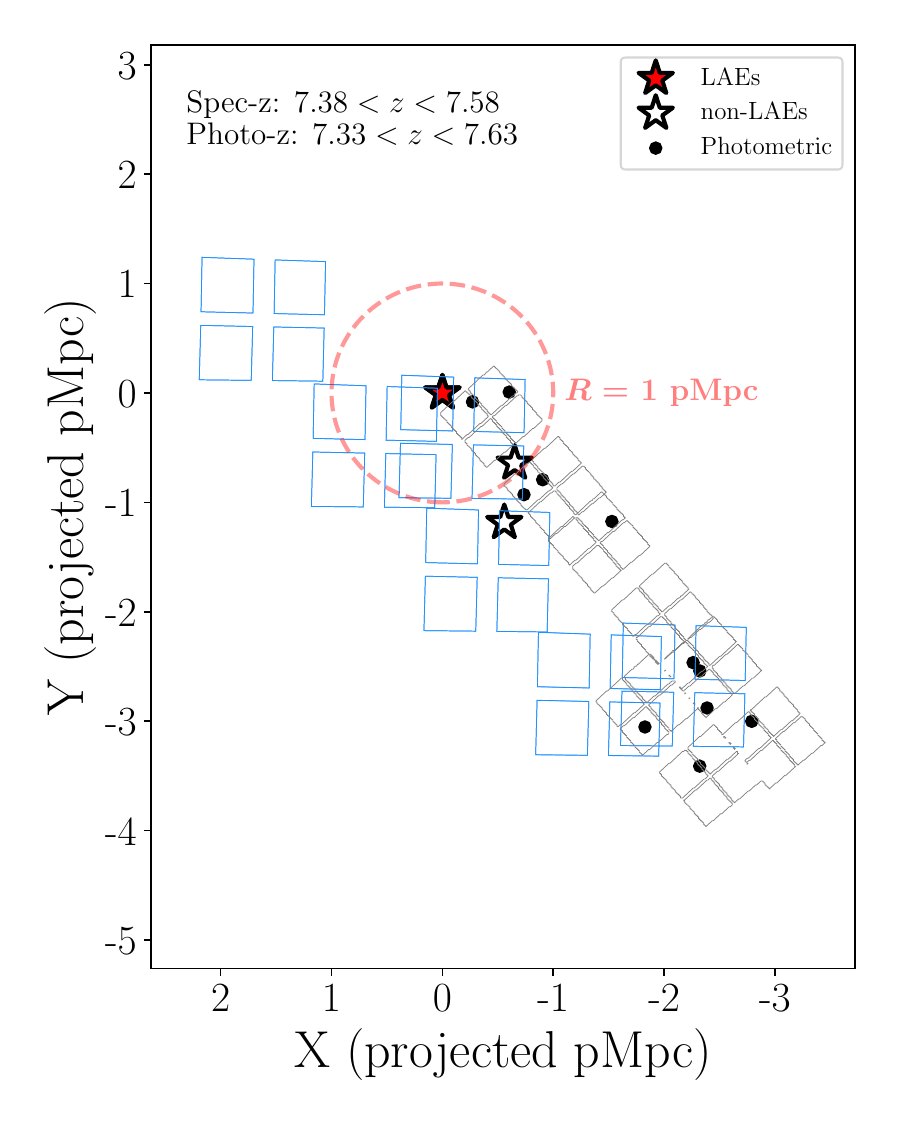}
\includegraphics[width=0.33\linewidth]{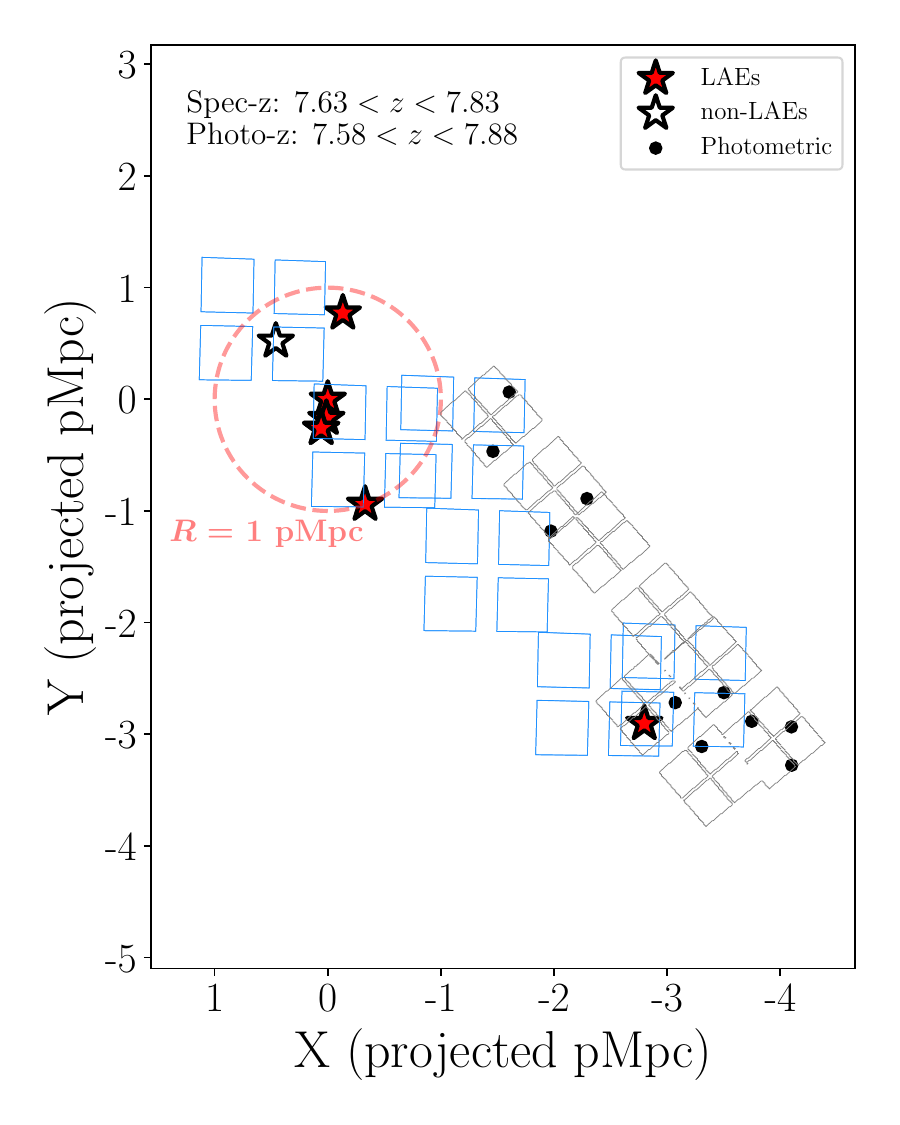}
\includegraphics[width=0.33\linewidth]{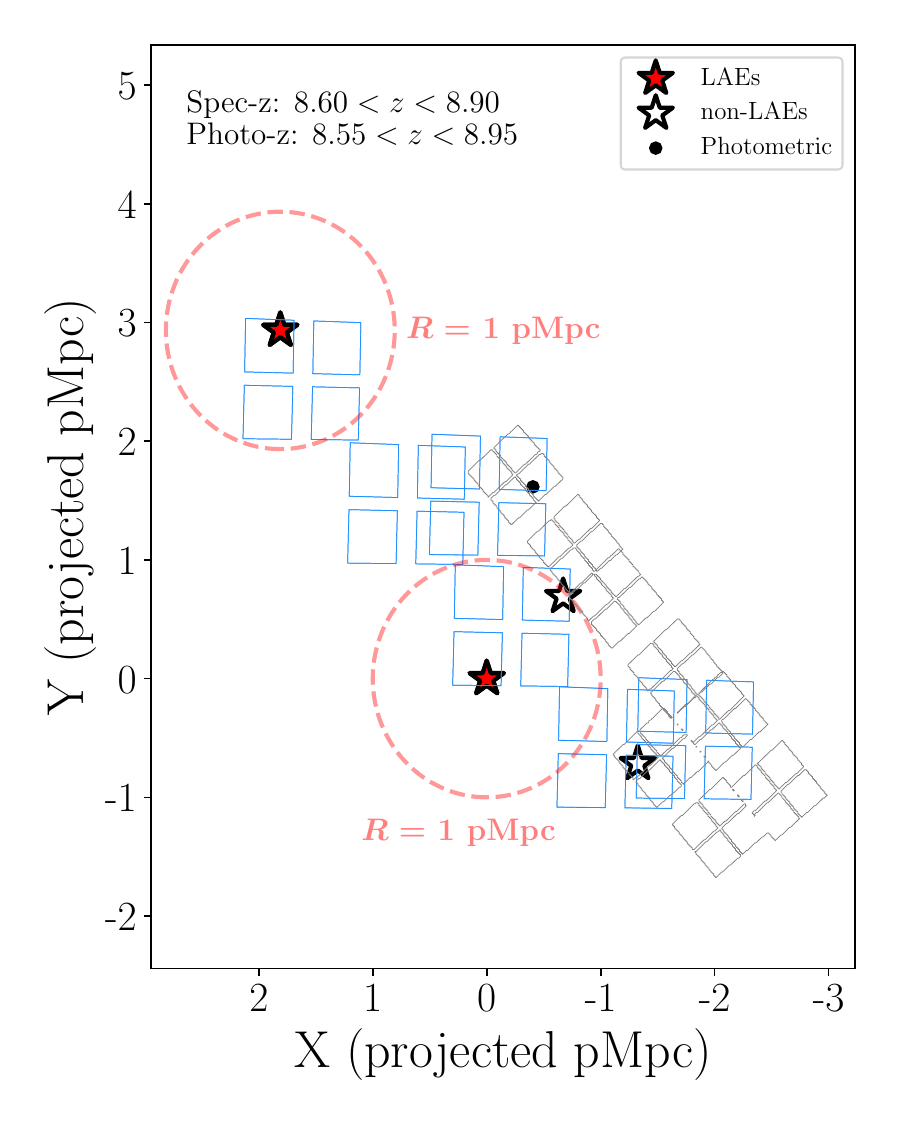}
\caption{Spatial distribution of spectroscopically confirmed galaxies at $z>7$ that are in the three groups of Ly$\alpha$ emitters in the EGS field ($7.38<z<7.58$, left panel; $7.63<z<7.83$, middle panel; $8.60<z<8.90$, right panel). The $1$~pMpc radius centered at the brightest Ly$\alpha$ emitter at each redshift bin is shown by red dashed circle. A $R=1$~pMpc ionized region at $z\simeq7$ allows a significant transmission of Ly$\alpha$ ($\sim30$~per~cent) at the systemic redshift and a stronger Ly$\alpha$ transmission ($\sim50$~per~cent) at Ly$\alpha$ velocity offset $=400$~km~s$^{-1}$ \citep[e.g.,][]{Mason2020,Endsley2022b,Qin2022}. Spectroscopic redshifts are from CEERS NIRSpec observations (this work, with footprint shown in blue) and literature \citep{Oesch2015,Zitrin2015,Roberts-Borsani2016,Tilvi2020,Jung2022,Larson2022}. Ly$\alpha$ emitters and non-Ly$\alpha$ emitters with spectroscopic redshift measurements are plotted by solid red stars and open stars, respectively. We overplot the photometric redshift $z>7$ sources in the three associations from \citet{Endsley2023a} and \citet{Whitler2023a} as black circles, which are selected based on the CEERS NIRCam imaging observed in June, 2022 (footprint shown in gray).}
\label{fig:distribution}
\end{center}
\end{figure*}

The EGS field contains three groups of Ly$\alpha$ emitting galaxies at $z>7$ (see Fig.~\ref{fig:distribution}). The visibility of Ly$\alpha$ in these associations has led to the suggestion that they trace rare ionized bubbles in the mostly neutral IGM. However the Ly$\alpha$ detections may alternatively reflect properties internal to the galaxies (i.e., efficient Ly$\alpha$ production). Prior to {\it JWST}, it was challenging to study these galaxies in any detail. Since redshift confirmation was mostly limited to those systems with strong Ly$\alpha$ emission, it was impossible to identify the extent to which Ly$\alpha$ visibility was enhanced in surrounding galaxies. In this section, we describe what CEERS spectroscopy has revealed about the galaxies with and without Ly$\alpha$ in the three associations.

\subsection{$z=7.5$ Association of Ly$\alpha$ Emitters} \label{sec:z7p5}

The CEERS dataset includes observations of CEERS-698, the brightest galaxy ($H_{160}=25.2$, M$_{\rm{UV}}=-21.9$) in the $z\simeq7.5$ association and two additional galaxies (CEERS-689 and CEERS-1163) with redshifts that place them within $3.1$~pMpc of CEERS-698. The latter two systems do not show Ly$\alpha$ emission. Below we summarize the properties of each of these galaxies.

{\bf CEERS-698.} CEERS-698 was first identified in \citet{Roberts-Borsani2016} as a UV-bright dropout (EGS-zs8-2) with an IRAC $[4.5]$ excess suggesting large [O~{\small III}]+H$\beta$ EW ($\simeq2564$~\AA; Table~\ref{tab:sed_fitting}), a young light-weighted age ($3.8$~Myr), and very efficient ionizing photon production ($\log{[\xi_{\rm{ion}}/\rm{erg}^{-1}\ \rm{Hz}]}=25.8$). The source was spectroscopically confirmed with Keck/MOSFIRE via detection of Ly$\alpha$ (EW $=9.3$~\AA) at $z_{\rm{Ly\alpha}}=7.477$ in \citet{Stark2017}, following tentative identification in \citet{Roberts-Borsani2016}. CEERS/NIRSpec observations detect a suite of rest-frame optical lines in CEERS-698, revealing line ratios that reflect the extreme ionizing nature of the SED. The large O32 ($14.72$) and Ne3O2 ($0.87$) ratios are rarely seen in normal star-forming galaxies at lower redshifts, pointing to gas with a very high ionization parameter ($\log{U}=-1.85$). Photoionization models suggest metal poor gas ($12+\log{(\rm{O/H})}=7.81$), consistent with the direct value implied by the tentative detection of [O~{\small III}]$\lambda4363$ (see Table~\ref{tab:photoionization}).

Ly$\alpha$ is detected with NIRSpec with similar flux and EW ($9.5$~\AA) to that measured previously with Keck. Using the systemic redshift derived from the rest-frame optical lines ($z_{\rm{sys}}=7.470$), we are able to compute a Ly$\alpha$ velocity offset of $545\pm184$~km~s$^{-1}$ (Fig.~\ref{fig:lya_offset}). The NIRSpec detections of Ly$\alpha$ and H$\beta$ indicate that the transmission of Ly$\alpha$ is very low ($0.045\pm0.015$), suggesting the majority of Ly$\alpha$ is scattered out of the NIRSpec micro-shutter. While such low Ly$\alpha$ escape fractions are common in $z\simeq2$ galaxies \citep[e.g.,][]{Hayes2011,Ciardullo2014,Matthee2016,Sobral2017}, we will show in Section~\ref{sec:discussion} that lower redshift galaxies with properties similar to CEERS-698 often transmit a much larger fraction of their Ly$\alpha$.

{\bf CEERS-689.} This is a multi-component galaxy with integrated H-band magnitude ($H_{\rm{160}}=24.9$, M$_{\rm{UV}}=-22.1$). The NIRSpec micro-shutter covers one of the clumps with $H_{\rm{160}}=26.2$. The strong rest-frame optical lines are detected ([O~{\small II}], [Ne~{\small III}], H$\gamma$, H$\beta$, [O~{\small III}]) indicating a systemic redshift of $z_{\rm{sys}}=7.545$. This places this galaxy $3.1$~pMpc away from CEERS-698. The ionization-sensitive rest-frame optical line ratios (O32 $=10.13$ and Ne3O2 $=0.71$) point to a high ionization parameter, which is also recovered in the photoionization model fits ($\log{U}=-2.02$). The gas-phase metallicity implied by the photoionization model fits is low ($12+\log{(\rm{O/H})}=7.84$), consistent with the direct method value suggested by the tentative detection of [O~{\small III}]$\lambda4363$ (see Table~\ref{tab:photoionization}). In spite of the intense ionizing nature of this source, we detect no Ly$\alpha$ with NIRSpec, suggesting relatively weak Ly$\alpha$ (EW $<26.1$~\AA) and correspondingly low transmission of Ly$\alpha$ through the micro-shutter ($<0.055$), with both corresponding to $3\sigma$ limits.

We note that this system was recently discussed in \citet{Jung2022} with an ID of z8\_32350. A faint emission feature (S/N $=4.6$) in Keck spectroscopy yielded tentative Ly$\alpha$ confirmation of this galaxy at a slightly higher redshift ($z=7.7759$). Since there are multiple clumps in this system, we have confirmed that the NIRSpec micro-shutter centroid is very close ($0.06$ arcsec) to the coordinates reported in \citet{Jung2022}. Given the confidence in the NIRSpec redshift (from numerous rest-frame optical lines), there is a disagreement with the redshift reported in \citet{Jung2022}.

{\bf CEERS-1163.} This fainter galaxy ($H_{160}=26.8$, M$_{\rm{UV}}=-20.2$) is detected in [O~{\small III}] and H$\beta$, indicating a systemic redshift of $z=7.448$ that places this system $1.2$~pMpc away from CEERS-698 and $3.6$~pMpc from CEERS-689. No Ly$\alpha$ is seen in the NIRSpec spectrum. Comparing the H$\beta$ flux and the Ly$\alpha$ upper limit, we find a $3\sigma$ upper limit on the Ly$\alpha$ escape fraction of $<0.145$. Deeper spectroscopy is required to provide a more stringent constraint on whether this faint galaxy may have enhanced Ly$\alpha$ emission.

\subsection{$z=7.7$ Association of Ly$\alpha$ Emitters} \label{sec:z7p7}

EGS-zs8-1 was the first galaxy identified in the $z=7.7$ association of Ly$\alpha$ emitters in the EGS. The source was spectroscopically confirmed at $z_{\rm{Ly\alpha}}=7.730$ by \citet{Oesch2015} following photometric selection as a bright ($H_{\rm{160}}=25.0$) galaxy with an IRAC excess in \citet{Roberts-Borsani2016}. Detection of the [C~{\small III}], C~{\small III}] doublet \citep{Stark2017} enabled confirmation of the Ly$\alpha$ velocity offset ($340$~km~s$^{-1}$). Subsequent work by \citet{Tilvi2020} and \citet{Jung2022} has identified 4 more galaxies at $z=7.63-7.83$ with confident ($>5\sigma$) Ly$\alpha$ detections. The CEERS observations provide rest-frame UV to optical spectra of CEERS-686, one of the Ly$\alpha$ emitters from \citet{Jung2022} (z8\_13573). CEERS spectra also present two new confirmations of galaxies at similar redshifts ($z=7.82$ and $z=7.76$), one of which shows Ly$\alpha$ (CEERS-1027) and one of which does not (CEERS-1023). We summarize the properties of these three galaxies below. 

{\bf CEERS-686} This is one of the fainter galaxies in the CEERS $z\gtrsim7$ spectroscopic sample ($H_{160}=26.4$, M$_{\rm{UV}}=-20.7$). \citet{Jung2022} present Keck/MOSFIRE detection of strong Ly$\alpha$ emission at $z_{\rm{Ly\alpha}}=7.7482$, revealing one of the largest Ly$\alpha$ EWs ($69.1^{+29.8}_{-19.9}$~\AA) known at $z>7$. The CEERS prism spectrum detects the [O~{\small III}] doublet and H$\beta$, confirming a systemic redshift of $z_{\rm{sys}}=7.74$ and placing this galaxy only $0.6$~pMpc away from EGS-zs8-1. We also detect Ly$\alpha$ (and underlying UV continuum) in the CEERS spectrum, with line flux similar to that measured with MOSFIRE. The implied Ly$\alpha$ EW ($41.9\pm1.6$\AA) is somewhat lower than the Keck measurement, but still consistent within the $1\sigma$ uncertainties. The Ly$\alpha$/H$\beta$ ratio in the NIRSpec spectrum indicates that a large fraction of the Ly$\alpha$ emission is transmitted through the NIRSpec micro-shutter ($0.325\pm0.028$). This confirms that the large Ly$\alpha$ EW is not simply due to efficient ionizing photon production, but the transmission is also enhanced. Such strong Ly$\alpha$ emitters tend to have a large fraction of their Ly$\alpha$ profile escaping near the systemic redshift \citep[e.g.,][]{Hashimoto2013,Erb2014,Shibuya2014,Tang2021b}, but owing to the low resolution of the prism spectrum, we are unable to extract a meaningful Ly$\alpha$ velocity offset for this galaxy.

{\bf CEERS-1027.} This moderately faint ($H_{150}=26.5$, M$_{\rm{UV}}=-20.6$) galaxy is covered in the NIRCam footprint, with an SED that indicates a young stellar population ($3.4$~Myr), efficient ionizing photon production ($\log{[\xi_{\rm{ion}}/\rm{erg}^{-1}\ \rm{Hz}]}=25.8$), and very strong nebular emission ([O~{\small III}]+H$\beta$ EW $=3276$~\AA). The NIRSpec spectrum reveals numerous emission lines, confirming the systemic redshift ($z_{\rm{sys}}=7.819$) and placing this galaxy $5.2$~pMpc from EGS-zs8-1. The line ratios of CEERS-1027 are the most extreme in the sample, with both O32 ($>30.22$) and Ne3O2 ($>1.80$) pointing to gas with very extreme ionization conditions. The \textsc{beagle} models reproduce the lines with very low metallicity ($12+\log{(\rm{O/H})}=7.58$) and high ionization parameter ($\log{U}=-1.31$). Tentative detection of [O~{\small III}]$\lambda4363$ reveals a similarly low metallicity ($12+\log{(\rm{O/H})}=7.74$) using the direct method. The UV spectrum reveals reasonably strong Ly$\alpha$ (EW $=20.4$~\AA) and lower S/N detections of C~{\small IV} (EW $=5.7$~\AA), He~{\small II} (EW $=3.9$~\AA), and C~{\small III}] (EW $=12.9$~\AA). As discussed in Section~\ref{sec:uv_lines}, the UV line ratios are consistent with powering by low metallicity stars.The Ly$\alpha$ emerges with a velocity offset of $323\pm18$~km~s$^{-1}$ with respect to the systemic redshift (Fig.~\ref{fig:lya_offset}). In spite of the large Ly$\alpha$ EW, we infer a relatively low escape fraction for Ly$\alpha$ ($0.085\pm0.018$). suggesting the vast majority of the line emission is not making its way through the NIRSpec micro-shutter.

{\bf CEERS-1023.} This galaxy is reasonably faint ($H_{160}=26.2$, M$_{\rm{UV}}=-20.9$) and red ($\beta=-0.9$), suggesting significant attenuation. We detect [O~{\small III}], H$\beta$, and [O~{\small II}] with NIRSpec, revealing a systemic redshift ($z_{\rm{sys}}=7.776$) that places this galaxy $2.0$~pMpc away from EGS-zs8-1. The observed value of O32 ($2.80$) is among the lowest in the sample, suggesting less extreme ionization conditions. If we were to apply a reddening correction consistent with the UV slope, the O32 ratio would be even lower (i.e., O32 $=2.31$ for $E(B-V)=0.16$). We do not detect Ly$\alpha$, with the $3\sigma$ upper limit implying a Ly$\alpha$ escape fraction of $<0.135$. While deeper spectroscopy will help put more robust constraints on the presence of Ly$\alpha$, the non-detection is not surprising given the very red colors seen in {\it HST}.

\subsection{$z=8.7$ Association of Ly$\alpha$ Emitters} \label{sec:z8p7}

The EGS contains two very bright galaxies with Ly$\alpha$ emission at $z\simeq8.7$ \citep{Zitrin2015,Larson2022}. These are among the highest redshift galaxies with robust Ly$\alpha$ detections. {\it HST} WFC3/IR imaging suggests that there may be an overdensity in the region \citep{Leonova2022}, potentially contributing the formation of an 
early ionized structure that is aiding in the visibility of Ly$\alpha$. The CEERS observations provide spectra of 
both $z\simeq8.7$ Ly$\alpha$ emitters (CEERS-1019 and CEERS-1029), while also confirming two additional galaxies (CEERS-23, CEERS-1025) at the same redshift. We describe the properties of the four galaxies below.

{\bf CEERS-1019.} This bright ($H_{150}=24.9$, M$_{\rm{UV}}=-22.4$) galaxy was identified in \citet{Roberts-Borsani2016} and spectroscopically confirmed with Ly$\alpha$ emission at $z_{\rm{Ly\alpha}}=8.683$ in \citet{Zitrin2015}. NIRCam imaging reveals it is a multi-component galaxy with three separate clumps, as is common for bright galaxies at these redshifts \citep[e.g.,][]{Chen2023}. The CEERS NIRSpec observations reveal numerous strong rest-frame optical lines, confirming a systemic redshift of $z_{\rm{sys}}=8.678$. The NIRCam SED indicates the galaxy is dominated by a very young stellar population ($4.9$~Myr) with powerful nebular emission ([O~{\small III}]+H$\beta$ EW $=2599$~\AA) and extremely efficient ionizing photon production ($\log{[\xi_{\rm{ion}}/\rm{erg}^{-1}\ \rm{Hz}]}=26.0$), though AGN can also contribute \citep{Larson2023}. A $4.6\sigma$ detection of N~{\small V}$\lambda1243$ has been reported in \citet{Mainali2018}, suggesting a hard ionizing spectrum which might be powered by AGN or fast radiative shocks. In NIRSpec spectra neither N~{\small V}$\lambda1238$ or N~{\small V}$\lambda1243$ is detected with $3\sigma$ upper limit of EW of $<4.8$~\AA, though the NIRSpec upper limit does not rule out the detection of N~{\small V}$\lambda1243$. Both O32 ($18.06$) and Ne3O2 ($1.08$) point to very high ionization gas. To reproduce the line ratios, the photoionization models require both low metallicity ($12+\log{(\rm{O/H})}=7.77$) and high ionization parameter ($\log{U}=-1.74$). The direct method gives a similar value for the metallicity ($12+\log{(\rm{O/H})}=7.73$). The UV shows low S/N detections of O~{\small III}]$\lambda1661$ (EW $=8.0\pm4.9$~\AA) and CIII] (EW $=10.9\pm6.0$~\AA). Ly$\alpha$ is seen in the NIRSpec data with a large velocity offset to the systemic redshift ($\Delta v_{\rm{Ly}\alpha}=458$~km~s$^{-1}$), but with moderately lower EW ($9.8\pm2.4$~\AA) than measured in \citet{Zitrin2015}. As discussed in Section~\ref{sec:opt_lines}, we adopt the NIRSpec value in our analysis owing to the impact of a sky line on the MOSFIRE spectrum. The escape fraction of CEERS-1019 is low ($f_{\rm{esc,Ly}\alpha}=0.035\pm0.010$), suggesting the majority of Ly$\alpha$ is scattered out of the NIRSpec micro-shutter.

{\bf CEERS-1029.} This bright galaxy ($H_{160}=25.6$, M$_{\rm{UV}}=-21.6$) was confirmed at $z_{\rm{Ly\alpha}}=8.665$ via detection of Ly$\alpha$ emission in \citet{Larson2022}. The Ly$\alpha$ EW ($4.7$~\AA) is very low, likely signalling that this galaxy transmits a very small fraction of its line emission. The NIRSpec spectrum confirms the [O~{\small III}] doublet, H$\beta$, and [O~{\small II}], revealing a systemic redshift of $z_{\rm{sys}}=8.610$ with a distance of $3.9$~pMpc away from CEERS-1019. This O32 ratio in CEERS-1029 ($4.81$) is not as large as most galaxies in our sample. The \textsc{beagle} models suggest that this galaxy is more metal-rich ($12+\log{(\rm{O/H})}=8.01$) than the other Ly$\alpha$ emitters, with a slightly lower ionization parameter ($\log{U}=-2.33$). The NIRSpec spectrum reveals a tentative line emission at observed frame $11758$~\AA, which recovers the Ly$\alpha$ emission detected in \citet{Larson2022} with nearly identical EW ($4.2\pm1.3$~\AA). The transmission of Ly$\alpha$ implied by the Ly$\alpha$/H$\beta$ ratio is low ($f_{\rm{esc,Ly}\alpha}=0.061\pm0.034$), as expected given the very low Ly$\alpha$ EW. The velocity offset between Ly$\alpha$ and the rest-frame optical lines is extremely large ($1938\pm162$~km~s$^{-1}$; Fig.~\ref{fig:lya_offset}). How Ly$\alpha$ is shifted to such large velocities is not entirely clear. Recent work has demonstrated that Ly$\alpha$ is very broad in similarly luminous reionization-era galaxies, with hints of multiple redshifted components \citep{Endsley2022b}. Based on comparison to [C~{\small II}] line profiles, these authors have speculated that such line profiles could arise if the different gas-rich clumps have significant proper motions with respect to one another. Along these lines, it is conceivable that the Ly$\alpha$ seen in CEERS-1029 could be backscattered off of a clump with significant proper motion. But regardless of the precise origin, the large velocity offset of the Ly$\alpha$ that escapes in CEERS-1029 will certainly facilitate its transmission through the galaxy and IGM, allowing $\sim70$~per~cent of the Ly$\alpha$ at this velocity offset to transmit through the neutral IGM even in very small ionized region ($R=0.1$~pMpc) as discussed in Section~\ref{sec:uv_lines}. 

{\bf CEERS-23.} This galaxy is very faint ($H_{150}=28.4$, M$_{\rm{UV}}=-18.9$) and is detected in [O~{\small III}] and H$\beta$. The systemic redshift measured from rest-frame optical emission lines is $z=8.881$, placing this galaxy $5.5$~pMpc away from CEERS-1019 and $8.6$~pMpc away from CEERS-1029. The NIRCam SED indicates a moderately large [O~{\small III}]+H$\beta$ EW ($1175$~\AA) and young stellar populations ($12$~Myr). We do not detect Ly$\alpha$ emission in the NIRSpec spectrum. By comparing its Ly$\alpha$ flux upper limit to H$\beta$ flux we put a $3\sigma$ upper limit on the Ly$\alpha$ escape fraction of $<0.422$. Thus deeper spectroscopy is required to constrain whether this very faint source has Ly$\alpha$ emission.

{\bf CEERS-1025.} This moderately bright galaxy ($H_{150}=26.2$, M$_{\rm{UV}}=-21.1$) is covered by NIRCam observations, with an SED indicating very young stellar populations ($1.6$~Myr) and very efficient ionizing photon production ($\log{[\xi_{\rm{ion}}/\rm{erg}^{-1}\ \rm{Hz}]}=25.8$). NIRSpec detections of rest-frame optical lines confirm this source at $z_{\rm{sys}}=8.715$, with a distance of $1.4$~pMpc from CEERS-1019 and $3.9$~pMpc from CEERS-1029. Line ratios reveal moderately large O32 ($8.12$), with photoionization models suggesting a combination of metal poor gas ($12+\log{(\rm{O/H})}=7.75$) and a reasonably high ionization parameter ($\log{U}=-2.13$). The NIRSpec spectrum does not extend blue enough to provide constraints on Ly$\alpha$ in this galaxy, so we cannot place an upper limit on the escape fraction of Ly$\alpha$.


\section{Discussion} \label{sec:discussion}

In the previous section, we described NIRSpec observations of galaxies in three separate associations of Ly$\alpha$ emitters in the EGS field at $z\gtrsim7$. In this section, we synthesize these new CEERS results and discuss implications for our understanding of the processes regulating Ly$\alpha$ visibility in the reionization era. 

We first focus on the five strongest $z\gtrsim7$ Ly$\alpha$ emitters (EW $>9$~\AA) in our CEERS spectroscopic sample (see Table~\ref{tab:lya_line})\footnote{This sample excludes the weaker Ly$\alpha$ emitting galaxy (CEERS-1029) identified in \citet{Larson2022}. In addition to showing lower EW Ly$\alpha$ emission ($4.2$~\AA), this galaxy also shows less extreme line ratios (O32 $=4.91$). This galaxy stands out in its exceptionally large velocity offset (see Section~\ref{sec:z8p7}), likely contributing to its visibility.}. This subset includes one of the highest redshift Ly$\alpha$ emitters (CEERS-1019; \citealt{Zitrin2015}), one of the largest EW Ly$\alpha$ emitters at $z\gtrsim7$ (CEERS-686; \citealt{Jung2022}), two newly-confirmed Ly$\alpha$ emitters (CEERS-1027, CEERS-44), and a previously-known Ly$\alpha$ emitting galaxy with strong rest-frame optical emission at $z=7.5$ (CEERS-698; \citealt{Roberts-Borsani2016,Stark2017}). The rest-frame optical line ratios of the $z\gtrsim7$ Ly$\alpha$ emitting galaxies are at the extreme end of the spectroscopic sample, with O32 values ($14.72$ to $>30$) that are only matched by nearby samples of LyC leakers \citep{Izotov2018,Flury2022}. The majority of Ly$\alpha$ emitting galaxies reported in \citet{Saxena2023b} also show similarly large O32 ratios (Fig.~\ref{fig:o32_r23}). For the subset of strong Ly$\alpha$ emitting galaxies with SED constraints, we find evidence for very large [O~{\small III}]+H$\beta$ EWs ($1768-3276$~\AA) and very young stellar populations ($3.4-10$~Myr of CSFH) likely formed in a recent burst or upturn in star formation. As a result of the young and hot stellar populations (and/or combined with AGN contributions; e.g., \citealt{Larson2023}), the ionizing photon production efficiency in these galaxies ($\log{[\xi_{\rm{ion}}/\rm{erg}^{-1}\rm{Hz}]}=25.7-26.0$) is $1.4-3\times$ the average value at $z\simeq7-8$ \citep{Endsley2023a,Prieto-Lyon2023}. Because ionizing photons are reprocessed into recombination lines, the large values of $\xi_{\rm{ion}}$ will translate into a higher-than-average production rate of Ly$\alpha$ photons (normalized to the SFR), likely contributing significantly to the visibility of Ly$\alpha$\footnote{A large $\xi_{\rm{ion}}$ can still contribute significantly to the Ly$\alpha$ visibility even if the galaxy is leaking a large fraction of LyC emission (e.g., $f_{\rm{esc,LyC}}\gtrsim0.2$). This is because the conditions producing a high $f_{\rm{esc}}$ (e.g., very young stellar populations) often produce a high $\xi_{\rm{ion}}$ and a very large intrinsic Ly$\alpha$ EW ($\simeq400$~\AA; e.g., \citealt{Saxena2023a}). Therefore, even if the observed Ly$\alpha$ EW is weakened due to the high $f_{\rm{esc}}$, its value can still be high \citep[e.g.,][]{Naidu2022}.} in these $z\gtrsim7.5$ systems.

\begin{figure}
\begin{center}
\includegraphics[width=\linewidth]{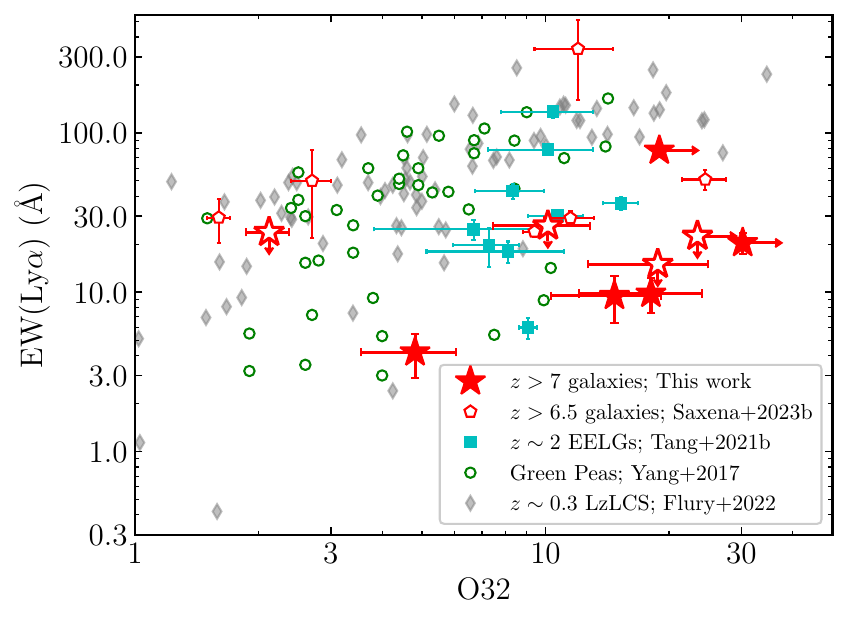}
\caption{Ly$\alpha$ EW as a function of O32 for CEERS NIRSpec galaxies at $z>7$ (red stars). Galaxies with Ly$\alpha$ emission detection are highlighted by solid stars. For those without Ly$\alpha$ emission detection, the upper limits are shown by open stars. We overplot data of $z>6.5$ Ly$\alpha$ emitters (open red pentagons) from \citet{Saxena2023b}, $z\sim2$ EELGs (cyan squares) from \citet{Tang2021b}, $z\sim0$ Green Peas (green open circles) from \citet{Yang2017}, and star-forming galaxies at $z\simeq0.2-0.4$ (grey diamonds) from LzLCS \citep{Flury2022} as a comparison. The $z>7$ galaxies tend to have on average lower Ly$\alpha$ EWs comparing to low redshift analogs at fixed O32.}
\label{fig:ewlya_o32}
\end{center}
\end{figure}

Lower redshift observations provide an additional baseline for understanding the expected Ly$\alpha$ strengths in galaxies with very large O32 ratios. In Fig.~\ref{fig:ewlya_o32}, we plot the Ly$\alpha$ EWs of three comparison samples. We consider $z\simeq2$ galaxies selected on [O~{\small III}] EWs \citep{Du2020,Tang2021b} and $z\simeq0.3$ galaxies from the Low-redshift Lyman Continuum Survey (LzLCS; \citealt{Flury2022}) and Green Pea (GP) database \citep{Cardamone2009,Yang2017}. The latter two Ly$\alpha$ samples were obtained with {\it HST}/COS spectroscopy, and the former was built with Keck and MMT spectroscopy. Each of the three lower redshift samples have rest-frame optical line ratios (i.e., O32) that are similar to those seen at $z\gtrsim7$, and they have [O~{\small III}] EWs indicating galaxies in the midst of strong bursts of star formation, also similar to our $z\gtrsim7$ galaxies. If we only include low redshift systems with O32 matched to the $z\gtrsim7$ Ly$\alpha$ emitters (O32 $>10$), we find intense Ly$\alpha$ emission becomes ubiquitous, with 26 of 28 showing Ly$\alpha$ EW $>30$~\AA. A further 22 of these ($79$~per~cent) have extremely strong Ly$\alpha$ with EW $>75$~\AA. The tendency for these galaxies to have strong Ly$\alpha$ is likely due in part to their efficient ionizing photon production, as we described above. 

\begin{figure}
\begin{center}
\includegraphics[width=\linewidth]{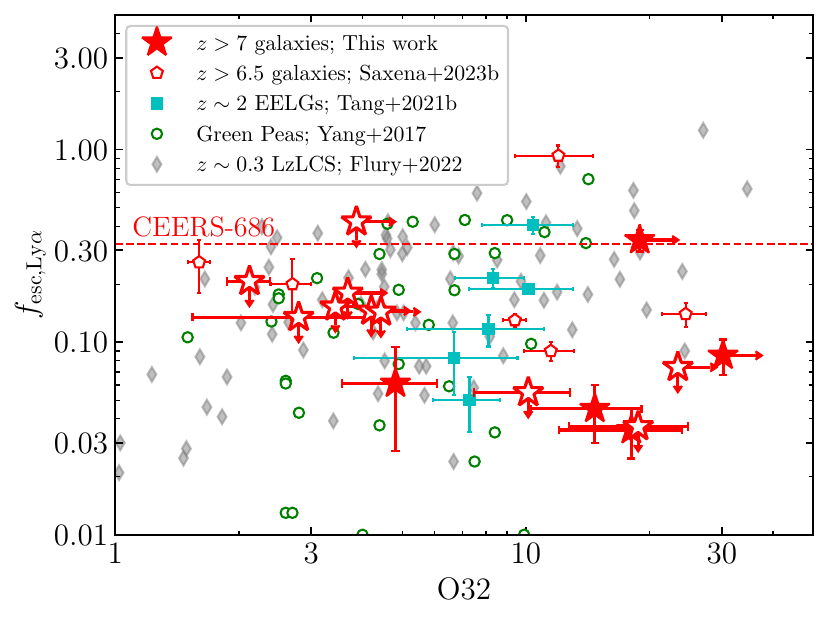}
\caption{Ly$\alpha$ escape fraction ($f_{\rm{esc,Ly}\alpha}$) versus O32 for CEERS NIRSpec galaxies at $z>7$ (red stars). Galaxies with Ly$\alpha$ emission detection are plotted as solid stars. For Ly$\alpha$ non-detected sources, upper limits are plotted as open stars. For CEERS-686 (i.e., z8\_13573 in \citealt{Jung2022}), which does not have [O~{\scriptsize II}] and hence an O32 measurement, its $f_{\rm{esc,Ly}\alpha}$ is shown by the red dashed line. We overplot data of $z>6.5$ Ly$\alpha$ emitters (open red pentagons) from \citet{Saxena2023b}, $z\sim2$ EELGs (cyan squares) from \citet{Tang2021b}, $z\sim0$ Green Peas (green open circles) from \citet{Yang2017}, and star-forming galaxies at $z\simeq0.2-0.4$ (grey diamonds) from LzLCS \citep{Flury2022} as a comparison. At fixed O32, $z>7$ galaxies have on average lower $f_{\rm{esc,Ly}\alpha}$ comparing to $z\sim0-3$ analogs.}
\label{fig:fesc_lya}
\end{center}
\end{figure}

But as we show in Fig.~\ref{fig:fesc_lya}, the low redshift large O32 galaxies also transmit a larger fraction of their Ly$\alpha$. The inferred Ly$\alpha$ escape fractions of the lower-$z$ comparison samples range from $0.09$ to $1$ for sources with O32 $>10$, with a median value of $0.29$. We also see large values if we just consider the $z\simeq2$ comparison sample with O32 $>6$ (median $f_{\rm{esc,Ly\alpha}}=0.15$), indicating that this is not a result limited to the $z\simeq0.3$ galaxies and is not only seen in the galaxies observed with the Cosmic Origins Spectrograph (COS) on {\it HST}. Based on these results, it seems likely that galaxies with recent upturns in star formation are often observed along low density channels that allow a larger fraction of Ly$\alpha$ to be transmitted. This may reflect the enhanced feedback that is associated with a recent burst \citep[e.g.,][]{Trebitsch2017,Rosdahl2018,Kimm2019,Barrow2020,Ma2020,Kakiichi2021}.

The CEERS spectroscopy presented in this paper has revealed that large O32 values ($>10$) are somewhat common at $z\gtrsim7$, particularly when sampling the very largest [O~{\small III}] EWs (see Fig.~\ref{fig:o32_o3hbew}). Based on the low redshift results described above, we expect to see strong Ly$\alpha$ emission in large O32 galaxies at $z\gtrsim7$, provided they are situated in ionized regions of the IGM where strong Ly$\alpha$ is effectively transmitted. If the bubbles are large enough and highly ionized, such that both resonant and damping wing absorption of Ly$\alpha$ is minimal \citep{Mason2020}, we may naively expect Ly$\alpha$ strengths to approach those seen locally, with a large percentage showing Ly$\alpha$ EW $>75$~\AA. While the $z\gtrsim7$ galaxies in the EGS stand out as having atypically strong Ly$\alpha$ for the reionization era, none are found with the intense Ly$\alpha$ line emission of lower redshift galaxies with identical rest-frame optical spectra. This offset is abundantly clear in Fig.~\ref{fig:ewlya_o32}, where the $z\gtrsim7$ Ly$\alpha$ emitters with O32 $>10$ are found shifted to lower Ly$\alpha$ EW (median $15$~\AA). The same trend is also seen in Fig.~\ref{fig:fesc_lya} where we see lower Ly$\alpha$ escape fractions at $z\gtrsim7$ compared to those at $z\simeq0-2$ with similar O32 values. We quantify the offset by comparing the Ly$\alpha$ EW distribution of galaxies at $z=0-2$ with O32 $>10$ to the Ly$\alpha$ EWs of $z>7$ LAEs that are in the three associations of LAEs discussed in Section~\ref{sec:lae} (CEERS-698 at $z=7.47$, CEERS-1027 at $z=7.82$, and CEERS-1019 at $z=8.68$). We focus on the associations, as these are likely to be ionized regions where Ly$\alpha$ transmission through the IGM is maximized. We compare the $z>7$ galaxies to the galaxies with O32 $>10$ at $z\sim0-2$ from LzLCS \citep{Flury2022}, GPs \citep{Yang2017}, and $z\sim2$ EELGs \citep{Tang2021b}. The median Ly$\alpha$ EW of $z=0-2$ galaxies with O32 $>10$ is $120$~\AA. In contrast, the $z>7$ galaxies with O32 $>10$ and Ly$\alpha$ detections in the three associations have Ly$\alpha$ EWs in the range $10-20$~\AA. We note that this does not include CEERS-686, a stronger Ly$\alpha$ emitter (EW $=41.9$~\AA) in the $z\simeq7.7$ group (see Section~\ref{sec:z7p7}) as it lacks a constraint on [O~{\small II}]. We will return to discuss this source later in the section. But overall, these results suggest that the Ly$\alpha$ emitters at $z\simeq7-9$ in the EGS field tend to be significantly weaker (in Ly$\alpha$ emission) than those with similar O32 indices in low redshift samples, with a typical downward shift of $6-12\times$ in the Ly$\alpha$ EW.

There are likely a variety of effects contributing to the absence of very strong Ly$\alpha$ emission in the $z\gtrsim7$ galaxies with large O32. Firstly, while we have only observed Ly$\alpha$ redshifted with respect to systemic, it is likely that some fraction of the photons may have been emitted in a blue peak, which would be resonantly absorbed by optically thick residual neutral gas in the ionized IGM \citep[e.g.,][]{Gunn1965,Mason2020}. In our comparison sample at $z\sim2$ about $30$~per~cent of the total Ly$\alpha$ flux is emitted in a blue peak, similar to Ly$\alpha$ emitter samples at $z\sim0-3$ which find $\simeq23$~per~cent of the flux is in a blue peak \citep{Hayes2021}. Thus, unless an unexpectedly large fraction of Ly$\alpha$ photons were emitted in a blue peak in our $z\gtrsim7$ sample, the resonant absorption should not be the dominant cause of the reduction in Ly$\alpha$ EW at $z\gtrsim7$. If the IGM at $z\gtrsim7$ is mostly neutral, depending on the size of the ionized region around the the three Ly$\alpha$ associations in the EGS (which span $5-10$~pMpc) we may expect significant attenuation from the damping wing of the surrounding IGM. 

We estimate the transmission of Ly$\alpha$ through the neutral IGM with different ionized region sizes at $z=7-9$. We first build an ``intrinsic'' Ly$\alpha$ profile (defined as the profile emerging from the ISM and CGM before being impacted by the neutral IGM) of $z\gtrsim7$ galaxies. We use the composite Ly$\alpha$ profile of EELGs at $z\simeq2-3$ in \citet{Tang2021b} as a baseline (Tang et al. in prep.), as this sample matches the large [O~{\small III}]+H$\beta$ EWs ($>1000$~\AA) and large O32 ratios ($>6$) seen in CEERS $z\gtrsim7$ galaxies but is situated at redshifts where the IGM is highly ionized and thus the impact of neutral IGM is minimal. We assume that the composite Ly$\alpha$ profile of $z\simeq2-3$ EELGs is the same as the Ly$\alpha$ profiles of $z>7$ galaxies before being attenuated by the neutral IGM. Then we compute the damping wing optical depth of Ly$\alpha$ at $z=7-9$ as a function of velocity offset from systemic \citep{Miralda-Escude1998} to the composite Ly$\alpha$ profile of $z\simeq2-3$ EELGs to estimate the IGM transmission. We consider a galaxy at $z=8$ that is situated in ionized regions with different radii, and we assume the IGM within the ionized region is completely ionized and outside is completely neutral. The results show that if the galaxy is situated in a relatively large ionized bubble ($R=1$~pMpc), about $30$~per~cent of the Ly$\alpha$ flux (emerging from the ISM and CGM) will transmit through the neutral IGM. If the galaxy is situated in smaller bubbles ($R=0.5$~pMpc and $0.1$~pMpc), the transmission of emerging Ly$\alpha$ flux will decrease to $20$~per~cent and $6$~per~cent, respectively. In order for neutral IGM to contribute significantly to the observed weak Ly$\alpha$ emission at $z\gtrsim7.5$ (a downward shift of $6-12\times$ in Ly$\alpha$ EW), the ionized regions surrounding the Ly$\alpha$ emitters must be smaller ($\lesssim0.5-1$~pMpc) than was previously thought. Such small bubbles may not be unexpected at such high redshifts where the IGM is likely significantly neutral \citep{Lu2023}. In this case, we would expect galaxies situated in between the two bright Ly$\alpha$ emitters to show much lower Ly$\alpha$ transmission. If the Ly$\alpha$ transmission is low in very small ionized bubbles, as suggested by the weak EWs, we can only see the reddest tail of the Ly$\alpha$ emission line which could help to explain why the Ly$\alpha$ velocity offsets of CEERS $z>7$ galaxies are large.

\begin{figure}
\begin{center}
\includegraphics[width=\linewidth]{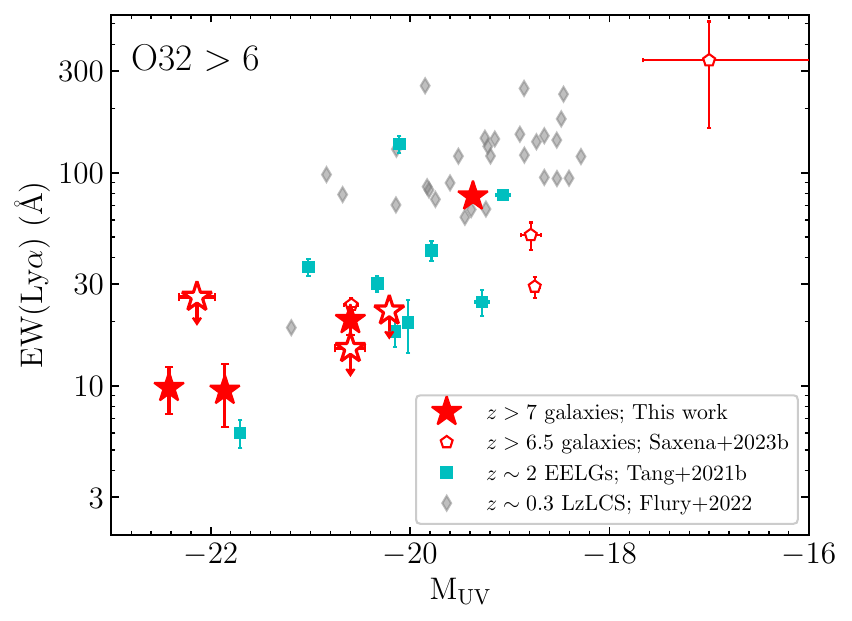}
\caption{Ly$\alpha$ EW as a function of M$_{\rm{UV}}$ for galaxies with large O32 ($>6$). Galaxies at $z\gtrsim7$ with Ly$\alpha$ emission detections are highlighted by solid stars. For those without Ly$\alpha$ emission detection, the upper limits are shown by open stars. We overplot data of $z>6.5$ Ly$\alpha$ emitters (open red pentagons) from \citet{Saxena2023b}, $z\sim2$ EELGs (cyan squares) from \citet{Tang2021b} and star-forming galaxies at $z\simeq0.2-0.4$ (grey diamonds) from LzLCS \citep{Flury2022} as a comparison. It is clear that current $z\gtrsim 7$ samples do not have the strongest Ly$\alpha$ emitters seen in lower redshift samples with similar M$_{\rm{UV}}$ and O32 except for the EW$_{\rm{Ly}\alpha}\simeq300$~\AA\ Ly$\alpha$ emitter JADES-GS-z7-LA (see also \citealt{Saxena2023a}).}
\label{fig:ewlya_muv}
\end{center}
\end{figure}

But other factors may be more important in driving down the Ly$\alpha$ EWs in the EGS field at $z\gtrsim7.5$. While the low redshift comparison samples are similar to the $z\gtrsim7$ galaxies in their rest-frame optical spectroscopic properties (i.e., O32, [O~{\small III}] EW), their UV luminosities are not comparable. The LzLCS galaxies have absolute magnitudes in the range of M$_{\rm{UV}}=-21.5$ to $-18.3$ with a median of M$_{\rm{UV}}=-19.9$. In contrast, the $z\gtrsim7$ galaxies in the Ly$\alpha$ associations are more luminous, spanning M$_{\rm{UV}}=-22.4$ to $-20.2$ with a median of $-21.1$. If galaxies with larger UV luminosities have more neutral hydrogen in their circumgalactic medium (CGM), we may expect more scattering of line photons within the galaxy, thereby decreasing the Ly$\alpha$ flux through the NIRSpec micro-shutter. This would also broaden and shift the lines redward, \citep[e.g.,][]{Verhamme2006,Yang2017}. To test the role of luminosity-dependent effects on the Ly$\alpha$ properties, we plot the Ly$\alpha$ EW as a function of absolute magnitude in Fig.~\ref{fig:ewlya_muv} for the $z\gtrsim7$ galaxies and the low redshift comparison samples. Several things stand out. Galaxies with the largest luminosities do tend to have weaker Ly$\alpha$ in both comparison and $z\gtrsim7$ samples. This trend is also seen in other $z\gtrsim7$ Ly$\alpha$ emitters (e.g., \citealt{Endsley2022b,Bunker2023,Saxena2023b}; see also Fig.~\ref{fig:lya_offset}) and in large Ly$\alpha$ emitter and Lyman break galaxy samples at lower redshifts \citep[e.g.,][]{Ouchi2008,Stark2010,Schaerer2011,Hashimoto2017,Oyarzun2017}. Yet at moderate (and lower) luminosities (M$_{\rm{UV}}\simeq-21$ to $-19$), there does appear to be a population of intense Ly$\alpha$ emitters (EW $>100$~\AA) that have yet to be observed at $z\gtrsim7$, even in regions where the IGM is thought to be ionized. If the bubbles in the EGS are very large, we may expect to see very intense Ly$\alpha$ once larger spectroscopic samples of lower luminosity galaxies are assembled. These sources should also present much lower velocity offsets than those of our current $z\gtrsim7$ sample (see Fig.~\ref{fig:lya_offset}).

Spectroscopic aperture effects must also be considered as a potential factor in the absence of strong Ly$\alpha$ at $z\gtrsim7$. The small area of the micro-shutters ($0.20''\times0.46''$) could potentially capture a smaller fraction of the resonantly scattered line emission. However we have demonstrated consistency between the NIRSpec and Keck results (with the only exception being CEERS-1019 where there is significant skyline contamination), so based on existing data, it does not appear that the small size of the micro-shutters is leading to significant losses relative to slit-based spectroscopic surveys. However it is conceivable that the {\it HST}/COS aperture may pick up more of the resonantly scattered Ly$\alpha$ emission from the $z\simeq0.3$ LzLCS sources. Future investigation with the integral field mode of NIRSpec or NIRISS is needed to help clarify the impact of the small size of the micro-shutters on Ly$\alpha$ properties.

Finally we consider two exceptions to the discussion above. CEERS-686 presents relatively strong Ly$\alpha$ (EW $=41.9$~\AA) and a Ly$\alpha$ escape fraction ($f_{\rm{esc,Ly\alpha}}=0.325$) that is very similar to what is seen in lower redshift Ly$\alpha$ emitters. This relatively faint galaxy (M$_{\rm{UV}}=-20.7$) is located $0.6$~pMpc behind the luminous galaxy EGS-zs8-1 (see Fig.~\ref{fig:distribution} and \S3.2). If EGS-zs8-1 is at the center of a moderately small ionized region ($\simeq1$~pMpc in size), the transmission of CEERS-686 will be enhanced owing to its position on the far side of the bubble (see discussion in \citealt{Jung2022}). The other exception to this discussion is CEERS-44, the strongest Ly$\alpha$ emitter in the NIRSpec sample (EW $=77.6$~\AA) at $z=7.10$. This galaxy is not in one of the known Ly$\alpha$ associations in the EGS. It is UV-faint (M$_{\rm{UV}}=-19.4$) with a very large O32 value ($>18.93$) and strong [O~{\small III}]+H$\beta$ emission is evident in its NIRCam SED (EW $=1786$~\AA). How Ly$\alpha$ is transmitted so readily in this galaxy is not immediately clear. There is no useful velocity offset constraint for this galaxy, as it was only observed with the low resolution prism. Future observations of the field are required to reveal whether there may be additional sources with similarly large Ly$\alpha$ transmission at this redshift.

Progress toward more quantitative estimates of the bubble sizes will require spectroscopic observations of the more numerous population of faint galaxies that span the distance between the known Ly$\alpha$ emitters in the EGS field. If the bubbles are very large (i.e., $>1$~pMpc) and the impact of the IGM on Ly$\alpha$ is minimal, we should expect to see very intense Ly$\alpha$ emission in a significant subset of faint galaxies. Such strong Ly$\alpha$ emitters should also present lower velocity offsets ($\lesssim200$~km~s${^{-1}}$; e.g., \citealt{Saxena2023a}) than have been observed in this field at $z\gtrsim7$ so far (Fig. ~\ref{fig:lya_offset}). Failure to detect faint galaxies with large Ly$\alpha$ EWs and small velocity offsets would give indications that the IGM is significantly impacting the Ly$\alpha$ transmission in these ionized regions. While current Ly$\alpha$ samples in this field lack faint galaxies (see Fig.~\ref{fig:ewlya_muv}), the availability of deep NIRCam photometry over much of the EGS field will allow fainter galaxies to be efficiently targeted with future spectroscopic campaigns with NIRSpec.


\section{Summary} \label{sec:summary}

We discuss new {\it JWST}/NIRSpec observations of $z\gtrsim7$ galaxies in the EGS field taken as part of the CEERS ERS program. This field has previously been shown to host three associations of Ly$\alpha$ emitting galaxies at redshifts ($z=7.5$, $7.7$ and $8.7$) where the IGM is likely substantially neutral. In this work, we have detected rest-frame UV to optical emission lines and confirmed redshifts of 21 galaxies at $z\gtrsim7$. Ly$\alpha$ emission lines have been detected for six galaxies in our sample, including four previously confirmed Ly$\alpha$ emitters \citep{Zitrin2015,Roberts-Borsani2016,Stark2017,Jung2022,Larson2022} and two newly detected Ly$\alpha$ emitters at $z>7$. Three (CEERS-1027, CEERS-686, CEERS-44) of these six galaxies present relatively high S/N ($>7$) Ly$\alpha$ detections and the other three (CEERS-1019, CEERS-1029, CEERS-698) show lower S/N ($\simeq3-4$) Ly$\alpha$ emission. By fitting the SEDs of a subset (12 out of 21) of our $z\gtrsim7$ galaxies with {\it JWST}/NIRCam or high S/N {\it HST}+{\it Spitzer} photometry, we demonstrate that these galaxies have large [O~{\small III}]+H$\beta$ EWs ($=879-3276$~\AA) that are more extreme than the average values of $z\sim7-9$ galaxies ($\simeq700-800$~\AA; e.g., \citealt{Labbe2013,DeBarros2019,Endsley2021a,Endsley2023a}). Using available emission line ratios, we discuss the properties of the ionizing spectra and nebular gas of galaxies in our $z\gtrsim7$ NIRSpec sample. We summarize our results below.

1. We present the O32 and Ne3O2 ratios of $z\gtrsim7$ CEERS NIRSpec galaxies, providing insight into the ionization state of the nebular gas. The $z\gtrsim7$ galaxies show that average O32 and Ne3O2 (O32 $=17.84$ and Ne3O2 $=0.89$ measured from the composite spectrum) are consistent with other {\it JWST} measurements at $z\gtrsim7$ in literature and $\sim10\times$ larger than typical star-forming galaxies at $z\sim2$. The O32 versus [O~{\small III}]+H$\beta$ EW of CEERS $z\gtrsim7$ galaxies is consistent with the the relation derived at lower redshifts, with the largest O32 values ($>10$) associated with the largest [O~{\small III}]+H$\beta$ EWs ($\gtrsim1500$~\AA). Our sample is biased toward strong [O~{\small III}] emission and therefore the large O32 ratios may not be representative of the full population. The Ly$\alpha$ emitters (EW$_{\rm{Ly}\alpha}>9$~\AA) in our sample show even larger O32 ($>14$) values than the non-LAEs, suggesting even more extreme ionization conditions in this population.

2. We interpret the ionized gas properties of our CEERS $z\gtrsim7$ spectroscopic sample by fitting rest-frame optical emission lines with the \textsc{beagle} tool. The model fits suggest high ionization parameters ($\log{U}=-2.59$ to $-1.31$ with a median of $-2.11$) and low gas-phase oxygen abundance ($12+\log{(\rm{O/H})}=7.45-8.22$ with a median of $7.84$). For the small subset galaxies with tentative auroral [O~{\small III}]$\lambda4363$ line detections, we derive very low gas-phase metallicities ($12+\log{(\rm{O/H})}=7.6-7.9$) that are consistent with the values inferred from photoionzation models. Deeper spectra will allow investigation of whether these properties extend to galaxies with lower [O~{\small III}] EW which are absent from the current spectroscopic sample.

3. We detect C~{\small III}] emission in three of the galaxies in the $z\gtrsim7$ CEERS NIRSpec sample. Each of these galaxies also shows very large O32 values ($>18$), consistent with the trend between C~{\small III}] EW and O32 found in lower redshift galaxies. We detect high ionization emission lines (C~{\small IV}, He~{\small II}) in one of our galaxies (CEERS-1027), indicating a very hard ionizing spectrum. The UV line ratios in this system are consistent with expectations for metal poor massive stars.

4. The CEERS NIRSpec sample at $z\gtrsim7$ contains 10 galaxies in the 3 Ly$\alpha$ emitter groups (at $z=7.5$, $7.7$, and $8.7$) in the EGS field. We identify one new Ly$\alpha$ emitter (CEERS-1027) situated $5$~pMpc from the brightest galaxy in the $z\simeq7.7$ association and demonstrate that not all galaxies at these redshifts show Ly$\alpha$. The strong Ly$\alpha$ emitters (EW $>9$~\AA) have among the largest O32 ratios in the entire sample ($14$ to $>30$) suggesting very high ionization parameters. Their SEDs suggest very large [O~{\small III}]+H$\beta$ EWs ($>1700$~\AA) indicating very young stellar populations ($<10$~Myr assuming CSFH) and efficient ionizing photon production efficiency ($\log{[\xi_{\rm{ion}}/\rm{erg}^{-1}\rm{Hz}]}=25.7-26.0$), potentially boosting the Ly$\alpha$ visibility.

5. The $z\gtrsim7$ CEERS NIRSpec galaxies present relatively low Ly$\alpha$ escape fractions and large Ly$\alpha$ velocity offsets. By comparing the Ly$\alpha$ flux to H$\beta$ flux, we 
derive Ly$\alpha$ escape fraction ranging from $0.035$ to $0.085$, suggesting that a very small fraction of the Ly$\alpha$ emerges through the NIRSpec micro-shutter. These values are lower than the typical $f_{\rm{esc,Ly}\alpha}$ of galaxies at lower redshifts with similarly large O32. The Ly$\alpha$ velocity offsets of Ly$\alpha$ emitting galaxies in our $z\gtrsim7$ sample range from $323$~km~s$^{-1}$ to $1938$~km~s$^{-1}$. Such large velocity offsets may help boost the transmission of Ly$\alpha$ through the IGM, contributing to their visibility.

6. We find that the Ly$\alpha$ emission in $z\gtrsim7$ galaxies is much weaker than the Ly$\alpha$ of lower redshift galaxies with similar rest-frame optical spectral properties. The Ly$\alpha$ EWs ($\simeq10-20$~\AA) of Ly$\alpha$ emitters in the 3 LAE associations at $z\simeq7.5-8.7$ is $6-12\times$ lower than the median Ly$\alpha$ EW ($=120$~\AA) of galaxies with similarly large O32 ($>10$) at $z\sim0-2$. Several effects likely contribute to this downturn of Ly$\alpha$ strength. If the ionized bubbles around the Ly$\alpha$ emitters at $z\gtrsim7$ are relatively small ($\lesssim1$~pMpc), we may expect a decrease of Ly$\alpha$ transmission even in ionized regions. Our $z\gtrsim7$ Ly$\alpha$ emitters are also on average $\sim1$~mag brighter than $z\sim0-2$ analogs. If brighter galaxies have more H~{\small I} in their CGM, the Ly$\alpha$ photons may also be more significantly scattered within galaxies, leading to weaker Ly$\alpha$ emission. Finally, the small area covered by NIRSpec MSA micro-shutter ($0.20''\times0.46''$) may capture smaller fraction of the resonantly scattered Ly$\alpha$ line emission than lower redshift galaxies, such as LzLCS sample which uses {\it HST}/COS. Future observations of Ly$\alpha$ in fainter galaxies that span these fields should enable more quantitative constraints on the bubble sizes and associated impact of IGM on Ly$\alpha$ transmission.


\section*{Acknowledgements}

The authors thank the anonymous referee for insightful comments. We thank the entire CEERS team for their effort designing and executing this program. MT acknowledges funding from the {\it JWST} Arizona/Steward Postdoc in Early galaxies and Reionization (JASPER) Scholar contract at the University of Arizona. DPS acknowledges support from the National Science Foundation through the grant AST-2109066. CAM and TYL acknowledge support by the VILLUM FONDEN under grant 37459. The Cosmic Dawn Center (DAWN) is funded by the Danish National Research Foundation under grant DNRF140. PS was generously supported by a Carnegie Fellowship through the Carnegie Observatories during the completion of this manuscript. LW acknowledges support from the National Science Foundation Graduate Research Fellowship under Grant No. DGE-2137419. The authors sincerely thank Ramesh Mainali for providing data of C~{\small III}] versus O32 relation at $z\sim0-3$, John Chisholm for providing data of LzLCS sample, and Jacopo Chevallard for providing access to the \textsc{beagle} tool used for SED fitting analysis.

This work is based on observations made with the NASA/ESA/CSA {\it James Webb Space Telescope}. The data were obtained from the Mikulski Archive for Space Telescopes at the Space Telescope Science Institute, which is operated by the Association of Universities for Research in Astronomy, Inc., under NASA contract NAS 5-03127 for {\it JWST}. These observations are associated with Early Release Science program 1345. Part of this work is based on observations taken by the 3D-HST Treasury Program (GO 12177 and 12328) with the NASA/ESA {\it Hubble Space Telescope} obtained from the Space Telescope Science Institute, which is operated by the Association of Universities for Research in Astronomy, Inc., under NASA contract NAS 5-26555. This work is based in part upon High Performance Computing (HPC) resources supported by the University of Arizona Technology and Research Initiative Fund (TRIF), University Information Technology Services (UITS), and Research, Innovation, and Impact (RII) and maintained by the UArizona Research Technologies department. 

This research made use of the following software: \textsc{astropy}, a community-developed core \textsc{python} package for Astronomy \citep{AstropyCollaboration2013,AstropyCollaboration2018,AstropyCollaboration2022}; \textsc{numpy} \citep{Harris2020}; \textsc{scipy} \citep{Virtanen2020}; \textsc{matplotlib} \citep{Hunter2007}; \textsc{source extractor} \citep{Bertin1996} via \textsc{sep} \citep{Barbary2016}; \textsc{beagle} \citep{Chevallard2016}; and \textsc{prospector} \citep{Johnson2021}.


\section*{Data Availability}

The {\it JWST}/NIRSpec spectra and NIRCam imaging data as well as {\it HST} images used in this work are available on the Mikulski Archive for Space Telescopes (\url{https://mast.stsci.edu/}). Other data underlying this article will be shared on reasonable request to the corresponding author.



\bibliographystyle{mnras}
\bibliography{ceers_lae} 


\appendix

\section{{\it JWST}/NIRSpec spectra of galaxies} \label{sec:appendix}

\setcounter{figure}{0}
\renewcommand{\thefigure}{A\arabic{figure}}

\begin{figure}
\begin{center}
\includegraphics[width=\linewidth]{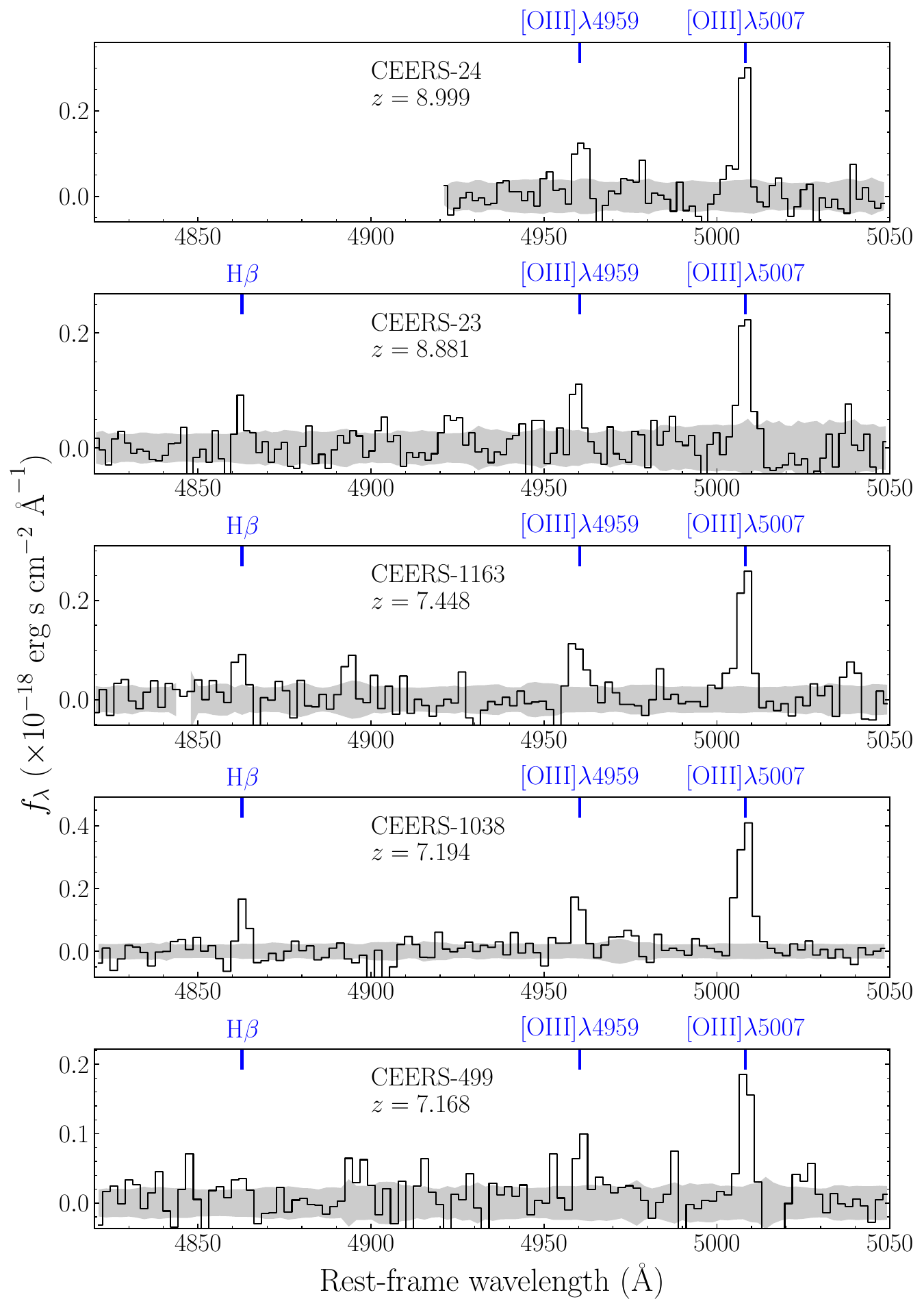}
\caption{Medium resolution ($R\sim1000$) NIRSpec 1D spectra of $z\gtrsim7$ galaxies with three or less detected emission lines at rest-frame optical ([O~{\small III}], H$\beta$; five objects). The spectra have been shifted to the rest frame. The grey shaded regions show the uncertainty.}
\label{fig:spectra_opt_1d_less}
\end{center}
\end{figure}

\begin{figure}
\begin{center}
\includegraphics[width=\linewidth]{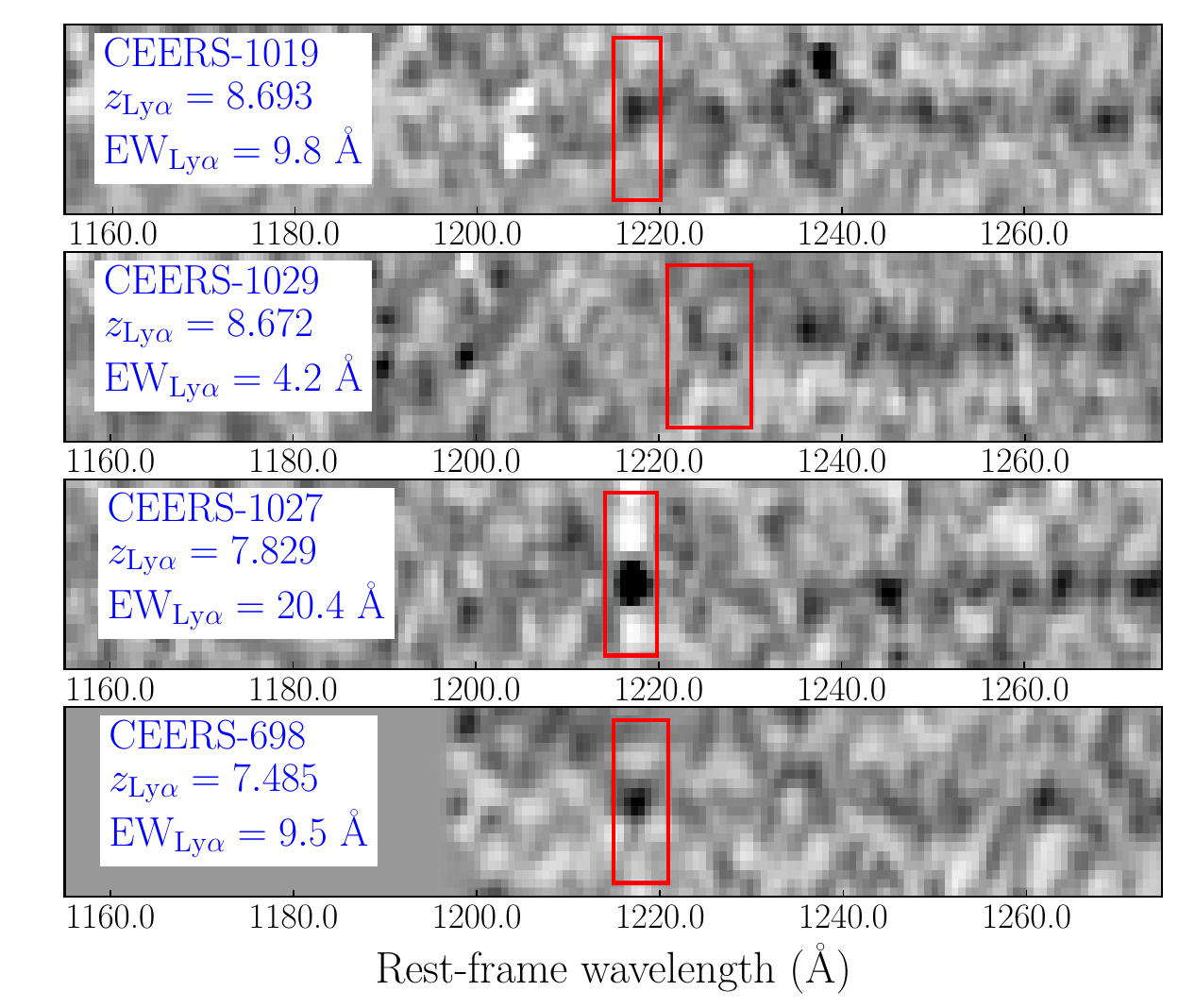}
\caption{2D NIRSpec spectra (medium resolution $R\sim1000$) of Ly$\alpha$ emission lines of $z\gtrsim7$ galaxies (four objects, see Table~\ref{tab:lya_line}). Lines are marked by red rectangles, with black-white showing positive-negative features.}
\label{fig:spectra_lya_2d}
\end{center}
\end{figure}

\begin{figure*}
\begin{center}
\includegraphics[width=\linewidth]{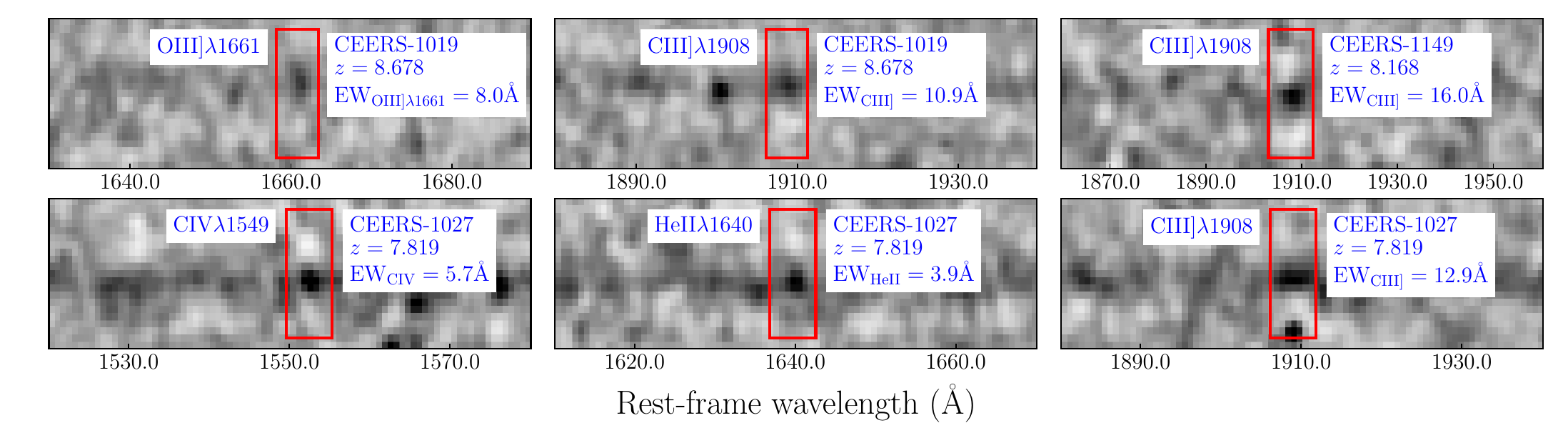}
\caption{2D NIRSpec spectra (medium resolution $R\sim1000$) of detected rest-frame UV emission lines of $z\gtrsim7$ galaxies. Lines are marked by red rectangles, with black-white showing positive-negative features. Unresolved C~{\scriptsize III}]~$\lambda1908$ emission lines have been detected in three galaxies: CEERS-1019, CEERS-1149, and CEERS-1027. Unresolved C~{\scriptsize IV}~$\lambda1549$ and He~{\scriptsize II}~$\lambda1640$ lines have also been detected in CEERS-1027, and O~{\scriptsize III}~$\lambda1661$ have been detected in CEERS-1019.}
\label{fig:spectra_uv_2d}
\end{center}
\end{figure*}


\bsp	
\label{lastpage}
\end{document}